\advance\hoffset by 5mm
\newif\ifPDFLaTeX
\expandafter\ifx\csname pdfoutput\endcsname\relax\else\PDFLaTeXtrue\fi
\documentclass[epj]{svjour}
\let\makeheadbox\relax
\ifPDFLaTeX
  \usepackage{type1cm}
  \usepackage[pdftex]{feynmp}
  \usepackage[pdftex,colorlinks]{hyperref} 
  \newcommand{\hepcite}[1]{\href{http://xxx.lanl.gov/abs/#1}{[arXiv:#1]}}
\else
  \usepackage{feynmp}
  \usepackage{hyperref} 
  \newcommand{\hepcite}[1]{[arXiv:#1]}
\fi
\usepackage{thophys}
\usepackage{array,amsmath,amssymb}
\allowdisplaybreaks
\DeclareMathOperator{\tr}{tr}
\newcommand{\dd}{\mathrm{d}}
\newcommand{\ii}{\mathrm{i}}
\newcommand{\ee}{\mathrm{e}}
\begin{document}
\begin{fmffile}{fghpics}
\fmfset{arrow_len}{2mm}
\title{Forests, Groves and Higgs Bosons}
\subtitle{Gauge Invariance Classes in Spontaneously Broken Gauge Theories}
\author{%
  T.~Ohl\thanks{\email{ohl@physik.uni-wuerzburg.de}}\inst{,1}\and
  C.~Schwinn\thanks{\email{schwinn@physik.uni-wuerzburg.de}}\inst{,1,2}}
\institute{Institut f\"ur Theoretische~Physik und Astrophysik,
  Universit\"at~W\"urzburg, Am~Hubland, D-97074~W\"urzburg, Germany
  \and
  Institut f\"ur Kernphysik, Darmstadt University of Technology,
  Schlo\ss{}gartenstr.~9 D-64289~Darmstadt, Germany}
\date{May 2003 (revised July 2003)}
\headnote{WUE-ITP-2003-004(rev), IKDA-2003-05(rev), hep-ph/0305334(rev)}
\abstract{%
  We determine the gauge invariance classes of tree level Feynman
  diagrams in spontaneously broken gauge theories, providing a proof
  for the formalism of gauge and flavor flips. We find new gauge
  invariance classes in theories with a nonlinearly realized scalar
  sector.  In unitarity gauge, the same gauge invariance classes
  correspond to a decomposition of the scattering amplitude into
  pieces that satisfy the relevant Ward Identities individually.
  In theories with a linearly realized scalar sector in $R_\xi$
  gauge, no additional nontrivial gauge invariance classes exist
  compared to the unbroken case.
  }
\maketitle
\section{Introduction}
The agreement of theoretical predictions derived from the Standard
Model~(SM) of
electroweak interactions and experiment has been established to an
impressive degree. The only missing ingredient is the Higgs boson that
has yet to be discovered. The electroweak SM can nevertheless only be
a low energy approximation to a more fundamental theory that should
become visible at TeV scale energies.  Indications on the nature of
the underlying 
theory should be found by experiments at the LHC or a future linear 
collider (see e.\,g.~\cite{ColliderStudies}). Channels with many
tagged particles open up at these experiments and challenge 
theorists to make precise predictions for processes with many
particles in the final state.

Assuming a Higgs boson will be found in
future experiments, determining its quantum numbers and couplings will
require the study of processes with many fermions in the final
state~\cite{ColliderStudies}.  Examples are the measurement of the
triple Higgs self coupling~\cite{Djouadi:1999} and of
the top-Higgs Yukawa coupling~\cite{TopHiggs} via associated top-Higgs
production.  In the latter example, there are five tree level diagrams
contributing to the signal process $e^+e^-\to t \bar t H$, while
almost forty-thousand diagrams contribute to the corresponding
observable eight-fermion final states like $e^+e^-\to b
\mu^+\nu_\mu \bar b d \bar u b \bar b $.
Existing calculations of such irreducible
backgrounds~\cite{Moretti:1999kx}
classify the contributing diagrams according to
their topology in order to perform the phase space integration,
arriving at a gauge invariant result only after the resulting
integrals are added up.

To disentangle signal and background diagrams in a gauge invariant
way, it is desirable to find separately gauge invariant subsets of
Feynman diagrams~(FDs), so called `gauge invariance classes'~(GICs). After a
classification of GICs in four fermion production
processes~\cite{Bardin:1994}, a systematic procedure to construct
minimal GICs or `groves' of tree level diagrams using the formalism of
`gauge and flavor flips' has been found in~\cite{Boos:1999}.  However,
a detailed discussion of GICs involving Higgs bosons has not yet been
given.

In this work, we will clarify the role of Higgs bosons in the GICs. We
provide a proof of the formalism of~\cite{Boos:1999} for Spontaneously
Broken Gauge Theories~(SBGTs), based on a
diagrammatic analysis of the Slavnov Taylor Identities~(STIs). We will
find that the brief discussion in~\cite{Boos:1999} is justified for
a linear realization of the symmetries of the SM and no new
nontrivial groves arise compared to an unbroken gauge theory. However,
additional groves appear in the case of nonlinearly realized
symmetries~\cite{Coleman:1969}. These groves are also consistent in
unitarity gauge, i.\,e.~the corresponding amplitudes satisfy the Ward
Identities (WIs).

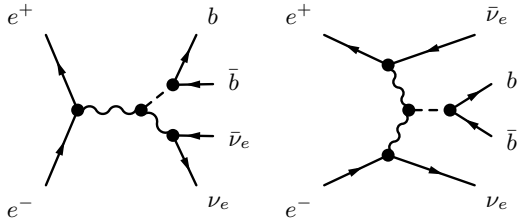
\begin{figure}[htbp]
\begin{equation*}
\fmfframe(6,7)(6,6){%
  \begin{fmfgraph*}(25,20)
    \fmfleft{i1,i2}
    \fmfright{o4,o5,o3,o6}
  \fmflabel{$e^-$}{i1}
  \fmf{fermion}{i1,v1}
  \fmfdot{v1}
  \fmflabel{$e^+$}{i2}
  \fmfdot{v1}
  \fmf{fermion}{v1,i2}
  \fmf{photon}{v5,v1}
  \fmfdot{v1,v5}
  \fmf{dashes}{v5,v20}
  \fmfdot{v5,v20}
  \fmflabel{$b$}{o6}
  \fmfdot{v20}
  \fmf{fermion,tension=0.5}{v20,o6}
  \fmflabel{$\bar b$}{o3}
  \fmfdot{v20}
  \fmf{fermion,tension=0.5}{o3,v20}
  \fmf{photon}{v21,v5}
  \fmfdot{v5,v21}
  \fmflabel{$\bar \nu_e$}{o5}
  \fmfdot{v21}
  \fmf{fermion,tension=0.5}{o5,v21}
  \fmflabel{$\nu_e$}{o4}
  \fmfdot{v21}
  \fmf{fermion,tension=0.5}{v21,o4}
  \end{fmfgraph*}}
\fmfframe(6,7)(6,6){%
  \begin{fmfgraph*}(25,20)
    \fmfleft{i1,i2}
    \fmfright{o4,o3,o6,o5}
  \fmflabel{$e^-$}{i1}
  \fmf{fermion}{i1,v1}
  \fmfdot{v1}
  \fmf{photon}{v4,v1}
  \fmfdot{v1,v4}
  \fmf{photon}{v16,v4}
  \fmfdot{v4,v16}
  \fmflabel{$e^+$}{i2}
  \fmfdot{v16}
  \fmf{fermion}{v16,i2}
  \fmflabel{$\bar \nu_e$}{o5}
  \fmfdot{v16}
  \fmf{fermion,tension=0.5}{o5,v16}
  \fmf{dashes}{v4,v17}
  \fmfdot{v4,v17}
  \fmflabel{$b$}{o6}
  \fmfdot{v17}
  \fmf{fermion,tension=0.5}{v17,o6}
  \fmflabel{${\bar b}$}{o3}
  \fmfdot{v17}
  \fmf{fermion,tension=0.5}{o3,v17}
  \fmflabel{$\nu_e$}{o4}
  \fmfdot{v1}
  \fmf{fermion,tension=0.5}{v1,o4}
  \end{fmfgraph*}}
\end{equation*}
\caption{\label{fig:higgs-strahlung}%
  Higgsstrahlung and Vector boson fusion}
\end{figure}

As an example, consider Higgs production via the two diagrams shown
in figure~\ref{fig:higgs-strahlung}. According to our results, the
Higgsstrahlung and the vector boson fusion diagram belong to different
GICs, both in the linear and the nonlinear representation. In a
nonlinear representation, however, both diagrams are gauge invariant
\emph{by themselves} (provided the electrons are taken as massless)
while in the linear representation they are part of larger GICs
including gauge boson exchange diagrams.

In section~\ref{sec:groves} we review GICs in unbroken gauge theories,
sketch the formalism of gauge and flavor flips and present the correct
flips for SBGTs. A summary of
our graphical notation for STIs is given in
section~\ref{sec:sti-graph} before we present the diagrammatic
derivation of GICs in section~\ref{sec:gi-classes}. The correct
definition of the gauge flips in SBGTs is discussed in
section~\ref{sec:sti-flips}, both for the linear and the nonlinear
realization. The structure of the GICs in SBGTs is analyzed in
section~\ref{sec:ssb-gics}.

\section{Gauge invariance classes and flips}
\label{sec:groves}
Before we turn to a formal derivation of the formalism from the
underlying STIs, we will briefly review existing
results for GICs in unbroken gauge theories and give the correct form
of the gauge flips in SBGTs.

Physical scattering amplitudes in gauge theories  satisfy the simple WI 
\begin{subequations}
\label{eq:gf-wi}
\begin{equation}
-\ii k_\mu\mathcal{M}^\mu(\text{in}+A\to\text{out})
=0
\end{equation}
in unbroken gauge theories and 
\begin{equation}
-\ii k_\mu\mathcal{M}^\mu(\text{in}+W\to\text{out})
=m_w\mathcal{M}(\text{in}+\phi\to\text{out}) 
\end{equation}
\end{subequations}
in SBGTs. For the application to tree level diagrams, it is a
sufficient requirement to define GICs as (minimal) subsets of FDs
that satisfy the WIs and are independent of the gauge fixing
parameters. Keeping in mind a future application to loop diagrams, we
will go further in our proof presented in section~\ref{sec:gi-classes}
and demand that the GICs satisfy the appropriate STIs when some of the
external particles are off their mass shell. However, the simpler
definition in terms of WIs suffices for the present introductory discussion.

\subsection{Gauge invariance classes in unbroken gauge theories}
\label{sec:qcd-groves}
As a trivial first example, consider the process $u\bar u \to
u\bar u$ in  QCD. Here a $s$- and a $t$-channel  diagram contribute:
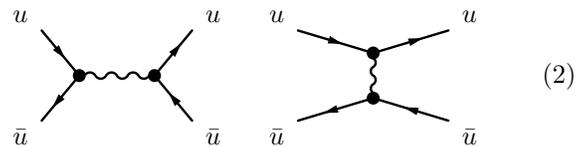
\begin{equation}
\label{eq:bhabbha}
  \parbox{20\unitlength}{
    \begin{fmfgraph*}(20,15)
      \fmftopn{t}{2}
      \fmfbottomn{b}{2}
      \fmflabel{$u$}{t1}
      \fmflabel{$u$}{t2}
      \fmflabel{$\bar u$}{b1}
      \fmflabel{$\bar u$}{b2}
      \fmf{fermion}{t1,t,b1}
      \fmf{photon}{v,t}
      \fmf{fermion}{b2,v,t2}
      \fmfdot{v,t}
    \end{fmfgraph*}}
  \qquad\qquad
  \parbox{20\unitlength}{
    \begin{fmfgraph*}(20,15)
      \fmftopn{t}{2}
      \fmfbottomn{b}{2}
      \fmflabel{$u$}{t1}
      \fmflabel{$u$}{t2}
      \fmflabel{$\bar u$}{b1}
      \fmflabel{$\bar u$}{b2}
      \fmf{fermion}{t1,t,t2}
      \fmf{photon}{v,t}
      \fmf{fermion}{b2,v,b1}
      \fmfdot{v,t}
    \end{fmfgraph*}}
\end{equation}
Both diagrams are separately gauge invariant (i.\,e.~are independent
of the gauge parameter in the gluon propagator), as can be seen
without calculation from the following observation: in the
process $u \bar u\to c \bar c$ where both fermion pairs belong to
different families, only the $s$-channel diagram appears, while in the
case of $u \bar c\to u \bar c$ only the $t$-channel diagram
appears. Since the scattering amplitudes for physical processes are gauge
invariant, both diagrams in~(\ref{eq:bhabbha}) must also be gauge
invariant by themselves.

This is a simple example of a `flavor selection rule'.  The separate
gauge invariance can be of course verified easily by an explicit
calculation, but it was shown in~\cite{Boos:1999} how the flavor
selection rule argument carries over to more complicated situations.

The argument leading to the flavor selection rules does not depend on
the existence of different flavors of quarks in the SM, since one can
always introduce fictitious additional generations, leading to a
conserved quantum number that has to be conserved along quark lines
passing through the diagrams. This allows to extend the formalism to
diagrams with an arbitrary number of external fermions.

As an application to a five point function, consider the amplitude for
the process $\bar qq \to \bar q q g$. Because of the flavor selection
rules, it contains two GICs, resulting from the insertion of the gluon
into the $s$- and $t$-channel diagrams for the process $\bar qq \to
\bar q q$. The GIC obtained by inserting the gluon in the
$s$-channel diagram is
\begin{equation}
\label{eq:qcd-grove}
G_s= \left\{
\begin{aligned}&\parbox{20\unitlength}{
 \begin{fmfgraph}(20,15)
\fmftopn{t}{2}
\fmfbottomn{b}{2}
\fmfright{r}
\fmf{fermion}{t1,t}
\fmf{fermion}{t,b1}
\fmf{photon}{v2,t}
\fmf{fermion,tension=2}{v1,t2}
\fmf{plain,tension=2}{v1,v2}
\fmf{fermion}{b2,v2}
\fmffreeze
\fmf{photon}{r,v1}
\fmfdot{v1,v2,t}
\end{fmfgraph}}\, ,\, 
\parbox{20\unitlength}{
\begin{fmfgraph}(20,15)
\fmftopn{t}{2}
\fmfbottomn{b}{2}
\fmfright{r}
\fmf{fermion}{t1,t,b1}
\fmf{photon}{v1,t}
\fmf{fermion}{v1,t2}
\fmf{plain,tension=2}{v1,v2}
\fmf{fermion,tension=2}{b2,v2}
\fmffreeze
\fmf{photon}{r,v2}
\fmfdot{v1,v2,t}
\end{fmfgraph}}\,,\, 
\parbox{20\unitlength}{
\begin{fmfgraph}(20,15)
\fmfbottomn{b}{2}
\fmftopn{t}{3}
\fmf{fermion}{b2,t,t3}
\fmf{photon}{v2,t}
\fmf{fermion,tension=2}{t1,v1}
\fmf{plain,tension=2}{v1,v2}
\fmf{fermion}{v2,b1}
\fmffreeze
\fmf{photon}{t2,v1}
\fmfdot{v1,v2,t}
\end{fmfgraph}}\,,\\
&\parbox{20\unitlength}{
\begin{fmfgraph}(20,15)
\fmftopn{t}{2}
\fmfbottomn{b}{3}
\fmfright{r}
\fmf{fermion}{b3,t,t2}
\fmf{photon}{v1,t}
\fmf{fermion}{t1,v1}
\fmf{plain,tension=2}{v1,v2}
\fmf{fermion,tension=2}{v2,b1}
\fmffreeze
\fmf{photon}{b2,v2}
\fmfdot{v1,v2,t}
\end{fmfgraph}}\, , \,
\parbox{20\unitlength}{
\begin{fmfgraph}(20,15)
\fmftopn{t}{3}
\fmfbottomn{b}{2}
\fmfright{r}
\fmf{fermion}{b2,t,t3}
\fmf{photon,tension=2}{v1,v2}
\fmf{photon,tension=2}{v2,t}
\fmf{fermion}{t1,v1,b1}
\fmffreeze
\fmf{photon}{t2,v2}
\fmfdot{v1,v2,t}
\end{fmfgraph}}
\end{aligned}\right\}
\end{equation}

The situation simplifies further in QED, because there is no triple photon
vertex. Indeed, it is well known in QED that the expression obtained from a
given FD by summing over all possible insertions of  a photon
along a charge carrying fermion line going through the diagram,
satisfies the WI by itself. Thus the amplitude for $e^+ e^- \to \mu^+
\mu^-\gamma$ can be separated further into two gauge invariant subsets:
\begin{equation}
\label{eq:qed-groves}
\begin{aligned}
G_s^{\text{FSR}}&= \left\{\parbox{20\unitlength}{
\begin{fmfgraph}(20,15)
\fmftopn{t}{2}
\fmfbottomn{b}{2}
\fmfright{r}
\fmf{fermion}{t1,t}
\fmf{fermion}{t,b1}
\fmf{photon}{v2,t}
\fmf{fermion,tension=2}{v1,t2}
\fmf{plain,tension=2}{v2,v1}
\fmf{fermion}{b2,v2}
\fmffreeze
\fmf{photon}{r,v1}
\fmfdot{v1,v2,t}
\end{fmfgraph}}\, ,\, 
\parbox{20\unitlength}{
\begin{fmfgraph}(20,15)
\fmftopn{t}{2}
\fmfbottomn{b}{2}
\fmfright{r}
\fmf{fermion}{t1,t,b1}
\fmf{photon}{v1,t}
\fmf{fermion}{v1,t2}
\fmf{plain,tension=2}{v2,v1}
\fmf{fermion,tension=2}{b2,v2}
\fmffreeze
\fmf{photon}{r,v2}
\fmfdot{v1,v2,t}
\end{fmfgraph}}\right\}\\
G_s^{\text{ISR}}&= \left\{\parbox{20\unitlength}{
\begin{fmfgraph}(20,15)
\fmfbottomn{b}{2}
\fmftopn{t}{3}
\fmf{fermion}{b2,t,t3}
\fmf{photon}{v2,t}
\fmf{fermion,tension=2}{t1,v1}
\fmf{plain,tension=2}{v1,v2}
\fmf{fermion}{v2,b1}
\fmffreeze
\fmf{photon}{t2,v1}
\fmfdot{v1,v2,t}
\end{fmfgraph}}\,,\, 
\parbox{20\unitlength}{
\begin{fmfgraph}(20,15)
\fmftopn{t}{2}
\fmfbottomn{b}{3}
\fmfright{r}
\fmf{fermion}{b3,t,t2}
\fmf{photon}{v1,t}
\fmf{fermion}{t1,v1}
\fmf{plain,tension=2}{v1,v2}
\fmf{fermion,tension=2}{v2,b1}
\fmffreeze
\fmf{photon}{b2,v2}
\fmfdot{v1,v2,t}
\end{fmfgraph}}\right\}
\end{aligned}
\end{equation}
This allows the separate treatment of initial-state and final-state
Bremsstrahlung. If we consider instead the process $e^+ e^- \to e^+
e^- \gamma$, i.\,e.~Bhabha scattering with an additional
Bremsstrahlungs-photon, we can appeal to the flavor selection rules
discussed above and see that we get altogether four GICs.

In order to construct minimal GICs of tree diagrams in non abelian
gauge theories systematically, the formalism of `gauge and flavor
flips' was introduced in~\cite{Boos:1999}.  An elementary `flavor
flip' is defined as an exchange of two diagrams in the set
\begin{multline}
\label{eq:flavor_flips}
F_4=\{ F_4^i | i = 1, 2, 3 \}=\\
\left\{
\parbox{15\unitlength}{
\begin{fmfgraph}(15,15)
\fmfleft{a,b}
\fmfright{f1,f2}
\fmf{fermion}{a,fwf1}
\fmf{photon}{fwf1,fwf2}
\fmf{fermion}{b,fwf2}
\fmf{fermion}{fwf1,f1}
\fmf{fermion}{fwf2,f2}
\fmfdot{fwf1}
\fmfdot{fwf2}
\end{fmfgraph}}\,,\,
\parbox{15\unitlength}{
\begin{fmfgraph}(15,15)
\fmfleft{a,b}
\fmfright{f1,f2}
\fmf{fermion}{a,fwf1}
\fmf{photon}{fwf1,fwf2}
\fmf{fermion}{b,fwf2}
\fmf{phantom}{fwf2,f2}
\fmf{phantom}{fwf1,f1}
\fmffreeze
\fmf{fermion}{fwf2,f1}
\fmf{fermion}{fwf1,f2}
\fmfdot{fwf1}
\fmfdot{fwf2}
\end{fmfgraph}}\, ,\, 
\parbox{15\unitlength}{
\begin{fmfgraph}(15,15)
\fmfleft{a,b}
\fmfright{f1,f2}
\fmf{fermion}{b,fwf,a}
\fmf{photon}{fwf,www}
\fmf{fermion}{f1,www,f2}
\fmfdot{fwf}
\fmfdot{www}
\end{fmfgraph}}
\right \}
\end{multline}
that is compatible with the Feynman rules.

In the example of the four point function, the two diagrams
in~(\ref{eq:bhabbha}) are connected by a flavor
flip~$F_4^1\leftrightarrow F_4^3$.  Flips between pairs of larger
diagrams are obtained by applying elementary flips to four particle
subdiagrams~\cite{Boos:1999}.

Elementary `gauge flips' are defined as exchanges of
diagrams in the sets
\begin{subequations}
\label{subeq:gauge_flips}
\begin{multline}\label{eq:4w-gf}
G_4 = \{ G_4^i | i = 1, 2, 3, 4 \}=\\
\left\{
\parbox{15\unitlength}{
\begin{fmfgraph}(15,15)
\fmfleft{a,b}
\fmfright{f1,f2}
\fmf{photon}{a,fwf1}
\fmf{photon}{fwf1,fwf2}
\fmf{photon}{fwf2,b}
\fmf{photon}{fwf1,f1}
\fmf{photon}{fwf2,f2}
\fmfdot{fwf1}
\fmfdot{fwf2}
\end{fmfgraph}}\,,\,
\parbox{15\unitlength}{
\begin{fmfgraph}(15,15)
\fmfleft{a,b}
\fmfright{f1,f2}
\fmf{photon}{a,fwf1}
\fmf{photon}{fwf1,fwf2}
\fmf{photon}{fwf2,b}
\fmf{phantom}{fwf2,f2}
\fmf{phantom}{fwf1,f1}
\fmffreeze
\fmf{photon}{fwf2,f1}
\fmf{photon,rubout}{fwf1,f2}
\fmfdot{fwf1}
\fmfdot{fwf2}
\end{fmfgraph}}\, ,\, 
\parbox{15\unitlength}{
\begin{fmfgraph}(15,15)
\fmfleft{a,b}
\fmfright{f1,f2}
\fmf{photon}{a,fwf}
\fmf{photon}{fwf,b}
\fmf{photon}{fwf,www}
\fmf{photon}{www,f1}
\fmf{photon}{www,f2}
\fmfdot{fwf}
\fmfdot{www}
\end{fmfgraph}}\, , \, 
\parbox{15\unitlength}{
\begin{fmfgraph}(15,15)
\fmfleft{a,b}
\fmfright{f1,f2}
\fmf{photon}{a,c}
\fmf{photon}{c,b}
\fmf{photon}{c,f1}
\fmf{photon}{c,f2}
\fmfdot{c}
\end{fmfgraph}}\right \}
\end{multline}
and
\begin{multline}
\label{eq:gauge_flips2}
G_{4,2F} = \{ G_{4,2F}^i | i = 1, 2, 3 \}=\\
\left\{
\parbox{15\unitlength}{
\begin{fmfgraph}(15,15)
\fmfleft{a,b}
\fmfright{f1,f2}
\fmf{fermion}{a,fwf1}
\fmf{fermion}{fwf1,fwf2}
\fmf{fermion}{fwf2,b}
\fmf{photon}{fwf1,f1}
\fmf{photon}{fwf2,f2}
\fmfdot{fwf1,fwf2}
\end{fmfgraph}}\quad,\quad
\parbox{15\unitlength}{
\begin{fmfgraph}(15,15)
\fmfleft{a,b}
\fmfright{f1,f2}
\fmf{fermion}{a,fwf1}
\fmf{fermion}{fwf1,fwf2}
\fmf{fermion}{fwf2,b}
\fmf{phantom}{fwf2,f2}
\fmf{phantom}{fwf1,f1}
\fmffreeze
\fmf{photon}{fwf2,f1}
\fmf{photon,rubout}{fwf1,f2}
\fmfdot{fwf1,fwf2}
\end{fmfgraph}}\quad,\quad 
\parbox{15\unitlength}{
\begin{fmfgraph}(15,15)
\fmfleft{a,b}
\fmfright{f1,f2}
\fmf{fermion}{a,fwf}
\fmf{fermion}{fwf,b}
\fmf{photon}{fwf,www}
\fmf{photon}{www,f1}
\fmf{photon}{www,f2}
\fmfdot{www,fwf}
\end{fmfgraph}}\right \}
\end{multline}
\end{subequations}
respectively.  E.\,g.~the diagrams in the GIC~(\ref{eq:qcd-grove}) are
connected by gauge flips of subdiagrams from~$G_{4,2F}$.

A set of diagrams connected by flavor and gauge flips is called
a `forest', while a set of diagrams connected by gauge flips only is
called a `grove'.  The example of the $\bar q q \to \bar q q g$
amplitude suggests that the groves can be identified with
GICs. Indeed it has been shown in~\cite{Boos:1999}, that the groves
are the \emph{minimal} GICs of FDs. Furthermore, the forest of FDs is
connected and consists of all FDs contributing to the
amplitude. Therefore the formalism can be used to implement a FD
generator~\cite{Ohl:bocages} that generates the groves \textit{en
passant}.  Further results and examples for the structure of groves 
in the electroweak SM can be found in~\cite{Boos:1999,Ohl:1999,Ondreka:2000}.

\subsection{Flips in spontaneously broken gauge theories}
\label{sec:ssb-flips}
In a SBGT, the role of Higgs bosons in gauge and flavor flips is not clear a
priori. Because of the presence of a $WWH$~vertex (where~$W$ denotes
an arbitrary massive gauge boson)
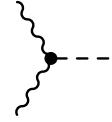
\begin{equation*}
\parbox{15\unitlength}{
\begin{fmfgraph}(15,15)
\fmfleft{a,b}
\fmfright{h}
\fmf{photon}{a,ghg,b}
\fmf{dashes}{ghg,h}
\fmfdot{ghg}
\end{fmfgraph}}
\end{equation*}
neutral Higgs bosons cannot be assigned a (fictitious) conserved
quantum number as we have done above to derive the flavor selection
rules. Thus it seems plausible that no new groves should appear
compared with unbroken gauge theories. This corresponds to  the
\textit{ad hoc} prescription given in~\cite{Boos:1999} for the construction of the
gauge flips, that proposes to treat Higgs bosons like gauge bosons.

We show in section~\ref{sec:sti-flips} that this intuitive argument is
essentially correct, provided a linear representation of the scalar
sector is used. A richer structure of the groves will emerge in
theories with a nonlinearly realized symmetry,

The appearance of new groves in nonlinear realizations is very
plausible if one considers the nonlinearly realized electroweak
SM~\cite{NL_SM}. Here the Higgs boson transforms trivially under
gauge transformations and can be removed from the theory without spoiling
gauge invariance. The trivial transformation law  implies that  the
diagrams without Higgs bosons form GICs by themselves, in
contrast to the linear parametrization. Therefore one can simplify the
elementary gauge flips by omitting the internal Higgs bosons. The case
of a more general Higgs sector with charged Higgs bosons requires more
careful considerations and is discussed in section~\ref{sec:nl-sti}.
 
Instead of~(\ref{eq:gauge_flips2}), the flips for $\bar f f \to WW$ are
in a linear representation of the symmetry: 
\begin{multline}
\label{eq:ffww-flips}
  {\tilde G}_{4,2F}
    = G_{4,2F} \cup \{ {\tilde G}_{4,2F}^4 \}
    = \{ {\tilde G}_{4,2F}^i | i = 1, 2, 3, 4 \} = \\
\left\{
\parbox{15\unitlength}{
\begin{fmfgraph}(15,15)
\fmfleft{a,b}
\fmfright{f1,f2}
\fmf{fermion}{a,fwf1}
\fmf{fermion}{fwf1,fwf2}
\fmf{fermion}{fwf2,b}
\fmf{photon}{fwf1,f1}
\fmf{photon}{fwf2,f2}
\fmfdot{fwf1,fwf2}
\end{fmfgraph}}\,,\,
\parbox{15\unitlength}{
\begin{fmfgraph}(15,15)
\fmfleft{a,b}
\fmfright{f1,f2}
\fmf{fermion}{a,fwf1}
\fmf{fermion}{fwf1,fwf2}
\fmf{fermion}{fwf2,b}
\fmf{phantom}{fwf2,f2}
\fmf{phantom}{fwf1,f1}
\fmffreeze
\fmf{photon}{fwf2,f1}
\fmf{photon,rubout}{fwf1,f2}
\fmfdot{fwf1,fwf2}
\end{fmfgraph}}\,,\, 
\parbox{15\unitlength}{
\begin{fmfgraph}(15,15)
\fmfleft{a,b}
\fmfright{f1,f2}
\fmf{fermion}{a,fwf}
\fmf{fermion}{fwf,b}
\fmf{photon}{fwf,www}
\fmf{photon}{www,f1}
\fmf{photon}{www,f2}
\fmfdot{www,fwf}
\end{fmfgraph}}\, ,\,
\parbox{15\unitlength}{
\begin{fmfgraph}(15,15)
\fmfright{a,b}
\fmfleft{f1,f2}
\fmf{photon}{a,fwf,b}
\fmf{dashes}{fwf,www}
\fmf{fermion}{f1,www,f2}
\fmfdot{www}
\fmfdot{fwf}
\end{fmfgraph}}\right \}
\end{multline}
As discussed above, the Higgs exchange diagram~${\tilde G}_{4,2F}^4$
is not present for nonlinear symmetries. Similarly in the linear
representation, the Higgs exchange diagrams have to be included in the
gauge flips for the four gauge boson amplitude,
cf.~(\ref{eq:gauge_flips1}), while they do not contribute in the
nonlinear representation.
 
As we will discuss in section~\ref{sec:nl-sti}, the simplifications in nonlinear
representations affect only the $WWH$~vertex, while the STI for the
$WHH$~vertex is similar in both realizations. Thus diagrams with
internal Higgs bosons cannot be left out of the gauge flips if a $WHH$~vertex
appears. An example is provided by the gauge flips for $\bar f f \to W
H$. Here the internal Higgs boson has to be included both in the
linear and nonlinear parametrization: 
\begin{multline}
\label{eq:h_flips_1}
{\tilde G}_{4,1H2F} = 
\{ {\tilde G}_{4,1H2F}^i | i = 1,2,3,4 \}=\\
\left\{
\parbox{15\unitlength}{
\begin{fmfgraph}(15,15)
\fmfleft{a,b}
\fmfright{f1,f2}
\fmf{fermion}{a,fwf1}
\fmf{fermion}{fwf1,fwf2}
\fmf{fermion}{fwf2,b}
\fmf{photon}{fwf1,f1}
\fmf{dashes}{fwf2,f2}
\fmfdot{fwf1,fwf2}
\end{fmfgraph}}\, ,\,
\parbox{15\unitlength}{
\begin{fmfgraph}(15,15)
\fmfleft{a,b}
\fmfright{f1,f2}
\fmf{fermion}{a,fwf1}
\fmf{fermion}{fwf1,fwf2}
\fmf{fermion}{fwf2,b}
\fmf{phantom}{fwf2,f2}
\fmf{phantom}{fwf1,f1}
\fmffreeze
\fmf{photon}{fwf2,f1}
\fmf{dashes,rubout}{fwf1,f2}
\fmfdot{fwf1,fwf2}
\end{fmfgraph}}\, ,\, 
\parbox{15\unitlength}{
\begin{fmfgraph}(15,15)
\fmfleft{a,b}
\fmfright{f1,f2}
\fmf{fermion}{a,fwf}
\fmf{fermion}{fwf,b}
\fmf{photon}{fwf,www}
\fmf{photon}{www,f1}
\fmf{dashes}{www,f2}
\fmfdot{www,fwf}
\end{fmfgraph}}\, , \,
\parbox{15\unitlength}{
\begin{fmfgraph}(15,15)
\fmfleft{a,b}
\fmfright{f1,f2}
\fmf{fermion}{a,fwf}
\fmf{fermion}{fwf,b}
\fmf{dashes}{fwf,www}
\fmf{photon}{www,f1}
\fmf{dashes}{www,f2}
\fmfdot{www,fwf}
\end{fmfgraph}}
\right \} 
\end{multline}
The complete set of gauge and flavor flips is given in
appendix~\ref{app:flips}, both for linear and nonlinear representations. 

To define the complete forest, additional types of flips have
to be introduced compared to the unbroken case. Four particle diagrams
with only external Higgs bosons and matter fields are found to be gauge
parameter independent by themselves, so we have to introduce another
class of flips, that plays a role similar to the flavor flips and
will be called `Higgs flips'. They consist of the four
diagrams~$\tilde H_{4,2F}$
contributing to the $\bar f f \to HH$ amplitude and the seven
diagrams~$\tilde H_4$ contributing to the four Higgs amplitude.  They are given
in~(\ref{subeq:higgs_flips}) in the appendix.

In a linear representation, the forest for a given set of
external diagrams is defined as the set of diagrams connected by
flavor, Higgs and gauge flips while the definition of the groves
remains as before.

In nonlinear realizations, one has to introduce yet another class
of flips that generate the diagrams not needed for the gauge flips,
i.\,e.~the exchange ${\tilde G}_{4,2F}^3\leftrightarrow {\tilde G}_{4,2F}^4$
from (\ref{eq:ffww-flips}) for the two fermion two gauge boson function
and flips from the diagrams of~(\ref{eq:4w-gf}) to
\begin{equation}
\left\{
\parbox{15\unitlength}{
\begin{fmfgraph}(15,15)
\fmfleft{a,b}
\fmfright{f1,f2}
\fmf{photon}{a,fwf1}
\fmf{dashes}{fwf1,fwf2}
\fmf{photon}{fwf2,b}
\fmf{photon}{fwf1,f1}
\fmf{photon}{fwf2,f2}
\fmfdot{fwf1}
\fmfdot{fwf2}
\end{fmfgraph}}\quad,\quad
\parbox{15\unitlength}{
\begin{fmfgraph}(15,15)
\fmfleft{a,b}
\fmfright{f1,f2}
\fmf{photon}{a,fwf1}
\fmf{dashes}{fwf1,fwf2}
\fmf{photon}{fwf2,b}
\fmf{phantom}{fwf2,f2}
\fmf{phantom}{fwf1,f1}
\fmffreeze
\fmf{photon}{fwf2,f1}
\fmf{photon,rubout}{fwf1,f2}
\fmfdot{fwf1}
\fmfdot{fwf2}
\end{fmfgraph}}\quad , \quad 
\parbox{15\unitlength}{
\begin{fmfgraph}(15,15)
\fmfleft{a,b}
\fmfright{f1,f2}
\fmf{photon}{a,fwf}
\fmf{photon}{fwf,b}
\fmf{dashes}{fwf,www}
\fmf{photon}{www,f1}
\fmf{photon}{www,f2}
\fmfdot{fwf}
\fmfdot{www}
\end{fmfgraph}}
\right \}
\end{equation}
for the $4 W$ function. These `Higgs exchange flips' have to be
included in the definition of the forest, if it is to remain
connected.

The structure of the groves in SBGTs is analyzed in
section~\ref{sec:ssb-gics}. Readers who are primarily interested in
applications of the formalism can jump to this section directly.  It
can be read independently from the derivation of the results in the
more formal sections~\ref{sec:sti-graph} to~\ref{sec:sti-flips}.

\section{Graphical notation for STIs}
\label{sec:sti-graph}
In the the diagrammatic derivation of the GICs in section~\ref{sec:gi-classes},
we will use the STIs for irreducible vertices as building blocks.
As a preparation, we need to set
up a notation for the STIs for the irreducible vertices. Our notation
for the BRS transformations and the STIs for Green's functions (GFs)
is reviewed in appendix~\ref{app:gf-sti}.

The symmetry of the effective action leads to the Zinn-Justin
equation~\cite{Zinn-Justin:1974} that implies the STIs for the
irreducible vertices. To derive the Zinn-Justin equation, one adds sources
$\Psi^\star$ for the BRS transforms of the fields $\Psi$ to the
effective action $\Gamma_0$
\begin{equation}
\label{eq:antifields}
 \Gamma=\Gamma_0+\sum_\Psi\int\dd^4x\,\tr\lbrack {\Psi^\star}
 (\delta_{\text{BRS}} \Psi)\rbrack 
\end{equation}
from which
the one particle irreducible vertices are obtained by taking
functional derivatives with respect to the classical fields
\begin{multline}
  \frac{\ii\delta^n\Gamma(\Phi_{cl})}
       {\delta\Phi_{cl}(x_1)\dots\delta\Phi_{cl}(x_n)}\equiv
    \Gamma_{\Phi_1\cdots\Phi_n}(x_1\dots x_n)\\
      = \Braket{\Phi(x_1)\dots\Phi(x_n)}^{\text{1PI}}
\end{multline}
The BRS invariance of the effective action in anomaly free theories
implies the Zinn-Justin equation
\begin{equation}\label{eq:sti}
\sum_\Psi\int\dd^4 x \frac{\delta_L \Gamma}{\delta \Psi^\star}\frac{\delta_R
  \Gamma}{\delta \Psi}+B\frac{\delta_R \Gamma}{\delta\bar c}=0\,.
\end{equation}
To derive the STI for the three point vertex, we take the derivative
of~(\ref{eq:sti}) with respect to two  fields $\Phi$ and one ghost
field and set the classical fields and sources to zero:
\begin{multline}
\label{eq:sti-irr-3point}
0=\sum_\Psi\int\dd^4 x \Biggl\{ \Gamma_{c_a \Psi^\star}\Gamma_{\Psi\Phi_1\Phi_2}\\
+\Bigl[\Gamma_{c_a\Psi^\star\Phi_1}\Gamma_{\Psi\Phi_2}+\frac{\delta
  B(x)}{\delta\Phi_1}\Gamma_{c_a\bar c\Phi_2} +\quad (1\leftrightarrow 2)\Bigr]\Biggr\}
\end{multline}
Repeating this procedure for an additional derivative, we can derive
the corresponding relation for the irreducible four point function, etc.

To see the physical content of these identities, we note that the
vertex $\Gamma_{c_a\Psi^\star}$ is only present for gauge bosons
and Goldstone bosons~(GBs).
Since these contributions are linear in the
fields, they do not require radiative corrections beyond the
renormalization of the Lagrangian and we get
to all orders in perturbation theory
\begin{multline}
\label{eq:fourier-linear-brs}
\text{F.T.}\sum_\Psi\int\dd^4 x \Gamma_{c_a\Psi^\star}\Gamma_{\Psi\dots}
=-\ii p_\mu \Gamma_{ W_a\dots}^\mu-m_{W_a}\Gamma_{ \phi_a\dots}\\
\equiv\Braket{ \mathcal{D}_a(p)\dots}^{\text{1PI}}
\end{multline}
where we have introduced the shorthand
$\mathcal{D}$ for the combination of the scalar gauge boson
component and the GB.
To illustrate the  STIs for the irreducible vertices, we introduce
the graphical notation
\begin{equation}
\label{eq:vertices-graph}
\begin{aligned}
\Gamma_{\Phi\Phi}&=\,
\parbox{11\unitlength}{
\begin{fmfgraph}(10,10)
\fmfleft{l}
\fmfright{r}
\fmf{zigzag}{l,r}
\end{fmfgraph}}
\\
\Gamma_{c\Psi^\star\Phi_1\dots\Phi_n}=&\quad\,
\parbox{16\unitlength}{
\begin{fmfgraph*}(15,15)
\fmfleftn{g}{3}
\fmfrightn{r}{2}
\fmf{ghost}{r1,m}
\fmf{plain}{g1,m,g3}
\fmf{phantom}{m,r2}
\fmffreeze
\fmf{plain,right=0.5}{m,r2}
\fmf{ghost,left=0.5}{m,r2}
\fmfv{decor.shape=square,decor.filled=empty,decor.size=5,label=$\Psi^*$,label.angle=0}{r2}
\fmfv{label=$c$,label.angle=0}{r1}
\fmfv{label=$\Phi_1$,label.angle=180}{g1}
\fmfv{label=$\vdots$,label.angle=0}{g2}
\fmfv{label=$\Phi_n$,label.angle=180}{g3}
\fmfblob{10}{m}
\end{fmfgraph*}}
\end{aligned}
\end{equation}
The STI~(\ref{eq:sti-irr-3point}) for the three point
function reads in this notation
\begin{equation}
\parbox{20\unitlength}{
\begin{fmfgraph}(15,15)
\fmfleft{l}
\fmfright{r1,r2}
\fmfpolyn{smooth,filled=shaded}{g}{3}
\fmf{double}{l,g1}
\fmf{plain}{r1,g2}
\fmf{plain}{r2,g3}
\end{fmfgraph}}
=\, \parbox{20\unitlength}{\begin{fmfgraph}(20,20)
\fmfleft{l}
\fmfright{r1,r2}
\fmf{plain}{r1,brs}
\fmf{ghost}{l,brs}
\fmf{plain,right=0.75}{brs,i}
\fmf{ghost,left=0.75}{brs,i}
\fmf{zigzag,tension=2}{i,r2}
\fmfv{decor.shape=circle,decor.filled=shaded,decor.size=10}{brs}
\fmfv{decor.shape=square,decor.filled=empty,decor.size=5}{i}
\end{fmfgraph}}+\, \parbox{20\unitlength}{\begin{fmfgraph}(20,20)
\fmfleft{l}
\fmfright{r1,r2}
\fmf{ghost}{l,brs}
\fmf{zigzag,tension=2}{r1,i}
\fmf{plain,right=0.5}{brs,i}
\fmf{ghost,left=0.5}{brs,i}
\fmf{plain}{brs,r2}
\fmfv{decor.shape=square,decor.filled=empty,decor.size=5}{i}
\fmfv{decor.shape=circle,decor.filled=shaded,decor.size=10}{brs}
\end{fmfgraph}}
\end{equation}
where we have not displayed the term $\propto \delta B/\delta\Phi$ that
will be discussed below. The STI for the four point
function is written as
\begin{equation}\label{eq:sti-irr-4point}
\begin{gathered}
\parbox{20\unitlength}{
\begin{fmfgraph}(20,20)
\fmfleft{g,r3}
\fmfrightn{r}{2}
\fmf{double}{g,b}
\begin{fmffor}{i}{1}{1}{3}
\fmf{plain}{r[i],b}
\end{fmffor}
\fmfblob{15}{b}
\end{fmfgraph}}
=\sum_{\Phi_i}
\parbox{35\unitlength}{
\fmfframe(0,0)(5,0){
\begin{fmfgraph*}(30,20)
\fmfleftn{g}{2}
\fmfrightn{r}{2}
\fmf{ghost}{g1,m}
\fmf{plain}{m,g2}
\fmf{phantom}{m,brs}
\fmf{plain,tension=3}{brs,b}
\begin{fmffor}{i}{1}{1}{2}
\fmf{plain}{r[i],b}
\end{fmffor}
\fmffreeze
\fmf{plain,right=0.75}{m,brs}
\fmf{ghost,left=0.75}{m,brs}
\fmfv{decor.shape=square,decor.filled=empty,decor.size=5}{brs}
\fmfblob{10}{m}
\fmfblob{10}{b}
\fmfv{l=$\Phi_i$,l.d=0.3}{g2}
\end{fmfgraph*}}}\\
+
\sum_{\Phi_i}
\parbox{25\unitlength}{
\fmfframe(0,5)(0,0){
\begin{fmfgraph*}(20,20)
\fmfleftn{g}{2}
\fmfrightn{r}{2}
\fmf{ghost}{g1,m}
\fmf{plain}{r2,m,r1}
\fmf{phantom,tension=2}{m,brs}
\fmf{zigzag,tension=2}{brs,g2}
\fmffreeze
\fmf{plain,right}{m,brs}
\fmf{ghost,left}{m,brs}
\fmfv{decor.shape=square,decor.filled=empty,decor.size=5}{brs}
\fmfblob{10}{m}
\fmfv{l=$\Phi_i$,l.d=0.1}{r2}
\end{fmfgraph*}}}
\end{gathered}
\end{equation}
The graphical representation of the fact that the irreducible two
point function is the inverse of the propagator
\begin{equation}
\int\dd^4y\,\Gamma_{\Phi\Phi}(x,y)D_{\Phi\Phi}(y,z)=-\delta^4(x,z)
\end{equation}
is given by 
\begin{equation}
\label{eq:inv-prop}
\parbox{15\unitlength}{
\begin{fmfgraph}(15,15)
\fmfleft{l}
\fmfright{r1,r2}
\fmf{plain,tension=1.5}{l,v}
\fmf{phantom}{v,r1}
\fmf{phantom}{v,r2}
\fmffreeze
\fmf{zigzag}{v,r1}
\fmf{ghost}{r2,r1}
\end{fmfgraph}}
\quad =\quad-\quad
\parbox{15\unitlength}{
\begin{fmfgraph}(15,15)
\fmfleft{l}
\fmfright{r1,r2}
\fmf{phantom}{l,r1}
\fmf{phantom}{r2,r1}
\fmffreeze
\fmf{ghost}{r2,l}
\end{fmfgraph}}
\end{equation}
This relation will be essential in relating the STIs for GFs and the STIs for irreducible vertices in the diagrammatical derivation of the GICs in
section \ref{sec:gi-classes}.
 
On tree level in a linear realization of the symmetry,
 the vertex functions with the insertions of sources for
the BRS transformed fields can be read from the BRS transformation.
The STI for the three point function~(\ref{eq:sti-irr-3point})
becomes
\begin{align}
\label{eq:sti-irr-3tree}
  &-\Braket{ \mathcal{D}_a(p)\Phi_i(k_i)\Phi_j(k_j)}^{\text{1PI}}\\
  &= T^a_{ki}\Braket{\Phi_k(p+k_i)\Phi_j(k_j)}^{\text{1PI}}
     + T^a_{kj}\Braket{\Phi_i(k_i)\Phi_k(p+k_j)}^{\text{1PI}}\nonumber
\end{align}
In the graphical notation~(\ref{eq:vertices-graph}) this is written as
\begin{equation}
\label{eq:sti-3graph}
\parbox{20\unitlength}{\fmfframe(5,0)(0,0){
\begin{fmfgraph*}(15,15)
\fmfleft{l}
\fmfright{r1,r2}
\fmfpolyn{smooth,filled=shaded}{g}{3}
\fmf{double}{l,g1}
\fmf{plain}{r1,g2}
\fmf{plain}{r2,g3}
\fmfv{l=$a$}{l}
\fmfv{l=$i$}{r1}
\fmfv{l=$j$}{r2}
\end{fmfgraph*}}}
= \parbox{20\unitlength}{\fmfframe(5,0)(0,0){\begin{fmfgraph*}(15,15)
\fmfleft{l}
\fmfright{r1,r2}
\fmf{plain}{r1,i}
\fmf{ghost}{l,i}
\fmf{zigzag}{i,r2}
\fmfv{l=$a$}{l}
\fmfv{decor.shape=square,decor.filled=empty,decor.size=5,label=$k$,la.di=7}{i}
\fmfv{l=$i$}{r1}
\fmfv{l=$j$}{r2}
\end{fmfgraph*}}}
 + \parbox{25\unitlength}{\fmfframe(5,3)(0,3){\begin{fmfgraph*}(15,15)
\fmfleft{l}
\fmfright{r1,r2}
\fmf{ghost}{l,i}
\fmf{zigzag}{r1,i}
\fmf{plain}{i,r2}
\fmfv{l=$a$}{l}
\fmfv{decor.shape=square,decor.filled=empty,decor.size=5,label=$k$,la.di=7}{i}
\fmfv{l=$i$}{r1}
\fmfv{l=$j$}{r2}
\end{fmfgraph*}}}
\end{equation}
For vertices involving gauge bosons or GBs we get
additional contributions from the term $\propto \delta B/\delta\Phi$
in the Zinn-Justin equation. Using the equation of motion of the
Na\-ka\-ni\-shi-Lautrup field $B$, we find that for every gauge boson or GB
that is not contracted, there appears a term in the
STI for three point functions
\begin{subequations}
\label{subeq:ghost-sti}
\begin{align}
\label{eq:2gb-sti}
 \ii \frac{1}{\xi}p_b^\nu \Braket{c_a(p_a)\bar c_b(p_b)\Phi_i(k_i)}^{\text{1PI}}
  &=\frac{1}{\xi}\,
\parbox{15\unitlength}{
\begin{fmfgraph}(15,15)
\fmfleft{l}
\fmfright{r1,r2}
\fmfpolyn{smooth,filled=shaded}{g}{3}
\fmf{ghost}{l,g1}
\fmf{ghost}{g2,r1}
\fmf{plain}{r2,g3}
\fmfv{decor.shape=square,decor.filled=full,decor.size=5}{r1}
\end{fmfgraph}}\\
m_{W_b}\Braket {c_a(p_a) \bar c_b(p_b)\Phi_i(k_i) }^{\text{1PI}}\,&=
\parbox{25\unitlength}{
\begin{fmfgraph}(15,15)
\fmfleft{l}
\fmfright{r1,r2}
\fmfpolyn{smooth,filled=shaded}{g}{3}
\fmf{ghost}{l,g1}
\fmf{ghost}{g2,r1}
\fmf{plain}{r2,g3}
\fmfv{decor.shape=cross,decor.size=5}{r1}
\end{fmfgraph}}\label{eq:phi-sti}
\end{align}
\end{subequations}

On
tree level in a linear representation of the symmetry, 
the terms with more than one derivative acting on the BRS 
transforms vanish so  the graphical representation of the STI for the 
four point functions~(\ref{eq:sti-irr-4point}) simplifies to
\begin{equation}
\label{eq:sti4-graph}
\parbox{15\unitlength}{
\begin{fmfgraph}(15,15)
\fmfleft{l1,l2}
\fmfright{r1,r2}
\fmf{double}{l1,g}
\fmf{plain}{l2,g}
\fmf{plain}{r1,g}
\fmf{plain}{r2,g}
\fmfv{decor.shape=circle,decor.filled=shaded,decor.size=15}{g}
\end{fmfgraph}}
=\, \parbox{15\unitlength}{\begin{fmfgraph}(15,15)
\fmfleft{l1,l2}
\fmfright{r1,r2}
\fmf{phantom}{l1,g}
\fmf{plain,tension=1}{i,l2}
\fmf{plain,tension=2.5}{i,g}
\fmf{plain}{r1,g}
\fmf{plain}{r2,g}
\fmfv{decor.shape=square,decor.filled=empty,decor.size=5}{i}
\fmfv{decor.shape=circle,decor.filled=shaded,decor.size=9}{g}
\fmffreeze
\fmf{ghost}{l1,i}
 \end{fmfgraph}}
+
\, \parbox{15\unitlength}{\begin{fmfgraph}(15,15)
\fmfleft{l1,l2}
\fmfright{r1,r2}
\fmf{phantom}{l1,g}
\fmf{plain}{l2,g}
\fmf{plain,tension=1}{r1,i}
\fmf{plain,tension=2.5}{i,g}
\fmf{plain}{r2,g}
\fmfv{decor.shape=square,decor.filled=empty,decor.size=5}{i}
\fmfv{decor.shape=circle,decor.filled=shaded,decor.size=9}{g}
\fmffreeze
\fmf{ghost}{l1,i}
 \end{fmfgraph}}+
\, \parbox{15\unitlength}{\begin{fmfgraph}(15,15)
\fmfleft{l1,l2}
\fmfright{r1,r2}
\fmf{phantom}{l1,g}
\fmf{plain}{l2,g}
\fmf{plain}{r1,g}
\fmf{plain,tension=1}{r2,i}
\fmf{plain,tension=2.5}{i,g}
\fmfv{decor.shape=square,decor.filled=empty,decor.size=5}{i}
\fmfv{decor.shape=circle,decor.filled=shaded,decor.size=9}{g}
\fmffreeze
\fmf{ghost,right=.3}{l1,i}
 \end{fmfgraph}}
\end{equation}
Here and in~(\ref{eq:sti5-graph}) below, additional terms from the
derivatives of the BRS-transforms appear for nonlinearly realized
symmetries and will be  discussed in subsection \ref{sec:nl-sti}.
In higher orders of perturbation theory and for nonlinearly realized 
symmetries, additional contributions from the term involving the 
Na\-ka\-ni\-shi-Lautrup field $B$ appear for external gauge and 
Goldstone bosons, similar to~(\ref{subeq:ghost-sti}). 

In a renormalizable theory, there is no five point vertex on tree level,
so taking four derivatives of~(\ref{eq:sti}) with respect to physical
fields we get
\begin{equation}
\label{eq:sti5-graph}
0=\sum_{i=1}^4 \parbox{15\unitlength}{
\begin{fmfgraph*}(15,15)
\fmfleftn{l}{2}
\fmfrightn{r}{3}
\fmf{plain}{l1,g}
\fmf{plain}{l2,g}
\fmf{plain}{r1,g}
\fmf{plain,tension=1}{r3,i}
\fmf{plain,tension=2}{i,g}
\fmffreeze
\fmf{ghost}{r2,i}
\fmfv{decor.shape=circle,decor.filled=shaded,decor.size=15}{g}
\fmfv{decor.shape=square,decor.filled=empty,decor.size=5,label=$i$,la.an=90}{i}
\end{fmfgraph*}}
\end{equation}
We will also need STIs in non renormalizable theories with nonlinearly
realized symmetries. Since tree level
calculations in effective field theories correspond by Weinberg's
power counting theorem to the lowest
order in the energy expansion, we don't consider additional
non renormalizable operators. 
This will no longer
suffice for the extension of the formalism to loop diagrams. Also, 
higher dimensional operators involving Goldstone bosons have to be included
to maintain gauge invariance in the nonlinear realization. 

\section{Graphical derivation of gauge invariance classes}
\label{sec:gi-classes}

\subsection{Definition of gauge invariance classes}
\label{sec:groves-def}
On tree level, the definition of GICs in terms of the WIs given in
section~\ref{sec:groves} is sufficient. However, for future applications in
loop calculations, the off-shell structure of the groves has to be
clarified as well. The natural extension of the definition of
section~\ref{sec:groves} is to demand that the GICs satisfy the appropriate
STIs (see appendix~\ref{app:gf-sti}) instead of the WIs.
To give a definition, we first have to clarify the notion of a subset of
diagrams satisfying a STI, including the definition of the set of
contact terms in the STI associated to the subset.

In this work. we will always consider GFs where the propagators of the
external particles are not amputated, because the STIs
are more familiar in this case. Identities for off shell amplitudes with amputated
external particles are more suitable for numerical calculations and
have been considered in detail in~\cite{Schwinn:2003}. The amputation
procedure adds notational complexity, but the conclusions remain
unchanged. 

To define the form of the contact terms, we introduce a mapping
$\mathcal{F}$ that maps every FD to the corresponding contact
terms. Note that this is a purely formal mapping and in general
it is not true that a contraction of a gauge boson in the original
diagram results in the contact terms generated by this mapping. In the
following, gray blobs denote a subdiagram, while white blobs
denote subamplitudes or subgroves, i.\,e.~sets of diagrams.
\begin{definition}
\begin{subequations}
\label{subeq:generate-contact}
The action of  $\mathcal{F}$ on a diagram with the insertion of a
gauge boson into an external line is given by
\begin{equation}
\quad\parbox{20\unitlength}{
\begin{fmfgraph}(20,20)
\fmfbottomn{v}{6}
\fmftop{t}
\fmf{phantom,tension=6}{t,v}
\begin{fmffor}{i}{1}{1}{5}
\fmf{plain}{v[i],v}
\end{fmffor}
\fmf{phantom}{v6,v}
\fmffreeze
\fmf{plain,tension=2}{v6,i}
\fmf{plain}{i,v}
\fmffreeze
\fmf{photon,left=0.5}{t,i}
\fmfdot{i}
\fmfv{d.sh=circle,d.f=30,d.size=25pt}{v}
\end{fmfgraph}} \,\xrightarrow{\mathcal{F}}\,
-\parbox{20\unitlength}{
\begin{fmfgraph}(20,20)
\fmfbottomn{v}{6}
\fmftop{t}
\fmf{phantom,tension=6}{t,v}
\begin{fmffor}{i}{1}{1}{6}
\fmf{plain}{v[i],v}
\end{fmffor}
\fmffreeze
\fmf{ghost,left=0.3}{t,v6}
\fmfv{decor.shape=square,decor.filled=empty,decor.size=5}{v6}
\fmfv{d.sh=circle,d.f=30,d.size=25pt}{v}
\end{fmfgraph}}
\end{equation}
The action of $\mathcal{F}$ on diagrams with an insertion of a gauge
boson  into an internal gauge boson line is defined as follows:
replace internal gauge bosons by ghosts in all possible ways until the
external particles are reached. In general, one original diagram can
correspond to more than one contact term, e.\,g.
\begin{multline}
  \parbox{25\unitlength}{
\begin{fmfgraph}(25,20)
\fmfbottomn{v}{6}
\fmftopn{t}{3}
\fmf{phantom,tension=3}{t1,v}
\fmf{phantom,tension=3}{t3,u}
\fmf{photon}{u,i,v}
\begin{fmffor}{i}{1}{1}{3}
\fmf{plain}{v[i],v}
\fmf{plain}{v[i+3],u}
\end{fmffor}
\fmfv{d.sh=circle,d.f=30,d.size=15pt}{v}
\fmfv{d.sh=circle,d.f=30,d.size=15pt}{u}
\fmffreeze
\fmf{photon}{t2,i}
\fmfdot{i}
\end{fmfgraph}}\,\xrightarrow{\mathcal{F}} \\
-\sum_{\Phi_i}
\parbox{30\unitlength}{\fmfframe(0,0)(0,5){%
\begin{fmfgraph*}(25,20)
\fmfbottomn{v}{6}
\fmftopn{t}{3}
\fmf{phantom,tension=3}{t1,v}
\fmf{phantom,tension=3}{t3,u}
\fmf{photon}{i,u}
\fmf{ghost}{i,v}
\begin{fmffor}{i}{1}{1}{3}
\fmf{plain}{v[i],v}
\fmf{plain}{v[i+3],u}
\end{fmffor}
\fmfv{d.sh=diamond,d.f=30,d.size=15pt}{v}
\fmfv{d.sh=circle,d.f=30,d.size=15pt}{u}
\fmffreeze
\fmf{ghost}{t2,i}
\fmf{ghost,left=0.5}{v,v3}
\fmfdot{i}
\fmfv{decor.shape=square,decor.filled=empty,decor.size=5,label=$\Phi_i$}{v3}
\end{fmfgraph*}}}\,-\,
\parbox{30\unitlength}{\fmfframe(0,0)(0,5){%
\begin{fmfgraph*}(25,20)
\fmfbottomn{v}{6}
\fmftopn{t}{3}
\fmf{phantom,tension=3}{t1,v}
\fmf{phantom,tension=3}{t3,u}
\fmf{photon}{i,v}
\fmf{ghost}{i,u}
\begin{fmffor}{i}{1}{1}{3}
\fmf{plain}{v[i],v}
\fmf{plain}{v[i+3],u}
\end{fmffor}
\fmfv{d.sh=diamond,d.f=30,d.size=15pt}{u}
\fmfv{d.sh=circle,d.f=30,d.size=15pt}{v}
\fmffreeze
\fmf{ghost}{t2,i}
\fmf{ghost,right=0.5}{u,v4}
\fmfdot{i}
\fmfv{decor.shape=square,decor.filled=empty,decor.size=5,label=$\Phi_i$}{v4}
\end{fmfgraph*}}}
\end{multline}
\end{subequations}
For each external gauge boson or GB, the inhomogeneous terms in the BRS
transformation laws~(\ref{eq:brs-gauge-graph})
and~(\ref{eq:brs-gold-graph}) have to be added.
\end{definition}
In this way we can associate a set of contact
diagrams to every FD. Conversely, replacing a ghost line by a gauge boson line and
the BRS transformed field by an external particle, we can associate
exactly one FD to each contact diagram. 

Since (in a linear $R_\xi$ gauge) to every vertex of the ghosts
corresponds a vertex of the gauge bosons, the contact terms
generated in that way from the complete set of FDs must indeed be
all the contact terms required by the STI.
Therefore it is sensible to define:
\begin{definition}
\label{def:sti}
  A subset of diagrams satisfies a STI if the contact terms obtained
  by the  mapping $\mathcal{F}$ agree with the result of contracting
  an external gauge boson.
\end{definition}
and
\begin{definition}
\label{def:gic}
  A GIC is a subset of FDs that satisfies the STIs and in addition becomes
  independent of the gauge parameter when all external particles are
  on-shell.
\end{definition}

\subsection{Definition of gauge flips}
\label{sec:flip-def}
As we will see below, we have to define the elementary gauge flips
as the minimal set of four point diagrams with a given set of
external particles and at least one external gauge boson,
\emph{satisfying the STI}. In a generic graphical notation, the gauge
flips are denoted as:
\begin{equation}
\label{eq:generic-flips}
\mathcal{G}_4=\left\{\parbox{15\unitlength}{
\begin{fmfgraph}(15,15)
\fmfleft{a,b}
\fmfright{f1,f2}
\fmf{photon}{a,fwf1}
\fmf{plain}{fwf1,fwf2}
\fmf{plain}{fwf2,b}
\fmf{plain}{fwf1,f1}
\fmf{plain}{fwf2,f2}
\fmfdot{fwf1}
\fmfdot{fwf2}
\end{fmfgraph}}\,,\,
\parbox{15\unitlength}{
\begin{fmfgraph}(15,15)
\fmfleft{a,b}
\fmfright{f1,f2}
\fmf{photon}{a,fwf1}
\fmf{plain}{fwf1,fwf2}
\fmf{plain}{fwf2,b}
\fmf{phantom}{fwf2,f2}
\fmf{phantom}{fwf1,f1}
\fmffreeze
\fmf{plain}{fwf2,f1}
\fmf{plain,rubout}{fwf1,f2}
\fmfdot{fwf1}
\fmfdot{fwf2}
\end{fmfgraph}}\, ,\, 
\parbox{15\unitlength}{
\begin{fmfgraph}(15,15)
\fmfleft{a,b}
\fmfright{f1,f2}
\fmf{photon}{a,fwf}
\fmf{plain}{fwf,b}
\fmf{plain}{fwf,www}
\fmf{plain}{www,f1}
\fmf{plain}{www,f2}
\fmfdot{fwf}
\fmfdot{www}
\end{fmfgraph}}\, , \, 
\parbox{15\unitlength}{
\begin{fmfgraph}(15,15)
\fmfleft{a,b}
\fmfright{f1,f2}
\fmf{photon}{a,c}
\fmf{plain}{c,b}
\fmf{plain}{c,f1}
\fmf{plain}{c,f2}
\fmfdot{c}
\end{fmfgraph}}\right\}
\end{equation}
The internal particles appearing in these diagrams are determined by
the requirement that the flips are the \emph{minimal} set of diagrams
satisfying the STIs. Of course, only diagrams allowed by the Feynman rules
have to be included in each application of~(\ref{eq:generic-flips}).

In $R_\xi$ gauge also the corresponding four point functions with some or
all external gauge bosons replaced by GBs appear as
subamplitudes in larger diagrams. It may happen, that the minimal GIC
for the gauge boson subamplitude does not coincide with the minimal
GIC for the GB subamplitude. In this case, the gauge
flips have to be defined in such a way that not only the gauge boson
amplitudes, but also all corresponding GB
amplitudes satisfy the STIs.

In the presence of quartic Higgs vertices, we will also need
elementary flips among five point functions:
\begin{equation}
\label{eq:generic-flips5}
\mathcal{G}_5=\left\{\,
\begin{aligned}
&\parbox{15\unitlength}{
\begin{fmfgraph}(15,15)
\fmfleftn{l}{2}
\fmfrightn{r}{3}
\fmf{dashes}{l1,g}
\fmf{dashes}{l2,g}
\fmf{dashes}{r1,g}
\fmf{plain,tension=2}{r3,i}
\fmf{dashes,tension=2}{i,g}
\fmffreeze
\fmf{photon}{r2,i}
\fmfdot{g,i}
\end{fmfgraph}}\quad,\quad
\parbox{15\unitlength}{
\begin{fmfgraph}(15,15)
\fmfleft{l1,l2}
\fmfright{r1,r2,r3}
\fmf{dashes}{l1,g}
\fmf{dashes}{l2,g}
\fmf{dashes}{r3,g}
\fmf{plain,tension=2}{r1,i}
\fmf{dashes,tension=2}{i,g}
\fmffreeze
\fmf{photon}{r2,i}
\fmfdot{g,i}
\end{fmfgraph}}\,,\\
&\parbox{15\unitlength}{
\begin{fmfgraph}(15,15)
\fmfleft{l1,l2,l3}
\fmfright{r1,r3}
\fmf{dashes}{l3,g}
\fmf{dashes}{r1,g}
\fmf{plain,tension=2}{l1,i}
\fmf{dashes,tension=2}{i,g}
\fmf{dashes}{g,r3}
\fmffreeze
\fmf{photon}{l2,i}
\fmfdot{g,i}
\end{fmfgraph}}\quad,\quad
\parbox{15\unitlength}{
\begin{fmfgraph}(15,15)
\fmfleft{l1,l2,l3}
\fmfright{r1,r2}
\fmf{dashes}{l1,g}
\fmf{dashes}{r1,g}
\fmf{dashes}{r2,g}
\fmf{plain,tension=2}{l3,i}
\fmf{dashes,tension=2}{i,g}
\fmffreeze
\fmf{photon}{l2,i}
\fmfdot{g,i}
\end{fmfgraph}}
\end{aligned}
\right\}
\end{equation}

\subsection{Gauge parameter independence}
\label{sec:gpi}
We will now show on tree level that the gauge parameter independence
of physical amplitudes is a consequence of the WIs of the theory.
To obtain gauge parameter independent amplitudes, the $\xi$ dependence
of the propagators must cancel among the gauge boson and the GB
exchange diagrams. To see how this works, we note that the gauge
boson propagator in $R_\xi$ gauge can be written as the propagator in
unitarity gauge plus a term proportional to the GB
propagator:
\begin{align}
\label{eq:rxi-uni-prop}
  \ii D_W^{\mu\nu}(q)&=\frac{1}{q^2-m_W^2}\left(g^{\mu\nu}-\frac{q^\mu
  q^\nu}{m_W^2}\right) +\frac{q^\mu q^\nu}{m_W^2}\frac{1}{q^2-\xi m_W^2}\nonumber\\
&=\ii D^{\mu\nu}_{W,U}-\frac{q^\mu q^\nu}{m_W^2}(\ii D_{\phi}) 
\end{align} 
If we consider a gauge boson that is exchanged between two
subamplitudes together with the corresponding GB, we see
that the gauge parameter dependence cancels between the unphysical
part of the gauge boson propagator and the GB propagator
if the subamplitudes satisfy the WI~(\ref{eq:gf-wi}).
However, in general we \emph{cannot} decompose a scattering matrix
element into a sum over subamplitudes, connected by \emph{one}
propagator:
\begin{equation}
\quad\parbox{20\unitlength}{
\begin{fmfgraph*}(20,20)
\fmfsurroundn{v}{6}
\begin{fmffor}{i}{1}{1}{6}
\fmf{plain}{v[i],v}
\end{fmffor}
\fmfv{d.sh=circle,d.f=empty,d.size=30pt,l=$N$,l.d=0}{v}
\end{fmfgraph*}}\neq \sum_{i+j=N+2}
\parbox{25\unitlength}{\begin{fmfgraph*}(25,20)
\fmfleftn{l}{3}
\fmfrightn{r}{3}
\begin{fmffor}{i}{1}{1}{3}
\fmf{plain}{l[i],i1}
\fmf{plain}{r[i],i2}
\end{fmffor}
\fmf{plain}{i1,i2}
\fmfv{d.sh=circle,d.f=empty,d.size=20pt,l=$i$,l.d=0}{i1}
\fmfv{d.sh=circle,d.f=empty,d.size=20pt,l=$j$,l.d=0}{i2}
\end{fmfgraph*}}
\end{equation}
For example, the grove $G_s$ from~(\ref{eq:qcd-grove}) cannot be factorized
into subamplitudes:
 \begin{equation}
G_s\neq\quad   
\parbox{20\unitlength}{
\begin{fmfgraph}(20,20)
\fmfleft{a,b}
\fmfright{f1,f2}
\fmftop{z}
\fmf{fermion,tension=1.0}{eae,a}
\fmf{photon,tension=1.0}{uwd,eae}
\fmf{fermion}{uwd,f1}
\fmf{fermion}{b,eae}
\fmf{fermion}{f2,uwd}
\fmfv{d.sh=circle,d.f=empty,d.size=20pt}{uwd}
\fmfdot{eae}
\fmffreeze
\fmf{photon}{z,uwd}
\end{fmfgraph}}\qquad +\qquad
\parbox{20\unitlength}{
\begin{fmfgraph}(20,20)
\fmfright{a,b}
\fmfleft{f1,f2}
\fmftop{z}
\fmf{fermion,tension=1.0}{eae,a}
\fmf{photon,tension=1.0}{uwd,eae}
\fmf{fermion}{uwd,f1}
\fmf{fermion}{b,eae}
\fmf{fermion}{f2,uwd}
\fmfv{d.sh=circle,d.f=empty,d.size=20pt}{uwd}
\fmfdot{eae}
\fmffreeze
\fmf{photon}{z,uwd}
\end{fmfgraph}}
\end{equation}
because the diagram
\begin{equation}\label{eq:double_diag}
\parbox{20\unitlength}{
\begin{fmfgraph}(20,15)
\fmftopn{b}{3}
\fmfbottomn{t}{2}
\fmfright{r}
\fmf{fermion}{t2,t,b3}
\fmf{photon,tension=2}{v1,v2}
\fmf{photon,tension=2}{v2,t}
\fmf{fermion}{t1,v1,b1}
\fmffreeze
\fmf{photon}{b2,v2}
\fmfdot{v1,v2,t}
\end{fmfgraph}}
\end{equation}
contributes to both subamplitudes and would be counted twice.

Nevertheless, this problem can be avoided if we consider an
infinitesimal change in the gauge parameter
\begin{equation*}
\xi = \xi_0+\delta\xi
\end{equation*}
and work to first order in $\delta\xi$. As usual, finite changes of
$\xi$ are generated by successive infinitesimal transformations.

Under an infinitesimal variation of $\xi$, the gauge boson propagator
changes as  (see~(\ref{eq:rxi-uni-prop}))
\begin{equation}
\label{eq:dxi-gauge}
  D_{W,\xi}^{\mu\nu}(q)
    = D_{W,\xi_0}^{\mu\nu}(q)
    - \frac{\ii q^\mu q^\nu}{(q^2-\xi_0 m_W^2)^2}\delta\xi
    + \mathcal{O}((\delta\xi)^2)
\end{equation}
We will  represent  this decomposition  graphically as
\begin{equation}
\label{eq:uni-xi-graph}
\begin{array}{rcl}
\parbox{15\unitlength}{
\begin{fmfgraph}(15,15)
\fmfleft{l}
\fmfright{r}
\fmf{photon}{l,r}  
\end{fmfgraph}}
&\quad=\quad
\parbox{15\unitlength}{
\begin{fmfgraph}(15,15)
\fmfleft{l}
\fmfright{r}
\fmf{gluon}{l,r}  
\end{fmfgraph}}
&\quad+\quad\parbox{15\unitlength}{
\begin{fmfgraph}(15,15)
\fmfleft{l}
\fmfright{r}
\fmf{dbl_dots}{l,r}  
\end{fmfgraph}}\\
 D_{W,\xi}^{\mu\nu}(q)&= D^{\mu\nu}_{W,\xi_0}&-q^\mu q^\nu D_{c,\xi_0}^2(q)\delta\xi 
\end{array}
\end{equation}
Similarly the GB propagator becomes
\begin{equation}
\label{eq:dxi-gold}
  D_{\phi\xi}
    = D_{\phi \xi_0}+m_W^2 \frac{\ii}{(q^2-\xi_0 m_W^2)^2}\delta\xi
    + \mathcal{O}((\delta\xi)^2)
\end{equation}
Inserting the decomposition~(\ref{eq:uni-xi-graph}) into the
diagram~(\ref{eq:double_diag}) we get two contributions
linear in $\delta\xi$. Therefore we can factorize the contributions
linear in $\delta \xi$:
\begin{equation}
\partial_\xi G_s=\quad   
\parbox{20\unitlength}{
\begin{fmfgraph}(20,20)
\fmfleft{a,b}
\fmfright{f1,f2}
\fmftop{z}
\fmf{fermion,tension=1.0}{a,eae}
\fmf{dbl_dots,tension=1.0}{uwd,eae}
\fmf{fermion}{f1,uwd}
\fmf{fermion}{eae,b}
\fmf{fermion}{uwd,f2}
\fmfv{d.sh=circle,d.f=empty,d.size=20pt}{uwd}
\fmfdot{eae}
\fmffreeze
\fmf{photon}{z,uwd}
\end{fmfgraph}}\qquad +\qquad
\parbox{20\unitlength}{
\begin{fmfgraph}(20,20)
\fmfright{a,b}
\fmfleft{f1,f2}
\fmftop{z}
\fmf{fermion,tension=1.0}{a,eae}
\fmf{dbl_dots,tension=1.0}{uwd,eae}
\fmf{fermion}{f1,uwd}
\fmf{fermion}{eae,b}
\fmf{fermion}{uwd,f2}
\fmfv{d.sh=circle,d.f=empty,d.size=20pt}{uwd}
\fmfdot{eae}
\fmffreeze
\fmf{photon}{z,uwd}
\end{fmfgraph}}
\end{equation}
Using the WIs for the subamplitudes, we see that the gauge parameter
dependence of the grove $G_s$ vanishes. 

For general amplitudes, the terms linear in $\delta\xi$ can be
factorized in a similar way. To see this, we regard the unphysical
propagators as new `particles' with the appropriate Feynman rules. The
$\mathcal{O}(\delta\xi)$ contribution to the GF consists of FDs where
the new particle appears \emph{exactly} once and no double counting
occurs. The same reasoning can be applied to the variation of the
GB propagator.

Therefore, the parts of the propagators linear in $\delta\xi$ connect
complete subamplitudes that satisfy the WIs and as a result the amplitude
is gauge parameter independent:
\begin{equation}
\label{eq:xi-factorize}
\partial_\xi \,\parbox{20\unitlength}{
\begin{fmfgraph*}(20,20)
\fmfsurroundn{v}{6}
\begin{fmffor}{i}{1}{1}{6}
\fmf{plain}{v[i],v}
\end{fmffor}
\fmfv{d.sh=circle,d.f=empty,d.size=30pt,l=$\xi$,l.d=0}{v}
\end{fmfgraph*}}
 = \sum_{i+j=N+2}
\parbox{25\unitlength}{\begin{fmfgraph*}(25,20)
\fmfleftn{l}{3}
\fmfrightn{r}{3}
\begin{fmffor}{i}{1}{1}{3}
\fmf{plain}{l[i],i1}
\fmf{plain}{r[i],i2}
\end{fmffor}
\fmf{dbl_dots}{i1,i2}
\fmfv{d.sh=circle,d.f=empty,d.size=20pt,l=$i$,l.d=0}{i1}
\fmfv{d.sh=circle,d.f=empty,d.size=20pt,l=$j$,l.d=0}{i2}
\end{fmfgraph*}}=0
\end{equation}
If the external particles are off-shell, we have to use the STIs
instead of the WIs and the GFs become $\xi$ dependent.

\subsection{Gauge invariance classes}
\label{sec:n-ppoint-groves}
We are now ready to show that the groves obtained by the gauge flips
defined as in section~\ref{sec:flip-def} are indeed the minimal GICs
according to definition~\ref{def:gic}.
For amplitudes without external gauge bosons, GICs are defined as
subsets of FDs that are gauge parameter independent when all external
particles are on their mass shell.

From~(\ref{eq:xi-factorize}) we know that it suffices that the parts
of the propagators linear in $\delta\xi$ connect subamplitudes
satisfying the WIs in order to obtain gauge parameter independent
quantities.  As induction hypothesis, we assume that it has been shown
that the $N-1$ particle
diagrams connected by gauge flips satisfy the STIs and are gauge
parameter independent. Therefore applying gauge flips to all
\emph{internal} gauge bosons of a $N$-point function ensures its gauge
parameter independence. Thus the case of amplitudes without external
gauge bosons is reduced to the discussion of amplitudes with
fewer external particles and external gauge bosons.

Now we consider the insertion of a gauge boson into a FD
with $N-1$ external particles.
We  pick out an arbitrary cubic vertex of the diagram and
insert the gauge boson into all three legs of the vertex and include a
quartic vertex (if allowed by the Feynman rules):
\begin{multline}
\label{eq:n-ampl-ternary}
\parbox{36\unitlength}{
      \begin{fmfgraph*}(35,20)
       \fmftop{x}
       \fmfbottomn{n}{6}
       \fmfleft{li}
       \fmfright{ri}
       \fmf{phantom,tension=8}{li,i}
       \fmf{phantom,tension=8}{ri,l}  
       \fmf{phantom,tension=6}{x,i}
       \fmf{phantom,tension=6}{x,l}        
       \fmf{phantom}{i,l}       
        \begin{fmffor}{i}{1}{1}{2}
         \fmf{plain,tension=3}{i,n[i]}
       \end{fmffor}        
       \begin{fmffor}{i}{5}{1}{6}
         \fmf{plain,tension=3}{l,n[i]}
       \end{fmffor}
       \fmf{plain,tension=1}{i,n}
       \fmf{plain,tension=1}{n,l}        
        \fmffreeze
        \begin{fmffor}{i}{3}{1}{4}
         \fmf{plain}{j,n[i]}
       \end{fmffor} 
        \fmf{plain}{n,j} 
      \fmfdot{n}
      \fmffreeze
      \fmf{photon}{x,y}
      \fmf{phantom,label=$\otimes$,la.di=0}{y,n}
      \fmfv{d.sh=circle,d.f=30,d.si=12}{i}
      \fmfv{d.sh=circle,d.f=30,d.si=12}{j}
      \fmfv{d.sh=circle,d.f=30,d.si=12}{l}
      \end{fmfgraph*}}\, \equiv\\
\parbox{36\unitlength}{
      \begin{fmfgraph}(35,20)
       \fmftop{x}
       \fmfbottomn{n}{6}
       \fmfleft{li}
       \fmfright{ri}
       \fmf{phantom,tension=8}{li,i}
       \fmf{phantom,tension=8}{ri,l}  
       \fmf{phantom,tension=6}{x,i}
       \fmf{phantom,tension=6}{x,l}        
       \fmf{phantom}{i,l}
        \begin{fmffor}{i}{1}{1}{2}
         \fmf{plain,tension=3}{i,n[i]}
       \end{fmffor}        
       \begin{fmffor}{i}{5}{1}{6}
         \fmf{plain,tension=3}{l,n[i]}
       \end{fmffor}
       \fmf{plain,tension=2}{i,y,n}
       \fmf{plain,tension=1}{n,l}        
        \fmffreeze
       \fmf{photon,tension=6}{x,y}
        \begin{fmffor}{i}{3}{1}{4}
         \fmf{plain}{j,n[i]}
       \end{fmffor} 
        \fmf{plain}{n,j} 
      \fmfdot{n,y}
      \fmfv{d.sh=circle,d.f=30,d.si=12pt}{i}
      \fmfv{d.sh=circle,d.f=30,d.si=12pt}{j}
      \fmfv{d.sh=circle,d.f=30,d.si=12pt}{l}
      \end{fmfgraph}}
+
\parbox{36\unitlength}{
      \begin{fmfgraph}(35,20)
       \fmftop{x}
       \fmfbottomn{n}{6}
        \fmfleft{li}
       \fmfright{ri}
       \fmf{phantom,tension=8}{li,i}
       \fmf{phantom,tension=8}{ri,l}   
       \fmf{phantom,tension=6}{x,i}
       \fmf{phantom,tension=6}{x,l}        
       \fmf{phantom}{i,l}       
        \begin{fmffor}{i}{1}{1}{2}
         \fmf{plain,tension=3}{i,n[i]}
       \end{fmffor}        
       \begin{fmffor}{i}{5}{1}{6}
         \fmf{plain,tension=3}{l,n[i]}
       \end{fmffor}
       \fmf{plain,tension=1}{i,n}
       \fmf{plain,tension=2}{n,y,l}        
        \fmffreeze
        \fmf{photon}{x,y}
        \begin{fmffor}{i}{3}{1}{4}
         \fmf{plain}{j,n[i]}
       \end{fmffor} 
        \fmf{plain}{n,j} 
      \fmfdot{n,y}
      \fmfv{d.sh=circle,d.f=30,d.si=12}{i}
      \fmfv{d.sh=circle,d.f=30,d.si=12}{j}
      \fmfv{d.sh=circle,d.f=30,d.si=12}{l}
      \end{fmfgraph}}\\
+\parbox{36\unitlength}{
      \begin{fmfgraph}(35,20)
       \fmftop{x}
       \fmfbottomn{n}{6}
        \fmfleft{li}
       \fmfright{ri}
       \fmfleft{li}
       \fmfright{ri}
       \fmf{phantom,tension=8}{li,i}
       \fmf{phantom,tension=8}{ri,l}  
       \fmf{phantom,tension=8}{li,i}
       \fmf{phantom,tension=8}{ri,l}  
       \fmf{phantom,tension=6}{x,i}
       \fmf{phantom,tension=6}{x,l}        
       \fmf{phantom}{i,l}       
        \begin{fmffor}{i}{1}{1}{2}
         \fmf{plain,tension=3}{i,n[i]}
       \end{fmffor}        
       \begin{fmffor}{i}{5}{1}{6}
         \fmf{plain,tension=3}{l,n[i]}
       \end{fmffor}
       \fmf{plain,tension=1}{i,n}
       \fmf{plain,tension=1}{n,l}        
        \fmffreeze
        \begin{fmffor}{i}{3}{1}{4}
         \fmf{plain}{j,n[i]}
       \end{fmffor} 
        \fmf{plain}{n,y,j} 
      \fmfdot{n,y}
      \fmffreeze
      \fmf{photon,rubout,left=0.3}{x,y}
      \fmfv{d.sh=circle,d.f=30,d.si=12}{i}
      \fmfv{d.sh=circle,d.f=30,d.si=12}{j}
      \fmfv{d.sh=circle,d.f=30,d.si=12}{l}
      \end{fmfgraph}}
+
\parbox{36\unitlength}{
      \begin{fmfgraph}(35,20)
       \fmftop{x}
       \fmfbottomn{n}{6}
       \fmfleft{li}
       \fmfright{ri}
       \fmf{phantom,tension=8}{li,i}
       \fmf{phantom,tension=8}{ri,l}  
       \fmf{phantom,tension=6}{x,i}
       \fmf{phantom,tension=6}{x,l}        
       \fmf{phantom}{i,l}       
        \begin{fmffor}{i}{1}{1}{2}
         \fmf{plain,tension=3}{i,n[i]}
       \end{fmffor}        
       \begin{fmffor}{i}{5}{1}{6}
         \fmf{plain,tension=3}{l,n[i]}
       \end{fmffor}
       \fmf{plain,tension=1}{i,n}
       \fmf{plain,tension=1}{n,l}        
        \fmffreeze
        \begin{fmffor}{i}{3}{1}{4}
         \fmf{plain}{j,n[i]}
       \end{fmffor} 
        \fmf{plain}{n,j} 
      \fmfdot{n}
      \fmffreeze
      \fmf{photon}{x,n}
      \fmfv{d.sh=circle,d.f=30,d.si=12}{i}
      \fmfv{d.sh=circle,d.f=30,d.si=12}{j}
      \fmfv{d.sh=circle,d.f=30,d.si=12}{l}
      \end{fmfgraph}}
\end{multline}
These diagrams are connected by the gauge
flips~$\mathcal{G}_4$~(\ref{eq:generic-flips}). The insertion of a
contracted gauge boson into a FD via a three point vertex can be 
evaluated using the STI~(\ref{eq:sti-3graph}) and the 
identity~(\ref{eq:inv-prop}):
\begin{multline}
\label{eq:n-point-internal}
\parbox{36\unitlength}{
      \begin{fmfgraph}(35,20)
       \fmftop{x}
       \fmfbottomn{n}{6}
       \fmfleft{li}
       \fmfright{ri}
       \fmf{phantom,tension=8}{li,i}
       \fmf{phantom,tension=8}{ri,l}  
       \fmf{phantom,tension=6}{x,i}
       \fmf{phantom,tension=6}{x,l}        
       \fmf{phantom}{i,l}
        \begin{fmffor}{i}{1}{1}{2}
         \fmf{plain,tension=3}{i,n[i]}
       \end{fmffor}        
       \begin{fmffor}{i}{5}{1}{6}
         \fmf{plain,tension=3}{l,n[i]}
       \end{fmffor}
       \fmf{plain,tension=2}{i,y,n}
       \fmf{plain,tension=1}{n,l}        
        \fmffreeze
       \fmf{double,tension=6}{x,y}
        \begin{fmffor}{i}{3}{1}{4}
         \fmf{plain}{j,n[i]}
       \end{fmffor} 
        \fmf{plain}{n,j} 
      \fmfdot{n,y}
      \fmfv{d.sh=circle,d.f=30,d.si=12pt}{i}
      \fmfv{d.sh=circle,d.f=30,d.si=12pt}{j}
      \fmfv{d.sh=circle,d.f=30,d.si=12pt}{l}
      \end{fmfgraph}}\\
=-\parbox{36\unitlength}{
      \begin{fmfgraph}(35,20)
       \fmftop{x}
       \fmfbottomn{n}{6}
       \fmfleft{li}
       \fmfright{ri}
       \fmf{phantom,tension=8}{li,i}
       \fmf{phantom,tension=8}{ri,l}  
       \fmf{phantom,tension=6}{x,i}
       \fmf{phantom,tension=6}{x,l}        
       \fmf{phantom}{i,l}
        \begin{fmffor}{i}{1}{1}{2}
         \fmf{plain,tension=3}{i,n[i]}
       \end{fmffor}        
       \begin{fmffor}{i}{5}{1}{6}
         \fmf{plain,tension=3}{l,n[i]}
       \end{fmffor}
       \fmf{plain,tension=2.5}{i,y}
        \fmf{plain,tension=1.5}{y,n}
       \fmf{plain,tension=1}{n,l}        
        \fmffreeze
       \fmf{ghost,tension=6}{x,y}
        \begin{fmffor}{i}{3}{1}{4}
         \fmf{plain}{j,n[i]}
       \end{fmffor} 
        \fmf{plain}{n,j} 
      \fmfdot{n}
       \fmfv{decor.shape=square,decor.filled=empty,decor.size=5}{y}
      \fmfv{d.sh=circle,d.f=30,d.si=12pt}{i}
      \fmfv{d.sh=circle,d.f=30,d.si=12pt}{j}
      \fmfv{d.sh=circle,d.f=30,d.si=12pt}{l}
      \end{fmfgraph}}
-\parbox{36\unitlength}{
      \begin{fmfgraph}(35,20)
       \fmftop{x}
       \fmfbottomn{n}{6}
       \fmfleft{li}
       \fmfright{ri}
       \fmf{phantom,tension=8}{li,i}
       \fmf{phantom,tension=8}{ri,l}  
       \fmf{phantom,tension=6}{x,i}
       \fmf{phantom,tension=6}{x,l}        
       \fmf{phantom}{i,l}
        \begin{fmffor}{i}{1}{1}{2}
         \fmf{plain,tension=3}{i,n[i]}
       \end{fmffor}        
       \begin{fmffor}{i}{5}{1}{6}
         \fmf{plain,tension=3}{l,n[i]}
       \end{fmffor}
       \fmf{phantom,tension=1}{i,n}
       \fmf{plain,tension=1}{n,l}        
        \fmffreeze
          \fmf{plain,tension=1}{i,y}
         \fmf{plain,tension=3}{y,n}
        \fmffreeze 
       \fmf{ghost,tension=6}{x,y}
        \begin{fmffor}{i}{3}{1}{4}
         \fmf{plain}{j,n[i]}
       \end{fmffor} 
        \fmf{plain}{n,j} 
      \fmfdot{n}
       \fmfv{decor.shape=square,decor.filled=empty,decor.size=5}{y}
      \fmfv{d.sh=circle,d.f=30,d.si=12pt}{i}
      \fmfv{d.sh=circle,d.f=30,d.si=12pt}{j}
      \fmfv{d.sh=circle,d.f=30,d.si=12pt}{l}
      \end{fmfgraph}}
\end{multline}
For internal gauge bosons the additional
pieces~(\ref{subeq:ghost-sti}) in the STIs contribute
terms of the form 
\begin{equation}
\label{eq:ghost-internal}
\parbox{41\unitlength}{
      \begin{fmfgraph}(40,20)
       \fmftop{x}
       \fmfbottomn{n}{8}
       \fmfleft{li}
       \fmfright{ri}
       \fmf{phantom,tension=8}{li,i}
       \fmf{phantom,tension=8}{ri,l}  
       \fmf{phantom,tension=6}{x,i}
       \fmf{phantom,tension=6}{x,l}        
       \fmf{phantom}{i,l}
        \begin{fmffor}{i}{1}{1}{2}
         \fmf{plain,tension=3}{i,n[i]}
       \end{fmffor}        
       \begin{fmffor}{i}{7}{1}{8}
         \fmf{plain,tension=3}{l,n[i]}
       \end{fmffor}
       \fmf{plain,tension=4}{n,y}
       \fmf{phantom,tension=2}{y,i}
       \fmf{plain,tension=4}{n,l}        
        \fmffreeze
       \fmf{ghost,tension=6}{x,y}
        \begin{fmffor}{i}{5}{1}{6}
         \fmf{plain}{j,n[i]}
       \end{fmffor} 
        \fmf{plain}{n,j} 
      \fmf{ghost,tension=2}{y,k}
       \fmf{photon,tension=2}{k,i}
      \fmfdot{n,y}
      \fmfv{decor.shape=square,decor.filled=full,decor.size=5}{k}
      \fmfv{d.sh=circle,d.f=30,d.si=12pt}{i}
      \fmfv{d.sh=circle,d.f=30,d.si=12pt}{j}
      \fmfv{d.sh=circle,d.f=30,d.si=12pt}{l}
      \end{fmfgraph}}
\end{equation}
and similarly for GBs. Here the subdiagram
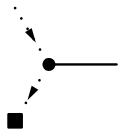
\begin{equation*}
\parbox{15\unitlength}{
\begin{fmfgraph}(15,15)
\fmfleft{l1,l2}
\fmfright{r}
\fmf{plain}{r,i}
\fmf{ghost}{l2,i,l1}
\fmfv{decor.shape=square,decor.filled=full,decor.size=5}{l1}
\fmfdot{i}
\end{fmfgraph}}
\end{equation*}
has to be understood as the insertion of a contracted ghost vertex, no
internal ghost propagator appears.  Note that the diagrams of this
form  only appear for particles that couple simultaneously to two gauge bosons
and to ghosts.

For the remaining diagrams  in~(\ref{eq:n-ampl-ternary}) involving
cubic vertices, we can repeat the same manipulations and find, using
the  STI of the four point vertex~(\ref{eq:sti4-graph}), that everything
cancels apart from the terms 
\begin{multline}
\label{eq:sti-int-n}
\parbox{36\unitlength}{
      \begin{fmfgraph*}(35,20)
       \fmftop{x}
       \fmfbottomn{n}{6}
       \fmfleft{li}
       \fmfright{ri}
       \fmf{phantom,tension=8}{li,i}
       \fmf{phantom,tension=8}{ri,l}  
       \fmf{phantom,tension=6}{x,i}
       \fmf{phantom,tension=6}{x,l}        
       \fmf{phantom}{i,l}       
        \begin{fmffor}{i}{1}{1}{2}
         \fmf{plain,tension=3}{i,n[i]}
       \end{fmffor}        
       \begin{fmffor}{i}{5}{1}{6}
         \fmf{plain,tension=3}{l,n[i]}
       \end{fmffor}
       \fmf{plain,tension=1}{i,n}
       \fmf{plain,tension=1}{n,l}        
        \fmffreeze
        \begin{fmffor}{i}{3}{1}{4}
         \fmf{plain}{j,n[i]}
       \end{fmffor} 
        \fmf{plain}{n,j} 
      \fmfdot{n}
      \fmffreeze
      \fmf{double}{x,y}
      \fmf{phantom,label=$\otimes$,la.di=0}{y,n}
      \fmfv{d.sh=circle,d.f=30,d.si=12}{i}
      \fmfv{d.sh=circle,d.f=30,d.si=12}{j}
      \fmfv{d.sh=circle,d.f=30,d.si=12}{l}
      \end{fmfgraph*}}\,=\,
-\parbox{36\unitlength}{
      \begin{fmfgraph}(35,20)
       \fmftop{x}
       \fmfbottomn{n}{6}
       \fmfleft{li}
       \fmfright{ri}
       \fmf{phantom,tension=8}{li,i}
       \fmf{phantom,tension=8}{ri,l}  
       \fmf{phantom,tension=6}{x,i}
       \fmf{phantom,tension=6}{x,l}        
       \fmf{phantom}{i,l}
        \begin{fmffor}{i}{1}{1}{2}
         \fmf{plain,tension=3}{i,n[i]}
       \end{fmffor}        
       \begin{fmffor}{i}{5}{1}{6}
         \fmf{plain,tension=3}{l,n[i]}
       \end{fmffor}
         \fmf{plain,tension=2.5}{i,y}
        \fmf{plain,tension=1.5}{y,n}
       \fmf{plain,tension=1}{n,l}        
        \fmffreeze
       \fmf{ghost,tension=6}{x,y}
        \begin{fmffor}{i}{3}{1}{4}
         \fmf{plain}{j,n[i]}
       \end{fmffor} 
        \fmf{plain}{n,j} 
      \fmfdot{n}
       \fmfv{decor.shape=square,decor.filled=empty,decor.size=5}{y}
      \fmfv{d.sh=circle,d.f=30,d.si=12pt}{i}
      \fmfv{d.sh=circle,d.f=30,d.si=12pt}{j}
      \fmfv{d.sh=circle,d.f=30,d.si=12pt}{l}
      \end{fmfgraph}}\\
- \parbox{36\unitlength}{
      \begin{fmfgraph}(35,20)
       \fmftop{x}
       \fmfbottomn{n}{6}
        \fmfleft{li}
       \fmfright{ri}
       \fmf{phantom,tension=8}{li,i}
       \fmf{phantom,tension=8}{ri,l}   
       \fmf{phantom,tension=6}{x,i}
       \fmf{phantom,tension=6}{x,l}        
       \fmf{phantom}{i,l}       
        \begin{fmffor}{i}{1}{1}{2}
         \fmf{plain,tension=3}{i,n[i]}
       \end{fmffor}        
       \begin{fmffor}{i}{5}{1}{6}
         \fmf{plain,tension=3}{l,n[i]}
       \end{fmffor}
       \fmf{plain,tension=1}{i,n}
         \fmf{plain,tension=1.5}{n,y}
        \fmf{plain,tension=2.5}{y,l}
        \fmffreeze
        \fmf{ghost}{x,y}
        \begin{fmffor}{i}{3}{1}{4}
         \fmf{plain}{j,n[i]}
       \end{fmffor} 
        \fmf{plain}{n,j} 
      \fmfdot{n}
       \fmfv{decor.shape=square,decor.filled=empty,decor.size=5}{y}
      \fmfv{d.sh=circle,d.f=30,d.si=12}{i}
      \fmfv{d.sh=circle,d.f=30,d.si=12}{j}
      \fmfv{d.sh=circle,d.f=30,d.si=12}{l}
      \end{fmfgraph}}
-\parbox{36\unitlength}{
       \begin{fmfgraph}(35,20)
       \fmftop{x}
       \fmfbottomn{n}{6}
        \fmfleft{li}
       \fmfright{ri}
       \fmfleft{li}
       \fmfright{ri}
       \fmf{phantom,tension=8}{li,i}
       \fmf{phantom,tension=8}{ri,l}  
       \fmf{phantom,tension=6}{x,i}
       \fmf{phantom,tension=6}{x,l}        
       \fmf{phantom}{i,l}       
        \begin{fmffor}{i}{1}{1}{2}
         \fmf{plain,tension=3}{i,n[i]}
       \end{fmffor}        
       \begin{fmffor}{i}{5}{1}{6}
         \fmf{plain,tension=3}{l,n[i]}
       \end{fmffor}
       \fmf{plain,tension=1}{i,n}
       \fmf{plain,tension=1}{n,l}        
        \fmffreeze
        \begin{fmffor}{i}{3}{1}{4}
         \fmf{plain}{j,n[i]}
       \end{fmffor} 
        \fmf{plain}{n,y,j} 
      \fmfdot{n}
       \fmfv{decor.shape=square,decor.filled=empty,decor.size=5}{y}
      \fmffreeze
      \fmf{ghost,left=0.3}{x,y}
      \fmfv{d.sh=circle,d.f=30,d.si=12}{i}
      \fmfv{d.sh=circle,d.f=30,d.si=12}{j}
      \fmfv{d.sh=circle,d.f=30,d.si=12}{l}
      \end{fmfgraph}}
\end{multline}
The contributions~(\ref{eq:ghost-internal}) exclusive to internal
gauge bosons and GBs will be discussed below.

To cancel the remaining diagrams in~(\ref{eq:sti-int-n}), we have to
`zoom in' into the blobs, insert the external gauge boson at the next
vertex and repeat the same procedure for the next vertices. This will
cancel the terms from~(\ref{eq:sti-int-n}) but leaves new terms of the
same form at the next vertices. This process can be iterated until
the external particles are reached. The remaining terms are the
contact terms of the STI for the GF with the ghost line going
through the diagram without interaction:
\begin{equation}
\parbox{36\unitlength}{
      \begin{fmfgraph}(35,20)
       \fmftop{x}
       \fmfbottomn{n}{6}
        \fmfleft{li}
       \fmfright{ri}
       \fmfleft{li}
       \fmfright{ri}
       \fmf{phantom,tension=8}{li,i}
       \fmf{phantom,tension=8}{ri,l}  
       \fmf{phantom,tension=6}{x,i}
       \fmf{phantom,tension=6}{x,l}        
       \fmf{phantom}{i,l}       
        \begin{fmffor}{i}{1}{1}{2}
         \fmf{plain,tension=3}{i,n[i]}
       \end{fmffor} 
       \fmf{plain,tension=3}{l,n5}
       \fmf{plain,tension=3}{l,n6}
       \fmf{plain,tension=1}{i,n}
       \fmf{plain,tension=1}{n,l}        
        \fmffreeze
        \begin{fmffor}{i}{3}{1}{4}
         \fmf{plain}{j,n[i]}
       \end{fmffor} 
        \fmf{plain}{n,j} 
      \fmfdot{n}
      \fmfv{decor.shape=square,decor.filled=empty,decor.size=5}{n6}
      \fmffreeze
      \fmf{ghost,left=0.5}{x,n6}
      \fmfv{d.sh=circle,d.f=30,d.si=12}{i}
      \fmfv{d.sh=circle,d.f=30,d.si=12}{j}
      \fmfv{d.sh=circle,d.f=30,d.si=12}{l}
      \end{fmfgraph}}
\end{equation}
Clearly, the procedure discussed above amounts to gauge-flipping
the external gauge boson through the original diagram. The contact
terms generated this way are among those generated by the mapping
$\mathcal{F}$~(\ref{subeq:generate-contact}) from diagrams with an
insertion of the gauge boson into an external leg.

The contraction of a gauge boson inserted adjacent to a quartic Higgs vertex  can
be treated analogously. This time the cancellation takes place because
of the STI~(\ref{eq:sti5-graph}). It involves the diagrams connected
by the five point flips ~(\ref{eq:generic-flips5}):
\begin{multline}\label{eq:5-contact}
\sum_{\mathcal{G}_5} \quad\parbox{20\unitlength}{
\begin{fmfgraph}(20,20)
\fmfleftn{l}{4}
\fmfrightn{r}{4}
\fmftopn{t}{5}
\fmfbottomn{b}{4}
\begin{fmffor}{i}{1}{1}{2}
\fmf{plain}{l[i],i1}
\fmf{plain}{b[i],i1}
\fmf{plain}{l[i+2],i2}
\fmf{plain}{t[i],i2}
\fmf{plain}{t[i+3],i3}
\fmf{plain}{r[i+2],i3}
\fmf{plain}{r[i],i4}
\fmf{plain}{b[i+2],i4}
\end{fmffor}
\fmf{dashes}{i1,g}
\fmf{dashes}{i2,g}
\fmf{dashes}{i4,g}
\fmf{plain,tension=2}{i3,i}
\fmf{dashes,tension=2}{i,g}
\fmffreeze
\fmf{double}{t3,i}
\fmfv{d.sh=circle,d.f=30,d.size=10pt}{i1,i2,i3,i4}
\fmfdot{g,i}
\end{fmfgraph}}
=\sum \quad \parbox{20\unitlength}{
\begin{fmfgraph}(20,20)
\fmfleftn{l}{4}
\fmfrightn{r}{4}
\fmftopn{t}{5}
\fmfbottomn{b}{4}
\begin{fmffor}{i}{1}{1}{2}
\fmf{plain}{l[i],i1}
\fmf{plain}{b[i],i1}
\fmf{plain}{l[i+2],i2}
\fmf{plain}{t[i],i2}
\fmf{plain}{t[i+3],i3}
\fmf{plain}{r[i+2],i3}
\fmf{plain}{r[i],i4}
\fmf{plain}{b[i+2],i4}
\end{fmffor}
\fmf{dashes}{i1,g}
\fmf{dashes}{i2,g}
\fmf{dashes}{i4,g}
\fmf{plain,tension=2.5}{i3,i}
\fmf{dashes,tension=1.5}{i,g}
\fmffreeze
\fmf{ghost}{t3,i}
\fmfv{d.sh=circle,d.f=30,d.size=10pt}{i1,i2,i3,i4}
\fmfdot{g}
\fmfv{decor.shape=square,decor.filled=empty,decor.size=5}{i}
\end{fmfgraph}}
 \end{multline}
Again, these terms can be can be canceled by iterating this procedure
at the next vertices in the blobs until the external particles are
reached. In a nonlinear realization of the symmetry, the cancellation of 
the contributions of higher vertices involving Goldstone bosons 
proceeds in the same way.

The cancellation mechanism just described involves only the STIs of
the vertices where the gauge boson is inserted and thus will be called
`first order cancellations'. As we have seen, the first order
cancellations do not require the introductions of diagrams of a different
topology.

Let us now turn to the cancellation of the terms of the
form~(\ref{eq:ghost-internal}). They will lead to the prescription to
flip also all internal gauge bosons and will be called `second order
cancellations' because they involve not only the STI for the vertex
with the gauge boson insertion but a STI for another subamplitude. In
contrast to the first order ones, they force us to introduce diagrams
of a different topology.

Combining the GB and gauge boson diagrams, the diagrams in
question have the form
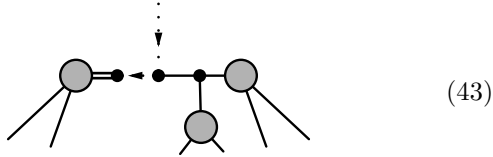
\begin{equation}
\parbox{41\unitlength}{
      \begin{fmfgraph}(40,20)
       \fmftop{x}
       \fmfbottomn{n}{8}
       \fmfleft{li}
       \fmfright{ri}
       \fmf{phantom,tension=8}{li,i}
       \fmf{phantom,tension=8}{ri,l}  
       \fmf{phantom,tension=6}{x,i}
       \fmf{phantom,tension=6}{x,l}        
       \fmf{phantom}{i,l}
        \begin{fmffor}{i}{1}{1}{2}
         \fmf{plain,tension=3}{i,n[i]}
       \end{fmffor}        
       \begin{fmffor}{i}{7}{1}{8}
         \fmf{plain,tension=3}{l,n[i]}
       \end{fmffor}
       \fmf{plain,tension=4}{n,y}
       \fmf{phantom,tension=2}{y,i}
       \fmf{plain,tension=4}{n,l}        
        \fmffreeze
       \fmf{ghost,tension=6}{x,y}
        \begin{fmffor}{i}{5}{1}{6}
         \fmf{plain}{j,n[i]}
       \end{fmffor} 
        \fmf{plain}{n,j} 
      \fmf{ghost,tension=2}{y,k}
       \fmf{double,tension=2}{k,i}
      \fmfdot{n,y,k}
      \fmfv{d.sh=circle,d.f=30,d.si=12pt}{i}
      \fmfv{d.sh=circle,d.f=30,d.si=12pt}{j}
      \fmfv{d.sh=circle,d.f=30,d.si=12pt}{l}
      \end{fmfgraph}}
\end{equation}
To proceed further, we have to add additional diagrams so that we can use a
STI for the subamplitude connected to the double line. We will
proceed by induction and assume that that the $N-1$ particle groves
satisfy the STI~(\ref{eq:sti-diag}) in the sense of
definition~\ref{def:sti}. After applying the gauge flips to the
subamplitude connected
to the double line, we can use the STI and obtain:
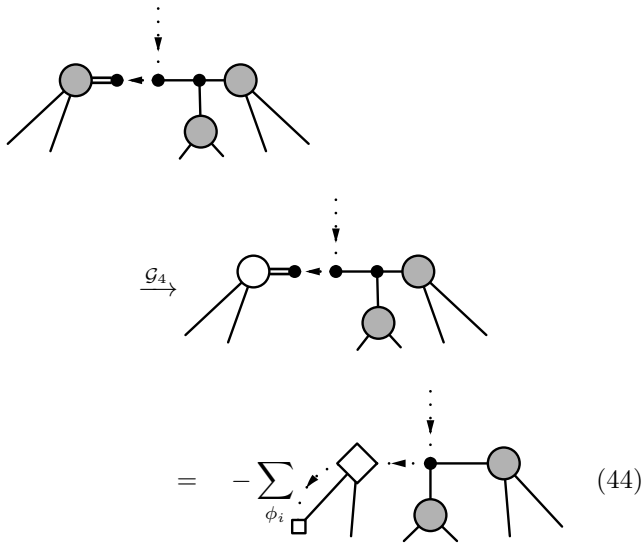
\begin{multline}
\parbox{41\unitlength}{
      \begin{fmfgraph}(40,20)
       \fmftop{x}
       \fmfbottomn{n}{8}
       \fmfleft{li}
       \fmfright{ri}
       \fmf{phantom,tension=8}{li,i}
       \fmf{phantom,tension=8}{ri,l}  
       \fmf{phantom,tension=6}{x,i}
       \fmf{phantom,tension=6}{x,l}        
       \fmf{phantom}{i,l}
        \begin{fmffor}{i}{1}{1}{2}
         \fmf{plain,tension=3}{i,n[i]}
       \end{fmffor}        
       \begin{fmffor}{i}{7}{1}{8}
         \fmf{plain,tension=3}{l,n[i]}
       \end{fmffor}
       \fmf{plain,tension=4}{n,y}
       \fmf{phantom,tension=2}{y,i}
       \fmf{plain,tension=4}{n,l}        
        \fmffreeze
       \fmf{ghost,tension=6}{x,y}
        \begin{fmffor}{i}{5}{1}{6}
         \fmf{plain}{j,n[i]}
       \end{fmffor} 
        \fmf{plain}{n,j} 
      \fmf{ghost,tension=2}{y,k}
       \fmf{double,tension=2}{k,i}
      \fmfdot{n,y,k}
      \fmfv{d.sh=circle,d.f=30,d.si=12pt}{i}
      \fmfv{d.sh=circle,d.f=30,d.si=12pt}{j}
      \fmfv{d.sh=circle,d.f=30,d.si=12pt}{l}
      \end{fmfgraph}}\\ 
\xrightarrow{\mathcal{G}_4}
\parbox{41\unitlength}{
      \begin{fmfgraph}(40,20)
       \fmftop{x}
       \fmfbottomn{n}{8}
       \fmfleft{li}
       \fmfright{ri}
       \fmf{phantom,tension=8}{li,i}
       \fmf{phantom,tension=8}{ri,l}  
       \fmf{phantom,tension=6}{x,i}
       \fmf{phantom,tension=6}{x,l}        
       \fmf{phantom}{i,l}
        \begin{fmffor}{i}{1}{1}{2}
         \fmf{plain,tension=3}{i,n[i]}
       \end{fmffor}        
       \begin{fmffor}{i}{7}{1}{8}
         \fmf{plain,tension=3}{l,n[i]}
       \end{fmffor}
       \fmf{plain,tension=4}{n,y}
       \fmf{phantom,tension=2}{y,i}
       \fmf{plain,tension=4}{n,l}        
        \fmffreeze
       \fmf{ghost,tension=6}{x,y}
        \begin{fmffor}{i}{5}{1}{6}
         \fmf{plain}{j,n[i]}
       \end{fmffor} 
        \fmf{plain}{n,j} 
      \fmf{ghost,tension=2}{y,k}
       \fmf{double,tension=2}{k,i}
      \fmfdot{n,y,k}
      \fmfv{d.sh=circle,d.f=empty,d.si=12pt}{i}
      \fmfv{d.sh=circle,d.f=30,d.si=12pt}{j}
      \fmfv{d.sh=circle,d.f=30,d.si=12pt}{l}
      \end{fmfgraph}} \\
=\quad- \sum_{\phi_i}\parbox{36\unitlength}{
      \begin{fmfgraph}(35,20)
       \fmftop{x}
       \fmfbottomn{n}{6}
        \fmfleft{li}
       \fmfright{ri}
       \fmfleft{li}
       \fmfright{ri}
       \fmf{phantom,tension=8}{li,i}
       \fmf{phantom,tension=8}{ri,l}  
       \fmf{phantom,tension=6}{x,i}
       \fmf{phantom,tension=6}{x,l}        
       \fmf{phantom}{i,l}       
        \begin{fmffor}{i}{5}{1}{6}
         \fmf{plain,tension=3}{l,n[i]}
       \end{fmffor} 
       \fmf{plain,tension=3}{i,n2}
       \fmf{plain,tension=3}{i,n1}
       \fmf{ghost,tension=1}{n,i}
       \fmf{plain,tension=1}{n,l}        
        \fmffreeze
        \fmf{ghost,right=0.5}{i,n1}
        \begin{fmffor}{i}{3}{1}{4}
         \fmf{plain}{j,n[i]}
       \end{fmffor} 
        \fmf{plain}{n,j} 
      \fmfdot{n}
       \fmfv{decor.shape=square,decor.filled=empty,decor.size=5}{n1}
      \fmffreeze
      \fmf{ghost}{x,n}
      \fmfv{d.sh=diamond,d.f=empty,d.si=15}{i}
      \fmfv{d.sh=circle,d.f=30,d.si=12}{j}
      \fmfv{d.sh=circle,d.f=30,d.si=12}{l}
      \end{fmfgraph}}
\end{multline}
Of course now we have to flip the external gauge boson
also  through the new diagrams.

Since by assumption the contact terms of the subamplitude are those
generated by the formal mapping $\mathcal{F}$, we see that the contact
terms are those corresponding to the diagrams
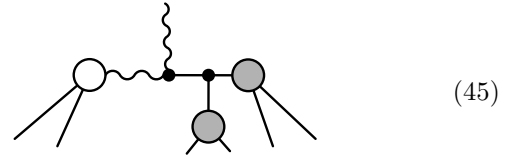
\begin{equation}
\parbox{41\unitlength}{
      \begin{fmfgraph}(40,20)
       \fmftop{x}
       \fmfbottomn{n}{8}
       \fmfleft{li}
       \fmfright{ri}
       \fmf{phantom,tension=6}{li,i}
       \fmf{phantom,tension=8}{ri,l}  
       \fmf{phantom,tension=6}{x,i}
       \fmf{phantom,tension=6}{x,l}        
       \fmf{phantom}{i,l}
        \begin{fmffor}{i}{1}{1}{2}
         \fmf{plain,tension=3}{i,n[i]}
       \end{fmffor}        
       \begin{fmffor}{i}{7}{1}{8}
         \fmf{plain,tension=3}{l,n[i]}
       \end{fmffor}
       \fmf{plain,tension=4}{n,y}
       \fmf{phantom,tension=2}{y,i}
       \fmf{plain,tension=4}{n,l}        
        \fmffreeze
       \fmf{photon,tension=6}{x,y}
        \begin{fmffor}{i}{5}{1}{6}
         \fmf{plain}{j,n[i]}
       \end{fmffor} 
        \fmf{plain}{n,j} 
      \fmf{photon}{y,i}
      \fmfdot{n,y}
      \fmfv{d.sh=circle,d.f=empty,d.si=12pt}{i}
      \fmfv{d.sh=circle,d.f=30,d.si=12pt}{j}
      \fmfv{d.sh=circle,d.f=30,d.si=12pt}{l}
      \end{fmfgraph}} 
\end{equation}
This shows that for the set of diagrams obtained by applying the gauge
flips, the contact terms that appear by contracting an external gauge
boson  are indeed the same ones that are assigned to every diagram by
the mapping $\mathcal{F}$ and therefore the STI is satisfied in the
sense of definition~\ref{def:sti}.
The sets of diagrams connected by gauge flips are indeed the
\emph{minimal} GICs since, by construction, an omission of a diagram
would lead to an violation of a WI. This concludes our proof. 

As an example, we discuss five point functions in a general
SBGT. We have seen in section~\ref{sec:qcd-groves} that the five point function
with four external matter fields can be decomposed into groves
like~(\ref{eq:qcd-grove}). Such a decomposition is no longer possible
if the five point amplitude contains more than one gauge external boson or
external Higgs bosons (in a linear representation of the
symmetry). This can be understood using the techniques of our proof:
if the external gauge boson is inserted into an external Higgs or
gauge boson leg, it can couple to an internal gauge boson so we have
to apply the gauge flips to obtain:
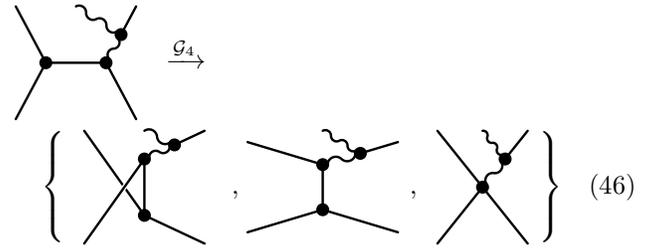
\begin{multline}\label{eq:5-internal-flips}
\parbox{21\unitlength}{
\begin{fmfgraph}(20,15)
\fmfright{a,b}
\fmfleft{f1,f2}
\fmftop{z}
\fmf{plain,tension=2.0}{eae,b}
\fmf{photon,tension=2.0}{ewn,eae}
\fmf{plain}{uwd,f1}
\fmf{plain}{ewn,uwd}
\fmf{plain}{a,ewn}
\fmf{plain}{f2,uwd}
\fmfdot{uwd}
\fmfdot{ewn}
\fmffreeze
\fmf{photon}{z,eae}
\fmfdot{eae}
\end{fmfgraph}}
\xrightarrow{\mathcal{G}_4}\\
\left\{\,
\parbox{21\unitlength}{
\begin{fmfgraph}(20,15)
\fmfright{a,b}
\fmftop{w}
\fmfleft{f1,f2}
\fmf{plain,tension=2}{b,i}
\fmf{photon,tension=2}{i,fwf1}
\fmf{plain}{fwf1,fwf2}
\fmf{plain}{fwf2,a}
\fmf{phantom}{fwf2,f1}
\fmf{phantom}{fwf1,f2}
\fmffreeze
\fmf{photon}{w,i}
\fmf{plain}{fwf2,f2}
\fmf{plain,rubout}{fwf1,f1}
\fmfdot{fwf1,fwf2,i}
\end{fmfgraph}}\, ,\, 
\parbox{21\unitlength}{
\begin{fmfgraph}(20,15)
\fmftop{a,b}
\fmfbottom{f1,f2}
\fmftop{w}
\fmf{plain,tension=2}{b,i}
\fmf{photon,tension=2}{i,fwf}
\fmf{plain}{fwf,a}
\fmf{plain}{fwf,www}
\fmf{plain}{www,f1}
\fmf{plain}{www,f2}
\fmffreeze
\fmf{photon}{w,i}
\fmfdot{fwf,www,i}
\end{fmfgraph}}\, ,\, 
\parbox{15\unitlength}{
\begin{fmfgraph}(15,15)
\fmfright{a,b}
\fmfleft{f1,f2}
\fmftop{w}
\fmf{plain,tension=2}{b,i}
\fmf{photon,tension=2}{i,c}
\fmf{plain}{c,a}
\fmf{plain}{c,f1}
\fmf{plain}{c,f2}
\fmffreeze
\fmf{photon}{w,i}
\fmfdot{c,i}
\end{fmfgraph}}
\right\}
\end{multline}
This brings in $t$- and $u$-channel diagrams and the quartic
vertices. Adding these diagrams and the corresponding GB
diagrams, we can use the STI for the four point function and a `second
order cancellation' can take place:
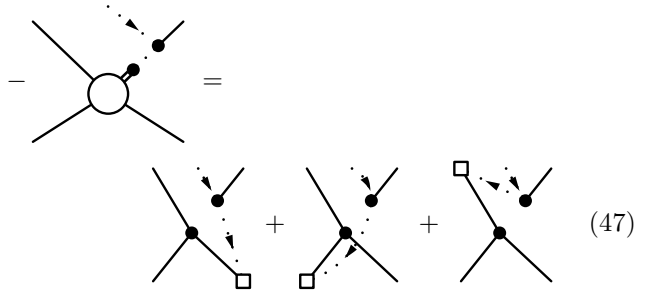
\begin{multline}
-\parbox{21\unitlength}{
\begin{fmfgraph}(20,20)
\fmftop{a,z,b}
\fmfbottom{f1,f2}
\fmf{plain,tension=2.0}{eae,b}
\fmf{dots,tension=2}{eae,p}
\fmf{double,tension=2.0}{p,x}
\fmf{plain,tension=1}{x,f1}
\fmf{plain,tension=0.66}{a,x}
\fmf{plain,tension=1}{f2,x}
\fmffreeze
\fmf{ghost}{z,eae}
\fmfdot{eae,p}
\fmfv{d.sh=circle,d.f=empty,d.size=15pt}{x}
\end{fmfgraph}}\,=\\
\parbox{15\unitlength}{
\begin{fmfgraph}(15,15)
\fmfright{a,b}
\fmftop{z}
\fmfleft{f1,f2}
\fmf{plain}{a,v}
\fmf{phantom}{b,i,v}
\fmf{plain}{f2,v,f1}
\fmffreeze
\fmf{ghost}{z,i,a}
\fmf{plain}{b,i}
\fmfdot{v,i}
\fmfv{decor.shape=square,decor.filled=empty,decor.size=5}{a}
\end{fmfgraph}}\, + \,
\parbox{15\unitlength}{
\begin{fmfgraph}(15,15)
\fmfright{a,b}
\fmftop{z}
\fmfleft{f1,f2}
\fmf{plain}{f1,v}
\fmf{phantom}{b,i,v}
\fmf{plain}{f2,v,a}
\fmffreeze
\fmf{ghost,left=0.3}{z,i,f1}
\fmf{plain}{b,i}
\fmfdot{v,i}
\fmfv{decor.shape=square,decor.filled=empty,decor.size=5}{f1}
\end{fmfgraph}}\,+\,
\parbox{15\unitlength}{
\begin{fmfgraph}(15,15)
\fmfright{a,b}
\fmftop{z}
\fmfleft{f1,f2}
\fmf{plain}{f2,v}
\fmf{phantom}{b,i,v}
\fmf{plain}{f1,v,a}
\fmffreeze
\fmf{ghost}{z,i,f2}
\fmf{plain}{b,i}
\fmfdot{v,i}
\fmfv{decor.shape=square,decor.filled=empty,decor.size=5}{f2}
\end{fmfgraph}}
\end{multline}
These are some of the contact terms obtained by applying the
mapping~(\ref{subeq:generate-contact}) to the diagrams
of~(\ref{eq:5-internal-flips}). Of course now we have to flip the
external gauge boson through all four diagrams
in~(\ref{eq:5-internal-flips}) and this will in general result in
\emph{all} diagrams of the amplitude.

We can also see the reason for the groves appearing in the $\bar q q
\to \bar q q g$ amplitude~(\ref{eq:qcd-grove}): Because of fermion
number conservation, diagrams of the form
of~(\ref{eq:5-internal-flips}) cannot be generated by the insertion of
a gauge boson into a fermion line. Since these are the only diagrams
that force us to perform the internal gauge flips that bring in the
$t$ and $u$ channel diagrams, we see again that the
grove~(\ref{eq:qcd-grove}) is indeed a GIC for external fermions.

\section{Definition of gauge flips in spontaneously broken gauge theories}
\label{sec:sti-flips}

\subsection{Gauge flips for linearly  realized symmetries}
\label{sec:lin-sti}
In order to apply the formalism of flips and groves to SBGTs, we have
to define the elementary flips of the theory. According
to~\cite{Boos:1999} and the results of section~\ref{sec:gi-classes}, the
gauge flips are given by the minimal sets of four point diagrams satisfying the
STIs. We first treat the case of the amplitude $\bar f f \to WW$ in
some detail, before extending the discussion to the remaining
elementary flips.

To see the origin of the subtleties in the definition of the gauge
flips including Higgs propagators, we point out a feature
of the conditions arising from the WIs of SBGTs. The evaluation of
the WI~(\ref{eq:gf-wi}) for the process $\bar f f \to WW$ in a general
SBGT~\cite{Schwinn:2003} leads to the same condition as in an unbroken
gauge theory, i.\,e.~the Lie algebra structure of the fermion-gauge
boson couplings. No such conditions arise for the Higgs couplings because
the Higgs exchange diagram $G_{4,2F}^4$ from~(\ref{eq:ffww-flips})
satisfies the WI by itself. The reason for this can be seen from
the STI for the $HWW$~vertex, resulting from the
STI~(\ref{eq:sti-irr-3tree}) with the additional ghost
term~(\ref{eq:2gb-sti}):
\begin{multline}
\label{eq:hww-sti}
- \Braket{ \mathcal{D}_a(p_a) W_b^\mu(p_b)H_i(p_i)}^{\text{1PI}}\\
=\ii \frac{1}{\xi}p_b^\mu\Braket{c_a(p_a)\bar c_b(p_b)H_i(p_i)}^{\text{1PI}}
\end{multline}
Here we have used that $W$-Higgs and $W$-GB mixing
vanishes at tree level.  Contracting with a physical polarization vector of the
gauge boson, the ghost term drops out and we find that a simple
WI~(\ref{eq:gf-wi}) for the $HWW$~vertex is valid even if the Higgs is
off shell.

For external particles off their mass shell, the
STI~(\ref{eq:hww-sti}) implies that the Higgs diagram in the $\bar f f
W W$ amplitude reproduces a ghost diagram from the STI:
\begin{equation} 
\parbox{15\unitlength}{
\begin{fmfgraph}(15,15)
  \fmfleft{f1,f2} \fmfright{A,H} \fmf{fermion}{f1,a1}
  \fmf{fermion}{a1,f2}
  \fmf{dashes}{a1,a2}
  \fmf{photon}{H,a2} \fmf{double}{a2,A}
\fmfdot{a1,a2}  
\end{fmfgraph}}\quad =\quad
\parbox{15\unitlength}{
\begin{fmfgraph}(15,15)
  \fmfleft{f1,f2} \fmfright{A,H} \fmf{fermion}{f1,a1}
  \fmf{fermion}{a1,f2}
  \fmf{dashes}{a1,a2}
  \fmf{ghost}{A,a2} \fmf{ghost}{a2,H}
\fmfdot{a1,a2}  
    \fmfv{decor.shape=square,decor.filled=full,decor.size=5}{H}
\end{fmfgraph}}
\end{equation}
According to the definitions in section~\ref{sec:groves-def}, this means that
this diagram satisfies the STI by itself. Therefore we would conclude
that the elementary gauge flips in a SBGT are still defined by~(\ref{eq:gauge_flips2}).
However, the recursive proof from section~\ref{sec:gi-classes}, requires
that the set of GB diagrams corresponding to  the elementary gauge
flips also satisfies the STIs by themselves. As we will now show,
this forces us to include the Higgs exchange diagram in the gauge
flips.

Considering the Higgs exchange diagram in the $\bar f f \to W \phi$
amplitude, we see, using the STI for the $WH\phi$~vertex, 
that there are additional contributions in this case: 
\begin{multline}\label{eq:gold-sti}
\parbox{15\unitlength}{
\begin{fmfgraph*}(15,15)
  \fmfleft{f1,f2} \fmfright{A,H} \fmf{fermion}{f1,a1}
  \fmf{fermion}{a1,f2}
  \fmf{dashes,label=$H$}{a1,a2}
  \fmf{dashes,label=$\phi$}{H,a2} \fmf{double}{a2,A}
\fmfdot{a1,a2}  
\end{fmfgraph*}}\quad =\\ 
\parbox{15\unitlength}{
\begin{fmfgraph}(15,15)
  \fmfleft{f1,f2} \fmfright{A,H} \fmf{fermion}{f1,a1}
  \fmf{fermion}{a1,f2}
  \fmf{dashes}{a1,a2}
  \fmf{ghost}{A,a2} \fmf{ghost}{a2,H}
\fmfdot{a1,a2}  
    \fmfv{decor.shape=cross,decor.size=5}{H}
\end{fmfgraph}}\quad -\quad
\parbox{15\unitlength}{
\begin{fmfgraph}(15,15)
  \fmfleft{f1,f2} \fmfright{A,H} \fmf{fermion}{f1,a,f2}
  \fmf{dashes}{H,i}
  \fmf{dashes,tension=2.5}{i,a} 
  \fmf{phantom}{a,A}
\fmffreeze
\fmf{ghost}{A,i}
\fmfdot{a}
\fmfv{decor.shape=square,decor.filled=empty,decor.size=5}{i}
\end{fmfgraph}}\quad -\quad
\parbox{15\unitlength}{
\begin{fmfgraph}(15,15)
  \fmfleft{f1,f2} \fmfright{A,H} \fmf{fermion}{f1,a1}
  \fmf{fermion}{a1,f2}
  \fmf{dashes}{a1,H}
   \fmf{phantom}{A,a1}
\fmffreeze
  \fmf{ghost}{A,H} 
\fmfdot{a1}  
\fmfv{decor.shape=square,decor.filled=empty,decor.size=5}{H}
\end{fmfgraph}}
\end{multline}
The first and the last diagram are contact terms required by the
STI. To cancel the second diagram, we are forced to add two more
diagrams:
\begin{equation*}
\parbox{15\unitlength}{
\begin{fmfgraph}(15,15)
\fmfleft{a,b}
\fmfright{f1,f2}
\fmf{fermion}{a,fwf1}
\fmf{fermion}{fwf1,fwf2}
\fmf{fermion}{fwf2,b}
\fmf{double}{fwf1,f1}
\fmf{dashes}{fwf2,f2}
\fmfdot{fwf1,fwf2}
\end{fmfgraph}}\quad ,\quad
\parbox{15\unitlength}{
\begin{fmfgraph}(15,15)
\fmfleft{a,b}
\fmfright{f1,f2}
\fmf{fermion}{a,fwf1}
\fmf{fermion}{fwf1,fwf2}
\fmf{fermion}{fwf2,b}
\fmf{phantom}{fwf2,f2}
\fmf{phantom}{fwf1,f1}
\fmffreeze
\fmf{double}{fwf2,f1}
\fmf{dashes}{fwf1,f2}
\fmfdot{fwf1,fwf2}
\end{fmfgraph}}
\end{equation*}
According to the general analysis in section~\ref{sec:gi-classes}, these
terms provide the remaining diagrams, so a cancellation because of the
STI
\begin{equation}
0=\quad
\parbox{15\unitlength}{
\begin{fmfgraph}(15,15)
  \fmfleft{f1,f2} \fmfright{A,H} \fmf{fermion}{f1,a,f2}
  \fmf{dashes}{H,i}
  \fmf{dashes,tension=3}{i,a} 
  \fmf{phantom}{a,A}
\fmffreeze
\fmf{ghost}{A,i}
\fmfdot{a}
\fmfv{decor.shape=square,decor.filled=empty,decor.size=5}{i}
\end{fmfgraph}}\quad +\quad
 \parbox{15\unitlength}{
\begin{fmfgraph}(15,15)
  \fmfleft{f1,f2} \fmfright{A,H} \fmf{fermion}{f1,a}
  \fmf{plain,tension=3}{a,i}\fmf{fermion}{i,f2}
  \fmf{dashes}{H,a} 
  \fmf{phantom}{a,A}
\fmffreeze
\fmf{ghost,left=0.5}{A,i}
\fmfdot{a}
\fmfv{decor.shape=square,decor.filled=empty,decor.size=5}{i}
\end{fmfgraph}}\quad +\quad
\parbox{15\unitlength}{
\begin{fmfgraph}(15,15)
  \fmfleft{f1,f2} \fmfright{A,H} \fmf{fermion}{f1,i}
  \fmf{plain,tension=3}{i,a}\fmf{fermion}{a,f2}
  \fmf{dashes}{H,a} 
  \fmf{phantom}{a,A}
\fmffreeze
\fmf{ghost}{A,i}
\fmfdot{a}
\fmfv{decor.shape=square,decor.filled=empty,decor.size=5}{i}
\end{fmfgraph}}
\end{equation}
takes place. Therefore, our definition in section~\ref{sec:flip-def} forces
us  to include the diagrams~${\tilde G}_{4,2F}^1$ and~${\tilde G}_{4,2F}^2$
in the gauge flips for $f\bar f \to WW$. But then also the
diagram~${\tilde G}_{4,2F}^3$ with a triple gauge boson vertex has to
be included and we 
obtain the correct set of flips given in~(\ref{eq:ffww-flips}).

This discussion can easily be generalized to the other elementary
gauge flips since the argument did depend only on the structure of the
STIs for the $WWH$ and $W\phi H$ vertices. Considering the remaining
vertices, we only used the fact that they satisfy the appropriate
STIs. Therefore the conclusions apply also for the other elementary
flips and we have to include the Higgs exchange diagrams in \emph{all}
gauge flips.  All flips including new elementary gauge flips for
gauge-Higgs boson four point functions are displayed in appendix~\ref{app:flips}.

\subsection{Flips for nonlinear realizations of symmetries}
\label{sec:nl-sti}
It is not obvious that the intuitive arguments of section~\ref{sec:ssb-flips}
for the simplifications of the gauge flips in a nonlinear realization
of the symmetry carry over to theories with a more complicated Higgs
sector where the Higgs bosons transform nontrivially under the unbroken
subgroup. To demonstrate that this is indeed the case, we will show
for a general nonlinearly realized symmetry that the STI for the
$WH\phi$~vertex becomes trivial at tree level. Therefore, according to
the discussion in the previous subsection, the Higgs exchange diagrams
have not to be included in the gauge flips without external Higgs
bosons.

We consider a symmetry group $G$ that is spontaneously broken down to
a subgroup $H\subset G$. The GBs can be used to parametrize the coset space
$G/H$ by introducing the exponential representation
\begin{equation}
 U(\phi) = \ee^{\frac{\ii}{f}\phi_a V^a}\in G/H
\end{equation}
where the $V^a$ are the broken generators. The generators of the
unbroken subgroup will be called $L^a$.

To derive the STIs, we need the BRS transformations of fields in the
nonlinear representation. They can be obtained in the usual way from
the infinitesimal gauge transformations by replacing the gauge
parameter $\omega$ by a ghost field. Under the gauge symmetry, the
GBs transform nonlinearly:
\begin{equation}
\label{eq:nl-dphi}
  \phi_a\to \phi'_a(\phi,\omega)
\end{equation}
On fermions, other matter fields and Higgs bosons, gauge
transformations of the full gauge group are realized as linear, $\phi$
dependent transformations of the unbroken subgroup:
\begin{equation}
\label{eq:nl-dpsi}
  \Phi'=\mathcal{H}(\phi,\omega)\Phi
\end{equation}
The explicit form of the functions $\phi'$ and $\mathcal{H}$ can be
found in the literature~\cite{Coleman:1969}.

For the derivation of the BRS transformations in theories with
nonlinear symmetries, we consider the
transformations~(\ref{eq:nl-dphi}) and~(\ref{eq:nl-dpsi}) for
infinitesimal parameters $\omega=\epsilon$. To linear order
in~$\epsilon$, we can write
\begin{equation}
\begin{aligned}
  U'(\phi,\epsilon) = U(\phi'(\phi,\epsilon))
    & = 1+\ii \mathcal{K}_a^b (\phi)\epsilon_b
          V^a +\mathcal{O}(\epsilon^2) \\ 
  \mathcal{H}(\phi,\epsilon)
    & = 1+\ii\Omega_a^b(\phi) \epsilon_b L^a +\mathcal{O}(\epsilon^2)
\end{aligned}
\end{equation}
We can now introduce the BRS transformations 
\begin{equation}
\begin{aligned}
\delta_{\text{BRS}} \phi_a&=f c_b \mathcal{K}_a^b (\phi) \\
\delta_{\text{BRS}} \Phi&=\ii c_a \Omega^a(\phi) \Phi
\end{aligned}
\end{equation}
with $\Omega^a=\Omega^a_bL^b$. Below, we will only need the first order in
$\phi$:
\begin{equation}
\begin{aligned}
\mathcal{K}^b_a(\phi)&=g_a^b+t_{a c}^{b}\phi_c+\mathcal{O}(\phi^2) \\
\Omega^a_{ij}(\phi)&= \mathcal{T}^a_{ij}+\mathcal{O}(\phi)
\end{aligned}
\end{equation}
We will still use a linear $R_\xi$ gauge fixing instead of a gauge
fixing function in terms of the $U$. Note that the gauge fixing term
in the nonlinear parametrization contains no Higgs-ghost interaction
since the Higgs bosons do not appear in the BRS transformation of the GBs.
 
Turning to  the STI for the $WH\phi$~vertex that is relevant for the 
discussion of gauge flips, after setting 
classical fields and sources to zero, the terms of higher order in $\phi$ drop
out and  we arrive at:
\begin{multline}\label{eq:phihw-sti-nl}
- \Braket{ \mathcal{D}_a(p_a)
  \phi_b(p_b)H_i(k_i)}^{\text{1PI}}\\
= f t^a_{cb}\Braket{\phi_c(p_a+p_b)H_i(k_i)}^{\text{1PI}}\\
+\ii \mathcal{T}^a_{ji}\Braket{\phi_b(p_b)H_j(-p_b)}^{\text{1PI}}
     \overset{\text{tree level}}{=}0
\end{multline}
On tree level, the right hand side vanishes because there is no
Higgs-GB mixing. Thus we can conclude that the diagram
\begin{equation}
\label{eq:3wphi-diag}
\parbox{15\unitlength}{
\begin{fmfgraph*}(15,15)
\fmftop{a,b}
\fmfbottom{f1,f2}
\fmf{photon}{a,fwf1}
\fmf{plain,label=$H$}{fwf1,fwf2}
\fmf{fermion}{fwf2,b}
\fmf{dashes,label=$\phi$}{fwf1,f1}
\fmf{fermion}{fwf2,f2}
\fmfdot{fwf1}
\fmfdot{fwf2}
\end{fmfgraph*}}
\end{equation}
satisfies the WI by itself and therefore the corresponding gauge
boson diagram $G_{4,2F}^4$  has not to be included in the gauge
flips. In higher orders of perturbation theory, the right hand side
of~(\ref{eq:phihw-sti-nl}) no longer vanishes since a $\phi-H$ mixing
is generated by loop diagrams.

No simplification compared to the linear case appears for the STI for
the $HHW$~vertex
\begin{align}
&- \Braket{ \mathcal{D}_a(p_a)
  H_i(k_i) H_j(k_j)}^{\text{1PI}}\\
&=\ii \mathcal{T}^a_{ki}\Braket{H_k(-p_j)H_j(p_j)}^{\text{1PI}}+\ii
  \mathcal{T}^a_{kj}\Braket{H_i(p_i)H_k(-p_i)}^{\text{1PI}}\nonumber 
\end{align}
This identity is similar to the linear case~\cite{Schwinn:2003} and,
as has already been stated in section~\ref{sec:ssb-flips}, Higgs exchange
diagrams with a $HHW$~vertex have to be included in the gauge flips
also in the nonlinear realization.

In contrast to a linear realization of the symmetry, 
derivatives of the BRS-transforms with respect to more than one field
contribute to the STIs
for vertices with more than three external particles involving GBs. 
For vertices with one GB, the STI~(\ref{eq:sti-irr-4point}) reads
\begin{multline}\label{eq:sti-irr-1phi}
  0 = \sum_\Psi\int\dd^4 x
         \Biggl\{ \Gamma_{c_a \Psi^\star}\Gamma_{\Psi\Phi_1\Phi_2\phi}\\
    + \Bigl[\Gamma_{c_a\Psi^\star\Phi_1}\Gamma_{\Psi\Phi_2\phi}
    + \Gamma_{c_a\Psi^\star\Phi_1\phi}\Gamma_{\Psi\Phi_2} 
    + (1\leftrightarrow 2)\Bigr]\Biggr\}
\end{multline}
In the STI for the $W\phi HH$ vertex, we get a contribution from 
the higher derivatives of the form 
\begin{equation*}
\Gamma_{\bar c_a H_k^* H_j \phi_b}\Gamma_{H_k H_i}
\end{equation*}
while the  similiar contributions to the STI for the $WW\phi H$ vertex
involve mixed two point functions that vanish on tree level. 
Therefore diagrams like
\begin{equation}
\parbox{15\unitlength}{
\begin{fmfgraph*}(15,15)
\fmftop{a,m,b}
\fmfbottom{f1,f2}
\fmf{photon,tension=0.5}{a,fwf1,m}
\fmf{plain,label=$H$}{fwf1,fwf2}
\fmf{fermion}{fwf2,b}
\fmf{dashes,label=$\phi$}{fwf1,f1}
\fmf{fermion}{fwf2,f2}
\fmfdot{fwf1}
\fmfdot{fwf2}
\end{fmfgraph*}}
\end{equation}
don't force us to include additional flips involving the internal Higgs 
bosons. The key feature of the nonlinear realization ensuring  this 
simplification is that there is no term~$\propto c\phi$ in
the BRS-transform of the Higgs field.

\section{Groves in spontaneously broken gauge theories}
\label{sec:ssb-gics}

\subsection{Groves for linearly realized symmetries}
\label{sec:ssb-groves}
To analyze the groves in SBGTs, we have implemented the flips
described in section~\ref{sec:ssb-flips} in the program
\texttt{bocages}~\cite{Ohl:bocages}. As an example for the structure
of the groves, we
consider the amplitude for the process $\bar f f \to f\bar f H$. The
only new features compared to the QCD example $\bar q q \to \bar q q
g$ from~(\ref{eq:qcd-grove}) are single diagram groves consisting of
diagrams without gauge bosons.  The remaining groves are similar to
the groves in QCD with the external gluon replaced by a Higgs boson.

Similarly, in the process $\bar f f \to \bar f f W$ there appear only
two groves, like in the case of QCD discussed in
section~\ref{sec:qcd-groves}. They include, of course, additional diagrams
involving Higgs bosons.

Apart from the one-diagram groves, the Higgs
flips~(\ref{subeq:higgs_flips}) do not lead to additional groves. If
there is at least one gauge boson in a diagram, the gauge flips can
always be used to arrive at diagrams with more than one internal gauge
boson. For example the diagrams
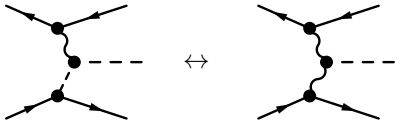
\begin{equation*}
\parbox{21\unitlength}{
  \begin{fmfgraph}(20,15)
    \fmfleft{i1,i2}
    \fmfright{o5,o3,o4}
    \fmf{fermion}{i1,v1}
  \fmfdot{v1}
  \fmf{dashes}{v1,v4}
  \fmfdot{v1,v4}
  \fmf{photon}{v16,v4}
  \fmfdot{v4,v16}
  \fmfdot{v16}
  \fmf{fermion}{v16,i2}
  \fmfdot{v16}
  \fmf{fermion,tension=0.5}{o4,v16}
  \fmfdot{v4}
  \fmf{dashes,tension=0.5}{o3,v4}
  \fmfdot{v1}
  \fmf{fermion,tension=0.5}{v1,o5}
  \end{fmfgraph}}\quad\leftrightarrow\quad
  \parbox{21\unitlength}{
\begin{fmfgraph}(20,15)
    \fmfleft{i1,i2}
    \fmfright{o5,o3,o4}
    \fmf{fermion}{i1,v1}
  \fmfdot{v1}
  \fmf{photon}{v1,v4}
  \fmfdot{v1,v4}
  \fmf{photon}{v16,v4}
  \fmfdot{v4,v16}
  \fmfdot{v16}
  \fmf{fermion}{v16,i2}
  \fmfdot{v16}
  \fmf{fermion,tension=0.5}{o4,v16}
  \fmfdot{v4}
  \fmf{dashes,tension=0.5}{o3,v4}
  \fmfdot{v1}
  \fmf{fermion,tension=0.5}{v1,o5}
  \end{fmfgraph}}
\end{equation*}
are  connected by a gauge flip from~(\ref{eq:h_flips_1}). 

It turns out that this is the generic structure: the only new groves
compared to the case of unbroken gauge theories consist of one diagram
each, where all internal particles are Higgs bosons. The remaining
diagrams fall in the same groves that have been discussed in
section~\ref{sec:qcd-groves}.

Since the Higgs-fermion Yukawa couplings are proportional to the
fermion masses, the couplings of Higgs bosons to light fermions are
usually set to zero in practical calculations.  Of course this is only a
consistent approximation if the masses of light fermions are
set to zero at the same time.
The set of diagrams obtained by neglecting the coupling
to light fermions in general does \emph{not} correspond to a gauge
invariant subset if the fermion masses are not set to zero. In
practice, the numerical instabilities caused by this inconsistency are
negligible but there are small corners in phase space where
they can become relevant.

\subsection{Groves for nonlinearly realized symmetries}
\label{sec:nl-groves}
In the case of nonlinearly realized symmetries, the gauge flips simplify
as we have discussed in section~\ref{sec:ssb-flips}, so there is a more
interesting structure of the groves than in the case of linearly
parametrized scalar sectors.

An especially interesting structure of the groves appears in theories
with a single, neutral Higgs boson and a nonlinearly realized
symmetry. This corresponds to a nonlinear realization of the minimal
electroweak SM~\cite{NL_SM}.  In this parametrization, the Higgs
boson is not connected to the symmetry breaking mechanism but merely
an additional matter particle. Even though the Higgs is no longer an
essential ingredient of the theory, the inclusion of the Higgs boson
in a nonlinearly realized theory is useful to study anomalous
Higgs-gauge boson couplings~\cite{He:2002}. For a Higgs boson that is
a singlet under the unbroken symmetry group, there is no
$HHW$~vertex. We see from the gauge flips in~(\ref{subeq:h_flips}),
that in this case the gauge flips `conserve' Higgs number in the sense that
only external Higgs bosons appear. Therefore gauge flips cannot change
the number of Higgs bosons and the groves can be classified according
to the number of internal Higgs bosons.

In theories with a general scalar sector, this `Higgs conservation
law' breaks down but the number of groves is still larger than in the
case of a linearly realized symmetry.

The reduced number of gauge flips has also consequences for unitarity
gauge. Unitarity gauge can equivalently be defined as the limit
$\xi\to\infty$ of the linearly realized theory in $R_\xi$ gauge or from
nonlinearly realized symmetries by transforming the GBs
away. Therefore it follows that the groves obtained from the reduced
sets of flips are also consistent in unitarity gauge, i.\,e.~they
satisfy the appropriate WIs. Although it is not sensible to speak of
`gauge invariance classes' in a fixed gauge, this result
nevertheless indicates that no numerical instabilities due to
violations of WIs will appear.

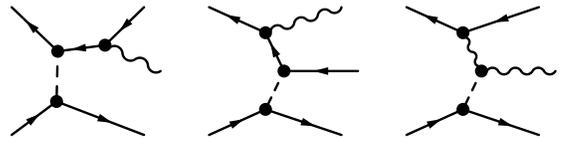
\begin{figure}[htbp]
\begin{equation*}
\parbox{23\unitlength}{
  \begin{fmfgraph}(22,17)
    \fmfleft{i1,i2}
    \fmfright{o5,o3,o4}
 \fmf{fermion}{i1,v1}
    \fmfdot{v1}
  \fmf{dashes}{v1,v4}
  \fmfdot{v1,v4}
  \fmfdot{v4}
  \fmf{fermion}{v4,i2}
  \fmf{fermion}{v17,v4}
  \fmfdot{v4,v17}
  \fmfdot{v17}
  \fmf{fermion,tension=0.5}{o4,v17}
  \fmfdot{v17}
  \fmf{photon,tension=0.5}{o3,v17}
  \fmfdot{v1}
  \fmf{fermion,tension=0.5}{v1,o5}
  \end{fmfgraph}}\quad
\parbox{23\unitlength}{
  \begin{fmfgraph}(22,17)
    \fmfleft{i1,i2}
    \fmfright{o5,o4,o3}
    \fmf{fermion}{i1,v1}
  \fmfdot{v1}
  \fmf{dashes}{v1,v4}
  \fmfdot{v1,v4}
  \fmf{fermion}{v4,v16}
  \fmfdot{v4,v16}
  \fmfdot{v16}
  \fmf{fermion}{v16,i2}
  \fmfdot{v16}
  \fmf{photon,tension=0.5}{o3,v16}
  \fmfdot{v4}
  \fmf{fermion,tension=0.5}{o4,v4}
  \fmfdot{v1}
  \fmf{fermion,tension=0.5}{v1,o5}
  \end{fmfgraph}}\quad
\parbox{23\unitlength}{%
  \begin{fmfgraph}(22,17)
    \fmfleft{i1,i2}
    \fmfright{o5,o3,o4}
    \fmf{fermion}{i1,v1}
  \fmfdot{v1}
  \fmf{dashes}{v1,v4}
  \fmfdot{v1,v4}
  \fmf{photon}{v16,v4}
  \fmfdot{v4,v16}
  \fmfdot{v16}
  \fmf{fermion}{v16,i2}
  \fmfdot{v16}
  \fmf{fermion,tension=0.5}{o4,v16}
  \fmfdot{v4}
  \fmf{photon,tension=0.5}{o3,v4}
  \fmfdot{v1}
  \fmf{fermion,tension=0.5}{v1,o5}
  \end{fmfgraph}}
 \end{equation*}
 \caption{Mixed grove for $\bar f f \to \bar f f W$ for a singlet Higgs }
\label{fig:4qw-grove-uni}
\end{figure}

As example, we consider the process $\bar f f \to \bar f f W$. Let us
discuss the case of a single, neutral Higgs boson first.  We find that
the amplitude for $\bar f f \to \bar f f W$ can be decomposed into six
groves instead of the two appearing in unbroken gauge theories and in
the linear parametrization of SBGTs. Two groves are `gauge groves'
consisting of five diagrams without internal Higgs boson that look
exactly like in QCD~(\ref{eq:qcd-grove}). The remaining groves are
`mixed groves' with one internal Higgs boson. An example of a mixed
grove is given in figure~\ref{fig:4qw-grove-uni}.
Here all diagrams are proportional to  the coupling of the Higgs to
one fermion pair, but this is not always the case as we will see
below.  In this example, two mixed groves correspond to one gauge
grove. This additional structure arises because there is no gauge flip
between the two diagrams 
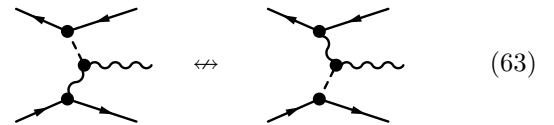
\begin{equation}
\label{eq:no-flip}
\parbox{21\unitlength}{ 
 \begin{fmfgraph}(20,15)
    \fmfleft{i1,i2}
    \fmfright{o5,o3,o4}
  \fmf{fermion}{i1,v1}
  \fmfdot{v1}
  \fmf{photon}{v4,v1}
  \fmfdot{v1,v4}
  \fmf{dashes}{v4,v16}
  \fmfdot{v4,v16}
  \fmfdot{v16}
  \fmf{fermion}{v16,i2}
  \fmfdot{v16}
  \fmf{fermion,tension=0.5}{o4,v16}
  \fmfdot{v4}
  \fmf{photon,tension=0.5}{o3,v4}
  \fmfdot{v1}
  \fmf{fermion,tension=0.5}{v1,o5}
  \end{fmfgraph}}\quad \nleftrightarrow\quad
\parbox{21\unitlength}{ 
 \begin{fmfgraph}(20,15)
    \fmfleft{i1,i2}
    \fmfright{o5,o3,o4}
  \fmf{fermion}{i1,v1}
  \fmfdot{v1}
  \fmf{dashes}{v4,v1}
  \fmfdot{v1,v4}
  \fmf{photon}{v4,v16}
  \fmfdot{v4,v16}
  \fmfdot{v16}
  \fmf{fermion}{v16,i2}
  \fmfdot{v16}
  \fmf{fermion,tension=0.5}{o4,v16}
  \fmfdot{v4}
  \fmf{photon,tension=0.5}{o3,v4}
  \fmfdot{v1}
  \fmf{fermion,tension=0.5}{v1,o5}
  \end{fmfgraph}} 
\end{equation}

\begin{figure}[htbp]
\begin{center}
\begin{multline*}
\fmfframe(6,7)(6,6){
  \begin{fmfgraph*}(25,20)
    \fmfleft{i1,i2}
    \fmfright{o5,o4,o3,o6}
  \fmflabel{$e^+$}{i1}
  \fmf{fermion}{i1,v1}
  \fmfdot{v1}
  \fmflabel{$e^-$}{i2}
  \fmfdot{v1}
  \fmf{fermion}{v1,i2}
  \fmf{photon,label=$\gamma/Z$,tension=2}{v5,v1}
  \fmfdot{v1,v5}
  \fmf{fermion}{v5,v20}
  \fmfdot{v5,v20}
  \fmf{dashes,tension=0.5}{v20,v80}
  \fmfdot{v20,v80}
  \fmflabel{$b$}{o6}
  \fmfdot{v80}
  \fmf{fermion,tension=0.5}{v80,o6}
  \fmflabel{$\bar b$}{o3}
  \fmfdot{v80}
  \fmf{fermion,tension=0.5}{o3,v80}
  \fmflabel{$t$}{o4}
  \fmfdot{v20}
  \fmf{fermion,tension=0.5}{v20,o4}
  \fmflabel{$\bar t$}{o5}
  \fmfdot{v5}
  \fmf{fermion,tension=0.5}{o5,v5}
  \end{fmfgraph*}}
\fmfframe(6,7)(6,6){%
  \begin{fmfgraph*}(25,20)
    \fmfleft{i1,i2}
    \fmfright{o4,o5,o3,o6}
  \fmflabel{$e^+$}{i1}
  \fmf{fermion}{i1,v1}
  \fmfdot{v1}
  \fmflabel{$e^-$}{i2}
  \fmfdot{v1}
  \fmf{fermion}{v1,i2}
  \fmf{photon,label=$\gamma/Z$,tension=2}{v5,v1}
  \fmfdot{v1,v5}
  \fmf{fermion}{v20,v5}
  \fmfdot{v5,v20}
  \fmf{dashes,tension=0.5}{v20,v80}
  \fmfdot{v20,v80}
  \fmflabel{$b$}{o6}
  \fmfdot{v80}
  \fmf{fermion,tension=0.5}{v80,o6}
  \fmflabel{$\bar b$}{o3}
  \fmfdot{v80}
  \fmf{fermion,tension=0.5}{o3,v80}
  \fmflabel{$\bar t$}{o5}
  \fmfdot{v20}
  \fmf{fermion,tension=0.5}{o5,v20}
  \fmflabel{$t$}{o4}
  \fmfdot{v5}
  \fmf{fermion,tension=0.5}{v5,o4}
  \end{fmfgraph*}}\\
\fmfframe(6,7)(6,6){%
  \begin{fmfgraph*}(25,20)
    \fmfleft{i1,i2}
    \fmfright{o4,o5,o3,o6}
  \fmflabel{$e^+$}{i1}
  \fmf{fermion}{i1,v1}
  \fmfdot{v1}
  \fmflabel{$e^-$}{i2}
  \fmfdot{v1}
  \fmf{fermion}{v1,i2}
  \fmf{photon,label=$Z$,tension=2}{v5,v1}
  \fmfdot{v1,v5}
  \fmf{dashes}{v5,v20}
  \fmfdot{v5,v20}
  \fmflabel{$b$}{o6}
  \fmfdot{v20}
  \fmf{fermion,tension=0.5}{v20,o6}
  \fmflabel{$\bar b$}{o3}
  \fmfdot{v20}
  \fmf{fermion,tension=0.5}{o3,v20}
  \fmf{photon}{v21,v5}
  \fmfdot{v5,v21}
  \fmflabel{$\bar t$}{o5}
  \fmfdot{v21}
  \fmf{fermion,tension=0.5}{o5,v21}
  \fmflabel{$t$}{o4}
  \fmfdot{v21}
  \fmf{fermion,tension=0.5}{v21,o4}
  \end{fmfgraph*}}
\fmfframe(6,7)(6,6){%
  \begin{fmfgraph*}(25,20)
    \fmfleft{i1,i2}
    \fmfright{o3,o6,o4,o5}
    \fmflabel{$e^+$}{i1}
  \fmf{fermion}{i1,v1}
  \fmfdot{v1}
  \fmf{fermion}{v1,v4}
  \fmfdot{v1,v4}
  \fmflabel{$e^-$}{i2}
  \fmfdot{v4}
  \fmf{fermion}{v4,i2}
  \fmf{photon,label=$\gamma /Z$}{v17,v4}
  \fmfdot{v4,v17}
  \fmflabel{$\bar t$}{o5}
  \fmfdot{v17}
  \fmf{fermion,tension=0.5}{o5,v17}
  \fmflabel{$t$}{o4}
  \fmfdot{v17}
  \fmf{fermion,tension=0.5}{v17,o4}
  \fmf{dashes}{v1,v5}
  \fmfdot{v1,v5}
  \fmflabel{$b$}{o6}
  \fmfdot{v5}
  \fmf{fermion,tension=0.5}{v5,o6}
  \fmflabel{$\bar b$}{o3}
  \fmfdot{v5}
  \fmf{fermion,tension=0.5}{o3,v5}
  \end{fmfgraph*}}\\
\fmfframe(6,7)(6,6){%
  \begin{fmfgraph*}(25,20)
    \fmfleft{i1,i2}
    \fmfright{o4,o5,o3,o6}
   \fmflabel{$e^+$}{i1}
  \fmf{fermion}{i1,v1}
  \fmfdot{v1}
  \fmf{fermion}{v1,v4}
  \fmfdot{v1,v4}
  \fmflabel{$e^-$}{i2}
  \fmfdot{v4}
  \fmf{fermion}{v4,i2}
  \fmf{dashes}{v4,v17}
  \fmfdot{v4,v17}
  \fmflabel{$b$}{o6}
  \fmfdot{v17}
  \fmf{fermion,tension=0.5}{v17,o6}
  \fmflabel{$\bar b$}{o3}
  \fmfdot{v17}
  \fmf{fermion,tension=0.5}{o3,v17}
  \fmf{photon}{v5,v1}
  \fmfdot{v1,v5}
  \fmflabel{$\bar t$}{o5}
  \fmfdot{v5}
  \fmf{fermion,tension=0.5}{o5,v5}
  \fmflabel{$t$}{o4}
  \fmfdot{v5}
  \fmf{fermion,tension=0.5}{v5,o4}
  \end{fmfgraph*}}
\end{multline*}
\end{center}
\caption{\label{fig:6f-grove}%
  Mixed grove in the $e^+e- \to b\bar b t\bar t$ amplitude} 
\end{figure}
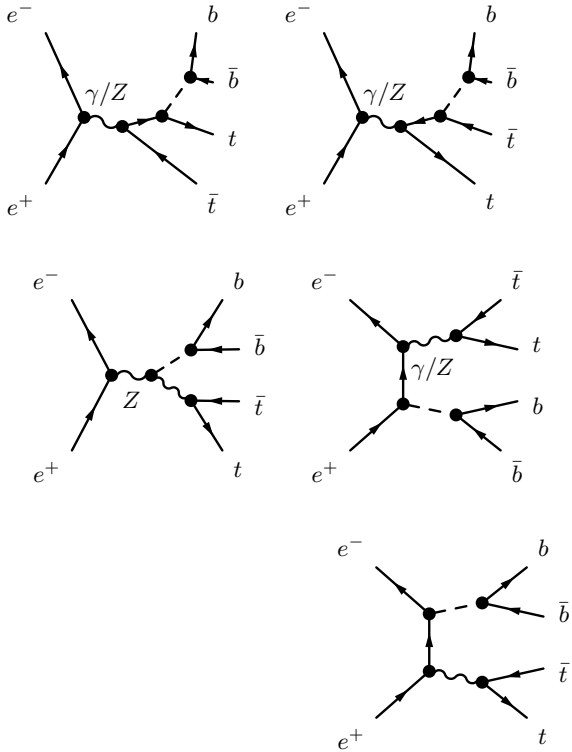

In a nonlinear realization of a more complicated Higgs sector, the
Higgs bosons transform according to a linear representation of the
unbroken subgroup and therefore may couple to the massless gauge boson
through $HHW$ vertices. This enables an indirect flip between the two
diagrams from~(\ref{eq:no-flip}) since the flips~(\ref{eq:h_flips_1})
have to be included in the gauge flips also for the nonlinear
parametrization
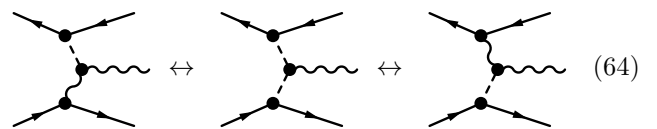
\begin{equation}
\label{eq:indirect-flip}
\parbox{21\unitlength}{ 
 \begin{fmfgraph}(20,15)
    \fmfleft{i1,i2}
    \fmfright{o5,o3,o4}
  \fmf{fermion}{i1,v1}
  \fmfdot{v1}
  \fmf{photon}{v4,v1}
  \fmfdot{v1,v4}
  \fmf{dashes}{v4,v16}
  \fmfdot{v4,v16}
  \fmfdot{v16}
  \fmf{fermion}{v16,i2}
  \fmfdot{v16}
  \fmf{fermion,tension=0.5}{o4,v16}
  \fmfdot{v4}
  \fmf{photon,tension=0.5}{o3,v4}
  \fmfdot{v1}
  \fmf{fermion,tension=0.5}{v1,o5}
  \end{fmfgraph}}\, \leftrightarrow\,
\parbox{21\unitlength}{ 
 \begin{fmfgraph}(20,15)
    \fmfleft{i1,i2}
    \fmfright{o5,o3,o4}
  \fmf{fermion}{i1,v1}
  \fmfdot{v1}
  \fmf{dashes}{v4,v1}
  \fmfdot{v1,v4}
  \fmf{dashes}{v4,v16}
  \fmfdot{v4,v16}
  \fmfdot{v16}
  \fmf{fermion}{v16,i2}
  \fmfdot{v16}
  \fmf{fermion,tension=0.5}{o4,v16}
  \fmfdot{v4}
  \fmf{photon,tension=0.5}{o3,v4}
  \fmfdot{v1}
  \fmf{fermion,tension=0.5}{v1,o5}
  \end{fmfgraph}}\, \leftrightarrow\,
\parbox{21\unitlength}{ 
 \begin{fmfgraph}(20,15)
    \fmfleft{i1,i2}
    \fmfright{o5,o3,o4}
  \fmf{fermion}{i1,v1}
  \fmfdot{v1}
  \fmf{dashes}{v4,v1}
  \fmfdot{v1,v4}
  \fmf{photon}{v4,v16}
  \fmfdot{v4,v16}
  \fmfdot{v16}
  \fmf{fermion}{v16,i2}
  \fmfdot{v16}
  \fmf{fermion,tension=0.5}{o4,v16}
  \fmfdot{v4}
  \fmf{photon,tension=0.5}{o3,v4}
  \fmfdot{v1}
  \fmf{fermion,tension=0.5}{v1,o5}
  \end{fmfgraph}} 
\end{equation}
We see that the `Higgs conservation' in the gauge flips breaks
down. Therefore the appearance of Higgs bosons charged under the
unbroken subgroup reduces the number of groves.  One finds that mixed
groves are still present, however, in general only \emph{one} mixed
grove corresponds to a gauge grove and they do not contain a fixed
number of Higgs bosons.  The same structure is found in amplitudes
with more external particles. Let us first consider the six-fermion
amplitude. For a single Higgs boson, 18 mixed groves are obtained by
inserting a fermion-antifermion pair via a Higgs boson in all possible
places in the gauge groves of the four fermion amplitude. An application to
the process $e^+ e^- \to b\bar b t \bar t$ that is relevant for the
measurement of the top Yukawa coupling is shown in
figure~\ref{fig:6f-grove}. Again all the diagrams in the mixed groves are
proportional to the coupling of the Higgs to one fermion pair. For a
general Higgs sector, because of the additional flips as
in~(\ref{eq:indirect-flip}) only six mixed groves remain, corresponding
to the gauge groves.

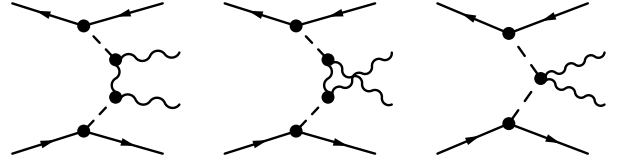
\begin{figure}[htbp]
\begin{center}
\begin{equation*}
\parbox{25\unitlength}{
  \begin{fmfgraph}(25,20)
    \fmfleft{i1,i2}
    \fmfright{o6,o4,o3,o5}
  \fmf{fermion}{i1,v1}
  \fmfdot{v1}
  \fmf{dashes}{v1,v4}
  \fmfdot{v1,v4}
  \fmf{photon}{v16,v4}
  \fmfdot{v4,v16}
  \fmf{dashes}{v16,v64}
  \fmfdot{v16,v64}
  \fmfdot{v64}
  \fmf{fermion}{v64,i2}
  \fmfdot{v64}
  \fmf{fermion,tension=0.5}{o5,v64}
  \fmfdot{v16}
  \fmf{photon,tension=0.5}{o3,v16}
  \fmfdot{v4}
  \fmf{photon,tension=0.5}{o4,v4}
  \fmfdot{v1}
  \fmf{fermion,tension=0.5}{v1,o6}
  \end{fmfgraph}}\quad
\parbox{25\unitlength}{
  \begin{fmfgraph}(25,20)
    \fmfleft{i1,i2}
    \fmfright{o6,o3,o4,o5}
  \fmf{fermion}{i1,v1}
  \fmfdot{v1}
  \fmf{dashes}{v1,v4}
  \fmfdot{v1,v4}
  \fmf{photon}{v16,v4}
  \fmfdot{v4,v16}
  \fmf{dashes}{v16,v64}
  \fmfdot{v16,v64}
  \fmfdot{v64}
  \fmf{fermion}{v64,i2}
  \fmfdot{v64}
  \fmf{fermion,tension=0.5}{o5,v64}
  \fmfdot{v16}
  \fmf{phantom,tension=0.5}{o4,v16}
  \fmfdot{v4}
  \fmf{phantom,tension=0.5}{o3,v4}
  \fmfdot{v1}
  \fmf{fermion,tension=0.5}{v1,o6}
  \fmffreeze
  \fmf{photon,tension=0.5}{o3,v16}
  \fmfdot{v4}
  \fmf{photon,tension=0.5}{o4,v4}
  \end{fmfgraph}}\quad
\parbox{25\unitlength}{
  \begin{fmfgraph}(25,20)
    \fmfleft{i1,i2}
    \fmfright{o6,o3,o4,o5}
  \fmf{fermion}{i1,v1}
  \fmfdot{v1}
  \fmf{dashes}{v1,v4}
  \fmfdot{v1,v4}
  \fmf{dashes}{v4,v16}
  \fmfdot{v4,v16}
  \fmfdot{v16}
  \fmf{fermion}{v16,i2}
  \fmfdot{v16}
  \fmf{fermion,tension=0.5}{o5,v16}
  \fmfdot{v4}
  \fmf{photon,tension=0.5}{o4,v4}
  \fmfdot{v4}
  \fmf{photon,tension=0.5}{o3,v4}
  \fmfdot{v1}
  \fmf{fermion,tension=0.5}{v1,o6}
  \end{fmfgraph}}
\end{equation*}
\end{center}
\caption{\label{fig:2hgrove}%
  Mixed grove with two Higgs bosons in the 4 fermion 2 gauge boson amplitude}
\end{figure}

As a final example we discuss the amplitude $\bar f f\to\bar f f
WW$. For a single Higgs boson, the Higgs number conservation leads to
the appearance of two mixed groves with two Higgs bosons. One
example is shown in figure~\ref{fig:2hgrove}.

Furthermore, apart from  two gauge groves (and several one diagram
groves), there are  ten mixed groves with one Higgs boson.  In general
it is not true that all mixed groves are proportional to one Higgs
coupling. An example is provided by the three diagrams
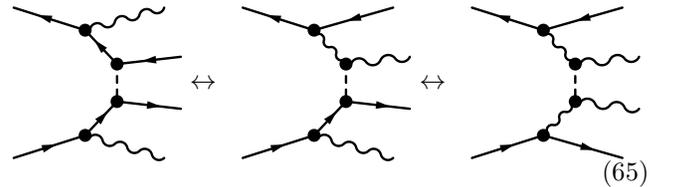
\begin{equation}
\parbox{25\unitlength}{
  \begin{fmfgraph}(25,20)
    \fmfleft{i1,i2}
    \fmfright{o3,o6,o5,o4}
  \fmf{fermion}{i1,v1}
  \fmfdot{v1}
  \fmf{fermion}{v1,v4}
  \fmfdot{v1,v4}
  \fmf{dashes}{v4,v16}
  \fmfdot{v4,v16}
  \fmf{fermion}{v16,v64}
  \fmfdot{v16,v64}
  \fmfdot{v64}
  \fmf{fermion}{v64,i2}
  \fmfdot{v64}
  \fmf{photon,tension=0.5}{o4,v64}
  \fmfdot{v16}
  \fmf{fermion,tension=0.5}{o5,v16}
  \fmfdot{v4}
  \fmf{fermion,tension=0.5}{v4,o6}
  \fmfdot{v1}
  \fmf{photon,tension=0.5}{o3,v1}
  \end{fmfgraph}}\leftrightarrow
\parbox{25\unitlength}{
  \begin{fmfgraph}(25,20)
    \fmfleft{i1,i2}
    \fmfright{o3,o6,o4,o5}
  \fmf{fermion}{i1,v1}
  \fmfdot{v1}
  \fmf{fermion}{v1,v4}
  \fmfdot{v1,v4}
  \fmf{dashes}{v4,v16}
  \fmfdot{v4,v16}
  \fmf{photon}{v64,v16}
  \fmfdot{v16,v64}
   \fmfdot{v64}
  \fmf{fermion}{v64,i2}
   \fmfdot{v64}
  \fmf{fermion,tension=0.5}{o5,v64}
   \fmfdot{v16}
  \fmf{photon,tension=0.5}{o4,v16}
   \fmfdot{v4}
  \fmf{fermion,tension=0.5}{v4,o6}
   \fmfdot{v1}
  \fmf{photon,tension=0.5}{o3,v1}
  \end{fmfgraph}} \leftrightarrow
\parbox{25\unitlength}{
  \begin{fmfgraph}(25,20)
    \fmfleft{i1,i2}
    \fmfright{o6,o3,o4,o5}
  \fmf{fermion}{i1,v1}
  \fmfdot{v1}
  \fmf{photon}{v4,v1}
  \fmfdot{v1,v4}
  \fmf{dashes}{v4,v16}
  \fmfdot{v4,v16}
  \fmf{photon}{v64,v16}
  \fmfdot{v16,v64}
  \fmfdot{v64}
  \fmf{fermion}{v64,i2}
  \fmfdot{v64}
  \fmf{fermion,tension=0.5}{o5,v64}
  \fmfdot{v16}
  \fmf{photon,tension=0.5}{o4,v16}
  \fmfdot{v4}
  \fmf{photon,tension=0.5}{o3,v4}
  \fmfdot{v1}
  \fmf{fermion,tension=0.5}{v1,o6}
  \end{fmfgraph}}
\end{equation}
that are connected by flips from~(\ref{eq:h_flips_1}) and that have no
Higgs coupling in common. Again, in a more general Higgs sector with
$HHW$ vertices, the `Higgs conservation law' breaks down and only two
mixed groves remain.

We have checked the results of this section numerically for processes
with up to eight external particles in the SM~\cite{Schwinn:2003}, using
the optimizing matrix element generator O'Mega~\cite{OMega}. We have
considered the `gauge groves' that can be obtained by setting all
Higgs couplings to zero. As expected, this is consistent in unitarity
gauge, i.\,e.~all WIs are satisfied. In $R_\xi$ gauge, the WIs for GFs
with four external particles are satisfied while the WIs with five external
particles are violated badly. This has to be expected, since according
to the discussion in section~\ref{sec:sti-flips} the four particle gauge
boson exchange diagrams satisfy the WI, but the four point diagrams with
external GBs that appear in the five point functions violate the
WIs. Starting from the six point functions, these violations of the five
point WIs cause inconsistencies between the matrix elements in
unitarity gauge and $R_\xi$ gauge.

\section{Summary and outlook}
We have given a new proof for the formalism
of flips~\cite{Boos:1999} for the determination of gauge invariant
subsets of FDs (groves). Our proof clarifies the precise definition
of gauge flips in SBGTs that has been applied to the classification
of the GICs. We found new GICs in theories with a nonlinearly
realized scalar sector. In this case the groves in theories with only
neutral Higgs bosons can be classified according to the number of
internal Higgs boson lines. These results are also relevant for
calculations in unitarity gauge. In theories with a linearly realized
scalar sector in $R_\xi$ gauge, no additional nontrivial groves
compared to the unbroken case exist. The applications of gauge flips
to loop diagrams is currently being studied~\cite{Ondreka:2003} and
the extension of our proof to loop diagrams, using the Feynman tree
theorem~\cite{Feynman:1963} is under investigation. Our approach might
also be useful for the extensions of groves to supersymmetric
theories, using the results of~\cite{Reuter:2002}.
 
\begin{acknowledgement}
We thank D.~Ondreka for useful discussions. 

This work has been supported by the Bundesministerium f\"ur Bildung
und Forschung Germany, (05HT1RDA/6).
\end{acknowledgement}

\appendix
\section{STIs for Green's functions}
\label{app:gf-sti}
In this appendix we set up a graphical notation for BRS
transformations and STIs for GFs. We work in  a general quantum
field theory, denoting the physical fields and GBs collectively by
$\Phi$:
\begin{equation*}
\Phi=\{\psi,W^\mu,H,\phi,\dots\}
\end{equation*}
All fields including ghosts and auxiliary fields will be denoted by
$\Psi$:
\begin{equation*}
\Psi=\{\Phi,c,\bar c,B\}
\end{equation*}
The STI for GFs reads
\begin{multline}
\label{eq:local-wi}
\braket{\text{out}|\tprod{B\, \Phi_1\dots\Phi_n}\text{in}}\\
=\sum_i (\pm)\braket{\text{out}|\tprod{\bar c\,\Phi_i\dots
   \delta_{\text{BRS}}\Phi_i\dots\Phi_n(y_n)}|\text{in}}
\end{multline}
and the equation of motion for the auxiliary field $B$ is
\begin{equation}\label{eq:b-eom}
B_a=-\frac{1}{\xi}(\partial_\mu W_a^\mu-\xi m_{W_a}\phi_a)
\end{equation} 
For the Higgs and GBs in a linear representation of the symmetry, we
parametrize the BRS transformations as\footnote{We use the convention
to include the 
gauge coupling constants in the generators of the gauge transformations.}
\begin{equation}\label{eq:brs-scalar}
\begin{aligned}
\delta_{\text{BRS}} H_i&=c_c (u^c_{ib}\phi_b -T^c_{ij}H_j)\\
\delta_{\text{BRS}} \phi_a&=-m_ac_a-c_c(t^c_{ab}\phi_b+ u^c_{ai}H_i) 
\end{aligned}
\end{equation}
To represent STIs diagrammatically, we will introduce the following
graphical notation:
\begin{align}\label{eq:brs-graph}
\delta_{\text{BRS}} \Phi_i &= T^a_{ij}c_a\Phi_j: \qquad
\parbox{15\unitlength}{
\begin{fmfgraph*}(10,15)
\fmfright{g1}
\fmftop{g2}
\fmfleft{l}
\fmf{plain,label=$\Phi_j$}{g1,g2}
\fmf{ghost,label=$c_a$}{l,g1}
\fmfv{decor.shape=square,decor.filled=empty,decor.size=5,label=$T^a_{ij}$}{g1}
\end{fmfgraph*}}\\
\label{eq:brs-gauge-graph}
\delta_{\text{BRS}} W^a_\mu &= \partial_\mu c^a+f^{abc}W_b c_c: \nonumber\\
&\parbox{20\unitlength}{
\fmfframe(0,0)(5,0){
\begin{fmfgraph*}(10,15)
\fmfright{g1}
\fmfleft{g2}
\fmf{ghost,label=$c_a$,la.si=left}{g2,g1}
\fmfv{decor.shape=square,decor.filled=full,decor.size=5,label=$\ii p^\mu$,la.di=7}{g1}
\end{fmfgraph*}}}\quad +\quad
\parbox{15\unitlength}{
\fmfframe(0,0)(5,0){
\begin{fmfgraph*}(10,15)
\fmfright{g1}
\fmftop{g2}
\fmfleft{l}
\fmf{photon,label=$W_b$}{g1,g2}
\fmf{ghost,label=$c_c$}{l,g1}
\fmfv{decor.shape=square,decor.filled=empty,decor.size=5,label=$f^{abc}$}{g1}
\end{fmfgraph*}}}\\
\label{eq:brs-gold-graph}
\delta_{\text{BRS}} \phi_a&=-m_{W_a}c_a-c_c(t^c_{ab}\phi_b+ u^c_{ai}H_i): \nonumber \\
&\parbox{25\unitlength}{
\fmfframe(0,0)(10,0){
\begin{fmfgraph*}(15,15)
\fmfright{g1}
\fmfleft{g2}
\fmf{ghost,label=$c_a$,la.si=left}{g2,g1}
\fmfv{decor.shape=cross,decor.size=5,label=$(-m_{W_a})$}{g1}
\end{fmfgraph*}}}
\qquad +\quad
\parbox{20\unitlength}{
\fmfframe(0,0)(5,0){
\begin{fmfgraph*}(10,15)
\fmfright{g1}
\fmftop{g2}
\fmfleft{l}
\fmf{dashes,label=$\phi_b$}{g1,g2}
\fmf{ghost,label=$c_c$}{l,g1}
\fmfv{decor.shape=square,decor.filled=empty,decor.size=5,label=$-t^{c}_{ab}$}{g1}
\end{fmfgraph*}}}\\
&+\quad\parbox{20\unitlength}{
\fmfframe(0,0)(5,0){
\begin{fmfgraph*}(10,15)
\fmfright{g1}
\fmftop{g2}
\fmfleft{l}
\fmf{dashes,label=$H_i$}{g1,g2}
\fmf{ghost,label=$c_c$}{l,g1}
\fmfv{decor.shape=square,decor.filled=empty,decor.size=5,label=$-u^{c}_{ai}$}{g1}
\end{fmfgraph*}}}\nonumber\\
\delta_{\text{BRS}} \bar c&= B :\qquad
\parbox{15\unitlength}{
\begin{fmfgraph*}(15,15)
\fmfleft{l}
\fmfright{r}
\fmf{double,label=$B$}{l,r}
\end{fmfgraph*}}
\end{align}

Because of the nonlinearity of the BRS transformations, these
transformations receive radiative corrections. The insertion of a BRS
transformed gauge field in a GF therefore is represented as
\begin{center}
\begin{equation*}
\braket{0|\Tprod{W_\mu\bar c \delta_{\text{BRS}} W_\nu}|0}= \quad 
\parbox{15\unitlength}{
\begin{fmfgraph*}(15,15)
\fmfleft{g1,g2}
\fmfright{b}
\fmf{ghost,label=$c$}{g2,y}
\fmf{ghost}{y,g1}
\fmf{photon,label=$W$}{y,b}
\fmfblob{10}{y}
\fmfv{decor.shape=square,decor.filled=full,decor.size=5}{g1}
\end{fmfgraph*}}
\quad+\quad
\parbox{15\unitlength}{
\begin{fmfgraph}(15,15)
\fmfleft{g1,g2}
\fmfright{b}
\fmf{photon}{b,m}
\fmf{ghost}{g2,m}
\fmf{ghost,left=0.5}{m,g1}
\fmffreeze
\fmf{photon,right=0.5}{m,g1}
\fmfv{decor.shape=square,decor.filled=empty,decor.size=5}{g1}
\fmfv{decor.shape=diamond,decor.filled=empty,d.si=13}{m}
\end{fmfgraph}}
\end{equation*}
 \end{center}
The second term consists of the tree level contribution
from~(\ref{eq:brs-gauge-graph}) plus loop corrections. The tree level
contribution is a disconnected diagram, therefore we denote the GFs
with insertions of BRS-transformed fields by diamond-shaped blobs to
distinguish them from connected GFs.

Using this graphical notation, we can represent the identities~(\ref{eq:local-wi}) as
\begin{equation}
\label{eq:sti-diag}
-\quad\parbox{21\unitlength}{
\begin{fmfgraph*}(20,20)
\fmfbottomn{v}{6}
\fmftop{t}
\fmf{double,tension=6}{t,v}
\begin{fmffor}{i}{1}{1}{6}
\fmf{plain}{v[i],v}
\end{fmffor}
\fmfv{d.sh=circle,d.f=empty,d.size=25pt,l=$N$,l.d=0}{v}
\end{fmfgraph*}}\quad
=\quad\sum_{\Phi_i}\quad
\parbox{30\unitlength}{\fmfframe(5,5)(5,5){%
\begin{fmfgraph*}(20,20)
\fmfbottomn{v}{6}
\fmftop{t}
\fmf{ghost,tension=8}{t,v}
\begin{fmffor}{i}{1}{1}{4}
\fmf{plain}{v[i],v}
\end{fmffor}
\fmf{phantom}{v,v5}
\fmf{plain}{v6,v}
\fmfv{d.sh=diamond,d.f=empty,d.size=25pt}{v}
\fmf{ghost,left}{v,v5}
\fmf{plain}{v,v5}
\fmfv{label=$\Phi_i$}{v5}
\fmfv{decor.shape=square,decor.filled=empty,decor.size=5}{v5}
\end{fmfgraph*}}}
\end{equation}
The `contact terms' on the right hand side of~(\ref{eq:local-wi}) give
rise to disconnected terms.   At tree level, the contact terms can be
written as a sum over factorized \emph{connected} diagrams,
interconnected only by the BRS-vertices:
\begin{equation}
\label{eq:contact-graph} 
\parbox{30\unitlength}{\fmfframe(5,5)(5,5){%
\begin{fmfgraph*}(20,20)
\fmfbottomn{v}{6}
\fmftop{t}
\fmf{ghost,tension=8}{t,v}
\begin{fmffor}{i}{1}{1}{6}
\fmf{plain}{v[i],v}
\end{fmffor}
\fmffreeze
\fmfv{d.sh=diamond,d.f=empty,d.size=25pt,l=$N$,l.di=0}{v}
\fmf{ghost,left}{v,v4}
\fmfv{decor.shape=square,decor.filled=empty,decor.size=5,label=$\Phi_i$}{v4}
\end{fmfgraph*}}} \overset{\text{tree level}}{=}\sum_{k+l=N+1}
\parbox{30\unitlength}{\fmfframe(5,5)(5,5){%
\begin{fmfgraph*}(20,20)
\fmfbottomn{v}{6}
\fmftopn{t}{3}
\fmf{phantom,tension=4}{t1,v}
\fmf{phantom,tension=4}{t3,u}
\fmf{phantom}{u,v}
\begin{fmffor}{i}{1}{1}{3}
\fmf{plain}{v[i],v}
\fmf{plain}{v[i+3],u}
\end{fmffor}
\fmfv{d.sh=circle,d.f=empty,d.size=15pt,l=$k$,l.d=0}{v}
\fmfv{d.sh=circle,d.f=empty,d.size=15pt,l=$l$,l.d=0}{u}
\fmffreeze
\fmf{phantom}{t2,u}
\fmf{ghost}{t2,v,v4}
\fmfv{decor.shape=square,decor.filled=empty,decor.size=5,label=$\Phi_i$}{v4}
\end{fmfgraph*}}} 
\end{equation}

Since the BRS transformation of the gauge
bosons~(\ref{eq:brs-gauge-graph}) and GBs~(\ref{eq:brs-gold-graph})
is inhomogeneous, those particles have to be treated separately in the
STI~(\ref{eq:local-wi}). The graphical representation of the STI with
one external gauge boson is
\begin{multline}
\label{eq:sti-gauge}
-\parbox{21\unitlength}{ 
\quad\begin{fmfgraph}(20,20)
\fmfbottomn{v}{6}
\fmftop{t}
\fmf{double,tension=8}{t,v}
\begin{fmffor}{i}{1}{1}{4}
\fmf{plain}{v[i],v}
\end{fmffor}
\fmf{plain}{v6,v}
\fmfv{d.sh=circle,d.f=empty,d.size=25pt}{v}
\fmf{photon}{v,v5}
\end{fmfgraph}}\quad = \quad
\parbox{21\unitlength}{ 
\begin{fmfgraph}(20,20)
\fmfbottomn{v}{6}
\fmftop{t}
\fmf{ghost,tension=8}{t,v}
\begin{fmffor}{i}{1}{1}{4}
\fmf{plain}{v[i],v}
\end{fmffor}
\fmf{ghost}{v,v5}
\fmf{plain}{v6,v}
\fmfv{d.sh=circle,d.f=empty,d.size=25pt}{v}
\fmfv{decor.shape=square,decor.filled=full,decor.size=5}{v5}
\end{fmfgraph}}\quad\\
 +\quad\sum_{\Phi_i}\qquad 
\parbox{21\unitlength}{ 
\begin{fmfgraph}(20,20)
\fmfbottomn{v}{6}
\fmftop{t}
\fmf{ghost,tension=8}{t,v}
\begin{fmffor}{i}{1}{1}{4}
\fmf{plain}{v[i],v}
\end{fmffor}
\fmf{plain}{v6,v}
\fmfv{d.sh=diamond,d.f=empty,d.size=25pt}{v}
\fmf{ghost,left}{v,v5}
\fmf{photon}{v,v5}
\fmfv{decor.shape=square,decor.filled=empty,decor.size=5}{v5}
\end{fmfgraph}}
\end{multline}
with an obvious generalization to more external gauge bosons and
GBs.

\section{Explicit form of flips}
\label{app:flips}
In this appendix we summarize the flips for SBGTs, both in the linear
and  nonlinear representation of the scalar sector.

\subsection{Gauge flips}
 
\begin{subequations}
\label{subeq:h_flips}

The gauge flips for the four gauge boson function in the linear
realization are:
\begin{multline}
\label{eq:gauge_flips1}
{\tilde G}_4 = \{ {\tilde G}_4^i | i = 1,\dots,7 \}=\\ 
\left\{
\begin{aligned}
&\parbox{15\unitlength}{
\begin{fmfgraph}(15,15)
\fmfleft{a,b}
\fmfright{f1,f2}
\fmf{photon}{a,fwf1}
\fmf{photon}{fwf1,fwf2}
\fmf{photon}{fwf2,b}
\fmf{photon}{fwf1,f1}
\fmf{photon}{fwf2,f2}
\fmfdot{fwf1}
\fmfdot{fwf2}
\end{fmfgraph}}\,,\,
\parbox{15\unitlength}{
\begin{fmfgraph}(15,15)
\fmfleft{a,b}
\fmfright{f1,f2}
\fmf{photon}{a,fwf1}
\fmf{photon}{fwf1,fwf2}
\fmf{photon}{fwf2,b}
\fmf{phantom}{fwf2,f2}
\fmf{phantom}{fwf1,f1}
\fmffreeze
\fmf{photon}{fwf2,f1}
\fmf{photon}{fwf1,f2}
\fmfdot{fwf1}
\fmfdot{fwf2}
\end{fmfgraph}}\, ,\, 
\parbox{15\unitlength}{
\begin{fmfgraph}(15,15)
\fmfleft{a,b}
\fmfright{f1,f2}
\fmf{photon}{a,fwf}
\fmf{photon}{fwf,b}
\fmf{photon}{fwf,www}
\fmf{photon}{www,f1}
\fmf{photon}{www,f2}
\fmfdot{fwf}
\fmfdot{www}
\end{fmfgraph}}\, , \,
\parbox{15\unitlength}{
\begin{fmfgraph}(15,15)
\fmfleft{a,b}
\fmfright{f1,f2}
\fmf{photon}{a,c}
\fmf{photon}{c,b}
\fmf{photon}{c,f1}
\fmf{photon}{c,f2}
\fmfdot{c}
\end{fmfgraph}}\,,\\
&\parbox{15\unitlength}{
\begin{fmfgraph}(15,15)
\fmfleft{a,b}
\fmfright{f1,f2}
\fmf{photon}{a,fwf1}
\fmf{dashes}{fwf1,fwf2}
\fmf{photon}{fwf2,b}
\fmf{photon}{fwf1,f1}
\fmf{photon}{fwf2,f2}
\fmfdot{fwf1}
\fmfdot{fwf2}
\end{fmfgraph}}\,,\,
\parbox{15\unitlength}{
\begin{fmfgraph}(15,15)
\fmfleft{a,b}
\fmfright{f1,f2}
\fmf{photon}{a,fwf1}
\fmf{dashes}{fwf1,fwf2}
\fmf{photon}{fwf2,b}
\fmf{phantom}{fwf2,f2}
\fmf{phantom}{fwf1,f1}
\fmffreeze
\fmf{photon}{fwf2,f1}
\fmf{photon}{fwf1,f2}
\fmfdot{fwf1}
\fmfdot{fwf2}
\end{fmfgraph}}\, , \,
\parbox{15\unitlength}{
\begin{fmfgraph}(15,15)
\fmfleft{a,b}
\fmfright{f1,f2}
\fmf{photon}{a,fwf}
\fmf{photon}{fwf,b}
\fmf{dashes}{fwf,www}
\fmf{photon}{www,f1}
\fmf{photon}{www,f2}
\fmfdot{fwf}
\fmfdot{www}
\end{fmfgraph}}
\end{aligned}
\right\}
\end{multline}
For nonlinear realizations, the Higgs exchange diagrams ${\tilde
G}_4^5$, ${\tilde G}_4^6$ and~${\tilde G}_4^7$ are not present in the
gauge flips.

The flips for $\bar f f \to WW$ are in the linear representation:
\begin{multline}
{\tilde G}_{4,2F} = \{ {\tilde G}_{4,2F}^i | i = 1,2,3,4 \}=\\ 
\left\{
\parbox{15\unitlength}{
\begin{fmfgraph}(15,15)
\fmfleft{a,b}
\fmfright{f1,f2}
\fmf{fermion}{a,fwf1}
\fmf{fermion}{fwf1,fwf2}
\fmf{fermion}{fwf2,b}
\fmf{photon}{fwf1,f1}
\fmf{photon}{fwf2,f2}
\fmfdot{fwf1,fwf2}
\end{fmfgraph}}\,,\,
\parbox{15\unitlength}{
\begin{fmfgraph}(15,15)
\fmfleft{a,b}
\fmfright{f1,f2}
\fmf{fermion}{a,fwf1}
\fmf{fermion}{fwf1,fwf2}
\fmf{fermion}{fwf2,b}
\fmf{phantom}{fwf2,f2}
\fmf{phantom}{fwf1,f1}
\fmffreeze
\fmf{photon}{fwf2,f1}
\fmf{photon}{fwf1,f2}
\fmfdot{fwf1,fwf2}
\end{fmfgraph}}\,,\, 
\parbox{15\unitlength}{
\begin{fmfgraph}(15,15)
\fmfleft{a,b}
\fmfright{f1,f2}
\fmf{fermion}{a,fwf}
\fmf{fermion}{fwf,b}
\fmf{photon}{fwf,www}
\fmf{photon}{www,f1}
\fmf{photon}{www,f2}
\fmfdot{www,fwf}
\end{fmfgraph}}\, ,\,
\parbox{15\unitlength}{
\begin{fmfgraph}(15,15)
\fmfright{a,b}
\fmfleft{f1,f2}
\fmf{photon}{a,fwf}
\fmf{photon}{fwf,b}
\fmf{dashes}{fwf,www}
\fmf{fermion}{www,f1}
\fmf{fermion}{f2,www}
\fmfdot{www}
\fmfdot{fwf}
\end{fmfgraph}}\right \}
\end{multline}
Again, the Higgs exchange diagram~${\tilde G}_{4,2F}^4$
is not present for nonlinear symmetries. 

The gauge $\bar f f \to W H$ flips are for linear and nonlinear realizations:
\begin{multline}
{\tilde G}_{4,1H2F} = \{ {\tilde G}_{4,1H2F}^i | i = 1,2,3,4 \}=\\ 
\left\{
\parbox{15\unitlength}{
\begin{fmfgraph}(15,15)
\fmfleft{a,b}
\fmfright{f1,f2}
\fmf{fermion}{a,fwf1}
\fmf{fermion}{fwf1,fwf2}
\fmf{fermion}{fwf2,b}
\fmf{photon}{fwf1,f1}
\fmf{dashes}{fwf2,f2}
\fmfdot{fwf1,fwf2}
\end{fmfgraph}}\, ,\,
\parbox{15\unitlength}{
\begin{fmfgraph}(15,15)
\fmfleft{a,b}
\fmfright{f1,f2}
\fmf{fermion}{a,fwf1}
\fmf{fermion}{fwf1,fwf2}
\fmf{fermion}{fwf2,b}
\fmf{phantom}{fwf2,f2}
\fmf{phantom}{fwf1,f1}
\fmffreeze
\fmf{photon}{fwf2,f1}
\fmf{dashes}{fwf1,f2}
\fmfdot{fwf1,fwf2}
\end{fmfgraph}}\, ,\, 
\parbox{15\unitlength}{
\begin{fmfgraph}(15,15)
\fmfleft{a,b}
\fmfright{f1,f2}
\fmf{fermion}{a,fwf}
\fmf{fermion}{fwf,b}
\fmf{photon}{fwf,www}
\fmf{photon}{www,f1}
\fmf{dashes}{www,f2}
\fmfdot{www,fwf}
\end{fmfgraph}}\, , \,
\parbox{15\unitlength}{
\begin{fmfgraph}(15,15)
\fmfleft{a,b}
\fmfright{f1,f2}
\fmf{fermion}{a,fwf}
\fmf{fermion}{fwf,b}
\fmf{dashes}{fwf,www}
\fmf{photon}{www,f1}
\fmf{dashes}{www,f2}
\fmfdot{www,fwf}
\end{fmfgraph}}
\right \} 
\end{multline}
A new feature in SBGTs are the $3W$-$1H$ flips that have the same form in
linear and nonlinear realizations:
\begin{multline}
{\tilde G}_{4,1H} = \{ {\tilde G}_{4,1H}^i | i = 1,\dots,6 \}=\\ 
\left\{\begin{aligned}
&\parbox{15\unitlength}{
\begin{fmfgraph}(15,15)
\fmfleft{a,b}
\fmfright{f1,f2}
\fmf{photon}{a,fwf1}
\fmf{photon}{fwf1,fwf2}
\fmf{photon}{fwf2,b}
\fmf{dashes}{fwf1,f1}
\fmf{photon}{fwf2,f2}
\fmfdot{fwf1}
\fmfdot{fwf2}
\end{fmfgraph}},
\parbox{15\unitlength}{
\begin{fmfgraph}(15,15)
\fmfleft{a,b}
\fmfright{f1,f2}
\fmf{photon}{a,fwf1}
\fmf{photon}{fwf1,fwf2}
\fmf{photon}{fwf2,b}
\fmf{phantom}{fwf2,f2}
\fmf{phantom}{fwf1,f1}
\fmffreeze
\fmf{dashes}{fwf2,f1}
\fmf{photon}{fwf1,f2}
\fmfdot{fwf1}
\fmfdot{fwf2}
\end{fmfgraph}}, 
\parbox{15\unitlength}{
\begin{fmfgraph}(15,15)
\fmfleft{a,b}
\fmfright{f1,f2}
\fmf{photon}{a,fwf}
\fmf{photon}{fwf,b}
\fmf{photon}{fwf,www}
\fmf{dashes}{www,f1}
\fmf{photon}{www,f2}
\fmfdot{www}
\fmfdot{fwf}
\end{fmfgraph}}\\
&
\parbox{15\unitlength}{
\begin{fmfgraph}(15,15)
\fmfleft{a,b}
\fmfright{f1,f2}
\fmf{photon}{a,fwf1}
\fmf{dashes}{fwf1,fwf2}
\fmf{photon}{fwf2,b}
\fmf{dashes}{fwf1,f1}
\fmf{photon}{fwf2,f2}
\fmfdot{fwf1}
\fmfdot{fwf2}
\end{fmfgraph}},
\parbox{15\unitlength}{
\begin{fmfgraph}(15,15)
\fmfleft{a,b}
\fmfright{f1,f2}
\fmf{photon}{a,fwf1}
\fmf{dashes}{fwf1,fwf2}
\fmf{photon}{fwf2,b}
\fmf{phantom}{fwf2,f2}
\fmf{phantom}{fwf1,f1}
\fmffreeze
\fmf{dashes}{fwf2,f1}
\fmf{photon}{fwf1,f2}
\fmfdot{fwf1}
\fmfdot{fwf2}
\end{fmfgraph}}, 
\parbox{15\unitlength}{
\begin{fmfgraph}(15,15)
\fmfleft{a,b}
\fmfright{f1,f2}
\fmf{photon}{a,fwf}
\fmf{photon}{fwf,b}
\fmf{dashes}{fwf,www}
\fmf{dashes}{www,f1}
\fmf{photon}{www,f2}
\fmfdot{www}
\fmfdot{fwf}
\end{fmfgraph}}
       \end{aligned}
\right \}
\end{multline}
The $2W2H$ flips are for linear symmetries:
\begin{multline}\label{eq:hhww-flips}
{\tilde G}_{4,2H} = \{ {\tilde G}_{4,2H}^i | i = 1,\dots,7 \}=\\ 
\left\{
\begin{aligned}
&\parbox{15\unitlength}{
\begin{fmfgraph}(15,15)
\fmfleft{a,b}
\fmfright{f1,f2}
\fmf{photon}{a,fwf1}
\fmf{photon}{fwf1,fwf2}
\fmf{photon}{fwf2,b}
\fmf{dashes}{fwf1,f1}
\fmf{dashes}{fwf2,f2}
\fmfdot{fwf1}
\fmfdot{fwf2}
\end{fmfgraph}},
\parbox{15\unitlength}{
\begin{fmfgraph}(15,15)
\fmfleft{a,b}
\fmfright{f1,f2}
\fmf{photon}{a,fwf1}
\fmf{photon}{fwf1,fwf2}
\fmf{photon}{fwf2,b}
\fmf{phantom}{fwf2,f2}
\fmf{phantom}{fwf1,f1}
\fmffreeze
\fmf{dashes}{fwf2,f1}
\fmf{dashes}{fwf1,f2}
\fmfdot{fwf1}
\fmfdot{fwf2}
\end{fmfgraph}},\parbox{15\unitlength}{
\begin{fmfgraph}(15,15)
\fmfleft{a,b}
\fmfright{f1,f2}
\fmf{photon}{a,fwf1}
\fmf{dashes}{fwf1,fwf2}
\fmf{photon}{fwf2,b}
\fmf{dashes}{fwf1,f1}
\fmf{dashes}{fwf2,f2}
\fmfdot{fwf1}
\fmfdot{fwf2}
\end{fmfgraph}} , 
\parbox{15\unitlength}{
\begin{fmfgraph}(15,15)
\fmfleft{a,b}
\fmfright{f1,f2}
\fmf{photon}{a,fwf1}
\fmf{dashes}{fwf1,fwf2}
\fmf{photon}{fwf2,b}
\fmf{phantom}{fwf2,f2}
\fmf{phantom}{fwf1,f1}
\fmffreeze
\fmf{dashes}{fwf2,f1}
\fmf{dashes}{fwf1,f2}
\fmfdot{fwf1}
\fmfdot{fwf2}
\end{fmfgraph}}\\
&\parbox{15\unitlength}{
\begin{fmfgraph}(15,15)
\fmfleft{a,b}
\fmfright{f1,f2}
\fmf{photon}{a,i1}
\fmf{photon}{i1,b}
\fmf{photon}{i1,i2}
\fmf{dashes}{i2,f1}
\fmf{dashes}{i2,f2}
\fmfdot{i1,i2}
\end{fmfgraph}} , 
\parbox{15\unitlength}{
\begin{fmfgraph}(15,15)
\fmfleft{a,b}
\fmfright{f1,f2}
\fmf{photon}{a,i1}
\fmf{photon}{i1,b}
\fmf{dashes}{i1,i2}
\fmf{dashes}{i2,f1}
\fmf{dashes}{i2,f2}
\fmfdot{i1,i2}
\end{fmfgraph}} , 
\parbox{15\unitlength}{
\begin{fmfgraph}(15,15)
\fmfleft{a,b}
\fmfright{f1,f2}
\fmf{photon}{a,c}
\fmf{photon}{c,b}
\fmf{dashes}{c,f1}
\fmf{dashes}{c,f2}
\fmfdot{c}
\end{fmfgraph}}
\end{aligned}
\right \}
\end{multline}
\end{subequations}
Here the diagram~${\tilde G}_{4,2H}^6$
is not included in the gauge flips for nonlinear realizations of the
symmetry.  

Finally we have  the $3HW$ flips that again have the same form in
linear and nonlinear realizations:
\begin{multline}
{\tilde G}_{4,3H} = \{ {\tilde G}_{4,3H}^i | i = 1,\dots,6 \}=\\ 
\left\{\begin{aligned}
&\parbox{15\unitlength}{
\begin{fmfgraph}(15,15)
\fmfleft{a,b}
\fmfright{f1,f2}
\fmf{dashes}{a,fwf1}
\fmf{dashes}{fwf1,fwf2}
\fmf{dashes}{fwf2,b}
\fmf{photon}{fwf1,f1}
\fmf{dashes}{fwf2,f2}
\fmfdot{fwf1}
\fmfdot{fwf2}
\end{fmfgraph}},
\parbox{15\unitlength}{
\begin{fmfgraph}(15,15)
\fmfleft{a,b}
\fmfright{f1,f2}
\fmf{dashes}{a,fwf1}
\fmf{dashes}{fwf1,fwf2}
\fmf{dashes}{fwf2,b}
\fmf{phantom}{fwf2,f2}
\fmf{phantom}{fwf1,f1}
\fmffreeze
\fmf{photon}{fwf2,f1}
\fmf{dashes}{fwf1,f2}
\fmfdot{fwf1}
\fmfdot{fwf2}
\end{fmfgraph}}, 
\parbox{15\unitlength}{
\begin{fmfgraph}(15,15)
\fmfleft{a,b}
\fmfright{f1,f2}
\fmf{dashes}{a,fwf}
\fmf{dashes}{fwf,b}
\fmf{dashes}{fwf,www}
\fmf{photon}{www,f1}
\fmf{dashes}{www,f2}
\fmfdot{www}
\fmfdot{fwf}
\end{fmfgraph}}\\
&
\parbox{15\unitlength}{
\begin{fmfgraph}(15,15)
\fmfleft{a,b}
\fmfright{f1,f2}
\fmf{dashes}{a,fwf1}
\fmf{photon}{fwf1,fwf2}
\fmf{dashes}{fwf2,b}
\fmf{photon}{fwf1,f1}
\fmf{dashes}{fwf2,f2}
\fmfdot{fwf1}
\fmfdot{fwf2}
\end{fmfgraph}},
\parbox{15\unitlength}{
\begin{fmfgraph}(15,15)
\fmfleft{a,b}
\fmfright{f1,f2}
\fmf{dashes}{a,fwf1}
\fmf{photon}{fwf1,fwf2}
\fmf{dashes}{fwf2,b}
\fmf{phantom}{fwf2,f2}
\fmf{phantom}{fwf1,f1}
\fmffreeze
\fmf{photon}{fwf2,f1}
\fmf{dashes}{fwf1,f2}
\fmfdot{fwf1}
\fmfdot{fwf2}
\end{fmfgraph}}, 
\parbox{15\unitlength}{
\begin{fmfgraph}(15,15)
\fmfleft{a,b}
\fmfright{f1,f2}
\fmf{dashes}{a,fwf}
\fmf{dashes}{fwf,b}
\fmf{photon}{fwf,www}
\fmf{photon}{www,f1}
\fmf{dashes}{www,f2}
\fmfdot{www}
\fmfdot{fwf}
\end{fmfgraph}}
       \end{aligned}
\right \}
\end{multline}

\subsection{Flavor and Higgs flips}
Higgs exchange has to be included in the flavor flips so they are given by
\begin{multline}
{\tilde F}_4 = \{ {\tilde F}_4^i | i = 1,\dots,6 \}=\\ 
\left\{
\begin{aligned}
\parbox{15\unitlength}{
\begin{fmfgraph}(15,15)
\fmfleft{a,b}
\fmfright{f1,f2}
\fmf{fermion}{a,fwf1}
\fmf{photon}{fwf1,fwf2}
\fmf{fermion}{b,fwf2}
\fmf{fermion}{fwf1,f1}
\fmf{fermion}{fwf2,f2}
\fmfdot{fwf1}
\fmfdot{fwf2}
\end{fmfgraph}},
\parbox{15\unitlength}{
\begin{fmfgraph}(15,15)
\fmfleft{a,b}
\fmfright{f1,f2}
\fmf{fermion}{a,fwf1}
\fmf{photon}{fwf1,fwf2}
\fmf{fermion}{b,fwf2}
\fmf{phantom}{fwf2,f2}
\fmf{phantom}{fwf1,f1}
\fmffreeze
\fmf{fermion}{fwf2,f1}
\fmf{fermion}{fwf1,f2}
\fmfdot{fwf1}
\fmfdot{fwf2}
\end{fmfgraph}} , 
\parbox{15\unitlength}{
\begin{fmfgraph}(15,15)
\fmfleft{a,b}
\fmfright{f1,f2}
\fmf{fermion}{a,fwf}
\fmf{fermion}{fwf,b}
\fmf{photon}{fwf,www}
\fmf{fermion}{f1,www}
\fmf{fermion}{www,f2}
\fmfdot{fwf}
\fmfdot{www}
\end{fmfgraph}}\\
\parbox{15\unitlength}{
\begin{fmfgraph}(15,15)
\fmfleft{a,b}
\fmfright{f1,f2}
\fmf{fermion}{a,fwf1}
\fmf{dashes}{fwf1,fwf2}
\fmf{fermion}{b,fwf2}
\fmf{fermion}{fwf1,f1}
\fmf{fermion}{fwf2,f2}
\fmfdot{fwf1}
\fmfdot{fwf2}
\end{fmfgraph}},
\parbox{15\unitlength}{
\begin{fmfgraph}(15,15)
\fmfleft{a,b}
\fmfright{f1,f2}
\fmf{fermion}{a,fwf1}
\fmf{dashes}{fwf1,fwf2}
\fmf{fermion}{b,fwf2}
\fmf{phantom}{fwf2,f2}
\fmf{phantom}{fwf1,f1}
\fmffreeze
\fmf{fermion}{fwf2,f1}
\fmf{fermion}{fwf1,f2}
\fmfdot{fwf1}
\fmfdot{fwf2}
\end{fmfgraph}} , 
\parbox{15\unitlength}{
\begin{fmfgraph}(15,15)
\fmfleft{a,b}
\fmfright{f1,f2}
\fmf{fermion}{a,fwf}
\fmf{fermion}{fwf,b}
\fmf{dashes}{fwf,www}
\fmf{fermion}{f1,www}
\fmf{fermion}{www,f2}
\fmfdot{fwf}
\fmfdot{www}
\end{fmfgraph}}
\end{aligned}
\right \}
\end{multline}
Finally there are `Higgs flips' for diagrams without external gauge
bosons that are gauge parameter independent by themselves:
\begin{subequations}
\label{subeq:higgs_flips}
\begin{multline}
{\tilde H}_4 = \{ {\tilde H}_4^i | i = 1,\dots,7 \} = \\
\left\{
\begin{aligned}
&\parbox{15\unitlength}{
\begin{fmfgraph}(15,15)
\fmfleft{a,b}
\fmfright{f1,f2}
\fmf{dashes}{a,fwf1}
\fmf{photon}{fwf1,fwf2}
\fmf{dashes}{fwf2,b}
\fmf{dashes}{fwf1,f1}
\fmf{dashes}{fwf2,f2}
\fmfdot{fwf1}
\fmfdot{fwf2}
\end{fmfgraph}},
\parbox{15\unitlength}{
\begin{fmfgraph}(15,15)
\fmfleft{a,b}
\fmfright{f1,f2}
\fmf{dashes}{a,fwf1}
\fmf{photon}{fwf1,fwf2}
\fmf{dashes}{fwf2,b}
\fmf{phantom}{fwf2,f2}
\fmf{phantom}{fwf1,f1}
\fmffreeze
\fmf{dashes}{fwf2,f1}
\fmf{dashes}{fwf1,f2}
\fmfdot{fwf1}
\fmfdot{fwf2}
\end{fmfgraph}} , 
\parbox{15\unitlength}{
\begin{fmfgraph}(15,15)
\fmfleft{a,b}
\fmfright{f1,f2}
\fmf{dashes}{a,fwf}
\fmf{dashes}{fwf,b}
\fmf{photon}{fwf,www}
\fmf{dashes}{www,f1}
\fmf{dashes}{www,f2}
\fmfdot{fwf}
\fmfdot{www}
\end{fmfgraph}} ,  
\parbox{15\unitlength}{
\begin{fmfgraph}(15,15)
\fmfleft{a,b}
\fmfright{f1,f2}
\fmf{dashes}{a,c}
\fmf{dashes}{c,b}
\fmf{dashes}{c,f1}
\fmf{dashes}{c,f2}
\fmfdot{c}
\end{fmfgraph}}\\
&\parbox{15\unitlength}{
\begin{fmfgraph}(15,15)
\fmfleft{a,b}
\fmfright{f1,f2}
\fmf{dashes}{a,fwf1}
\fmf{dashes}{fwf1,fwf2}
\fmf{dashes}{fwf2,b}
\fmf{dashes}{fwf1,f1}
\fmf{dashes}{fwf2,f2}
\fmfdot{fwf1}
\fmfdot{fwf2}
\end{fmfgraph}},
\parbox{15\unitlength}{
\begin{fmfgraph}(15,15)
\fmfleft{a,b}
\fmfright{f1,f2}
\fmf{dashes}{a,fwf1}
\fmf{dashes}{fwf1,fwf2}
\fmf{dashes}{fwf2,b}
\fmf{phantom}{fwf2,f2}
\fmf{phantom}{fwf1,f1}
\fmffreeze
\fmf{dashes}{fwf2,f1}
\fmf{dashes}{fwf1,f2}
\fmfdot{fwf1}
\fmfdot{fwf2}
\end{fmfgraph}} ,  
\parbox{15\unitlength}{
\begin{fmfgraph}(15,15)
\fmfleft{a,b}
\fmfright{f1,f2}
\fmf{dashes}{a,fwf}
\fmf{dashes}{fwf,b}
\fmf{dashes}{fwf,www}
\fmf{dashes}{www,f1}
\fmf{dashes}{www,f2}
\fmfdot{fwf}
\fmfdot{www}
\end{fmfgraph}}
\end{aligned}
\right \}
\end{multline}
and
\begin{multline}
\tilde H_{4,2F} = \{ \tilde H_{4,2F}^i | i = 1,2,3,4 \} = \\
\left\{
\parbox{15\unitlength}{
\begin{fmfgraph}(15,15)
\fmfleft{a,b}
\fmfright{f1,f2}
\fmf{fermion}{a,fwf1}
\fmf{fermion}{fwf1,fwf2}
\fmf{fermion}{fwf2,b}
\fmf{dashes}{fwf1,f1}
\fmf{dashes}{fwf2,f2}
\fmfdot{fwf1,fwf2}
\end{fmfgraph}}\, ,\,
\parbox{15\unitlength}{
\begin{fmfgraph}(15,15)
\fmfleft{a,b}
\fmfright{f1,f2}
\fmf{fermion}{a,fwf1}
\fmf{fermion}{fwf1,fwf2}
\fmf{fermion}{fwf2,b}
\fmf{phantom}{fwf2,f2}
\fmf{phantom}{fwf1,f1}
\fmffreeze
\fmf{dashes}{fwf2,f1}
\fmf{dashes}{fwf1,f2}
\fmfdot{fwf1,fwf2}
\end{fmfgraph}}\, ,\, 
\parbox{15\unitlength}{
\begin{fmfgraph}(15,15)
\fmfleft{a,b}
\fmfright{f1,f2}
\fmf{fermion}{a,fwf}
\fmf{fermion}{fwf,b}
\fmf{photon}{fwf,www}
\fmf{dashes}{www,f1}
\fmf{dashes}{www,f2}
\fmfdot{www,fwf}
\end{fmfgraph}}\, , \,
\parbox{15\unitlength}{
\begin{fmfgraph}(15,15)
\fmfleft{a,b}
\fmfright{f1,f2}
\fmf{fermion}{a,fwf}
\fmf{fermion}{fwf,b}
\fmf{dashes}{fwf,www}
\fmf{dashes}{www,f1}
\fmf{dashes}{www,f2}
\fmfdot{www,fwf}
\end{fmfgraph}}
\right \} 
\end{multline}
\end{subequations}


\end{fmffile}    
\end{document}